\documentclass{appolb}
\usepackage{cite,mcite}
\usepackage{graphicx}
\usepackage{atlasphysics}
\usepackage{xspace}
\usepackage{xcolor}
\usepackage{longtable}
\usepackage{booktabs}
\usepackage{multirow}
\usepackage{relsize}
\usepackage[small,nooneline,center]{subfigure}

\makeatletter
\g@addto@macro\bfseries{\boldmath}
\makeatother

\usepackage{amssymb}
\usepackage{amsmath}

\usepackage{setspace}
\usepackage{threeparttable}
\definecolor{todoboxcolor}{rgb}{1.0,0.6,0.6}

\definecolor{todoneboxcolor}{rgb}{1.0,0.9,0.9}

\newlength{\capindent}
\newlength{\capwidth}
\newlength{\figwidth}
\newcommand{\icaption}[2][!*!,!]{\hspace*{\capindent}%
  \begin{minipage}{\capwidth}
    \ifthenelse{\equal{#1}{!*!,!}}%
      {\caption{#2}}%
      {\caption[#1]{#2}}
      \vspace*{3mm}
  \end{minipage}}
\def\jimmy{\textsc{Jimmy}\xspace}
\def\herwig{\textsc{Herwig}\xspace}
\def\herwigjimmy{\herwig/\jimmy}
\def\pythia{\textsc{Pythia}\xspace}
\def\professor{\textsc{Professor}\xspace}
\def\rivet{Rivet\xspace}

\def\atlas{ATLAS\xspace}

\def\pythiasix{\pythia~\!6\xspace}
\def\pythiaeight{Pythia\,8\xspace}
\def\phojet{Phojet\xspace}

\def\herwigpp{Herwig\raisebox{0.2ex}{\smaller++}\xspace}

\def\rivet{Rivet\xspace}
\def\professor{Professor\xspace}

\def\ptmin{\ensuremath{p_\perp^\text{min}}\xspace}

\def\pt{\ensuremath{p_\perp}\xspace}

\def\qqbar{\Pquark-\APquark}

\def\sqrts{\ensuremath{\sqrt{s}}\xspace}

\DeclareRobustCommand{\lostar}[1]{\mbox{\ensuremath{\text{LO}\mspace{-0.5mu}*}}\xspace}
\DeclareRobustCommand{\lostst}[1]{\mbox{\ensuremath{\text{LO}\mspace{-0.5mu}*\mspace{-0.5mu}*}}\xspace}

\usepackage{bm}
\def\vec#1{\bm{#1}}

\usepackage{url}
\usepackage[section]{placeins}

\title{Soft QCD in ATLAS: measurements and\\ modelling of multi-parton interactions}
\author{Andy~Buckley\\PPE Group, School of Physics and Astronomy,\\ University of Edinburgh}
\begin{document}
\maketitle
\thispagestyle{empty}

\begin{abstract}
  Soft QCD contributes to all observables at the LHC, due to the presences of
  underlying event (UE) and pile-up in all events. Both these processes are
  dominated by multi-parton interactions (MPI), i.e. the result of proton
  collisions containing more than one partonic interaction due to collective and
  beam remnant effects. While there is undoubtedly interesting physics involved
  in MPI, the primary interest of LHC experiments is to characterise and model
  the behaviour of UE and pile-up sufficiently well that their influence may be
  cleanly subtracted in the process of searching for new physics signatures at
  7~TeV and beyond. I summarise the soft QCD measurements made by ATLAS using
  the 2010 and early 2011 datasets, and the use of this data to improve Monte
  Carlo generator models of MPI for use in forthcoming simulation campaigns.
\end{abstract}

\section{Introduction}

A consequence of doing physics at a hadron collider is that one has to
understand the incoming hadrons rather well. A multitude of LHC ``new physics''
processes are illustrated by diagrams in which a pair of partons neatly extract
themselves from their parent protons without consequence, the only inconvenience
being the loss of longitudinal momentum information thanks to the probabilistic
nature of parton distributions. However, real life is not so clean, in
particular because the rest of the proton constituents and the beam remnants
left after parton extraction cannot be so easily ignored. Including such effects
leads to a model in which multiple partonic interactions may occur in each
event, and where the influence of the colour charge flows associated with those
multiple scatterings and beam remnants can also be substantial.

Of course, the LHC was not built to run at 7~TeV (and eventually 14~TeV) to
provide greater insight into the soft structure of protons -- although that will
be a welcome consequence of its existence. The main reason for LHC
collaborations to be interested in multi-parton interactions is that they form
troublesome backgrounds in the core LHC task of searching for signatures of new
physics. These backgrounds occur in two forms: first, the additional parton
interactions in hard processes adds activity to that event. This additional
activity is, as a consequence of relative cross-sections, predominantly QCD
based and can change the energies and distributions of QCD jets, add new jets,
and fake electron signatures: this is the ``underlying event'' (UE).

The second way in which MPI can affect hard-process physics searches is via
pile-up: the overlay of multiple $pp$ interactions in a single
bunch-crossing. Again as a consequence of relative cross-sections, pile-up
events are overwhelmingly dominated by soft QCD scattering (minimum bias), and
are typically modelled as pure MPI scatterings with no ``hard'' process. As LHC
luminosities increase, the mean number of pile-up interactions per bunch
crossing (assuming a Poisson distribution) also increases from $\mu \sim 0.1$ in
the earliest LHC runs, to $\mu \sim 10$ in early 2011, $\mu \sim 30$ at the end
of the 2011 run, and eventually $\mu \gtrsim 100$ in the LHC luminosity upgrade
scenario. A typical \pt density contribution of 1\;GeV per unit in $\eta-\phi$
for each pile-up event can significantly change event characteristics at high
$\mu$.  Unlike UE, pile-up can be directly reduced by use of track-to-vertex
matching; this is far from a panacea, however, as the reduction of tracking
efficiency and increase in overlapping primary vertices at high
occupancy. Additionally, the need to account for the uncharged component of
pile-up activity and to use calorimeter elements out of the tracking acceptance
means that pile-up must be understood well enough that subtraction can be
attempted.

So both pile-up and UE at the LHC require a level of understanding of multiple
interactions in proton--proton collisions that was not established at the
Tevatron. In the remainder of this contribution I summarise the current state of
MPI modelling in Monte Carlo event generators and the degrees of freedom in
these models, present the ATLAS QCD measurements which are most useful in
constraining their free parameters, and show the resulting MC tunings which will
be used in the next year of ATLAS (and other) physics studies.

\section{Monte Carlo modelling of soft QCD}

MC event generators are a crucial tool for experimental particle physics.
Despite occasional protestations to the contrary, it is sufficiently difficult
to cleanly disentangle contributing processes in an experimental analysis that
some degree of reliance on simulation -- of Standard Model processes at least --
is virtually inevitable. Even ``data-driven'' background estimations often use
the data to fix the normalisation of simulated distributions.

\subsection{General-purpose MC event generators}

Event generators come in several forms, from parton-level codes which calculate
only total or perhaps differential cross-sections, to ``general-purpose'' codes
in which the partons are hadronised, hadrons are decayed, and MPI effects are
simulated. The real power of general-purpose generators from an
experimentalist's point of view is that they do not just produce asymptotic
distributions, but that they typically use sampling to produce physical-looking
simulated events. Fully exclusive simulation of this kind is key to the design
of detectors and analyses, and the \emph{unfolding} of detector effects on
measured observables~\cite{Wynne:1357548}.

In addition to MPI simulation, general-purpose generators enhance the matrix
element and phase space sampling of parton-level generators by using (matrix
element matched) perturbative parton showers to stochastically approximate full
QCD resummation, and by use of non-perturbative hadronisation models and hadron
decays to produce realistic particle kinematics and identified
multiplicities. As MPI and hadronisation are modelled phenomenologically rather
than from first principles, the bulk of event generator free parameters are
concentrated in these modelling areas. The process of parameter optimisation by
comparison to experimental data is known as ``tuning''.

\subsection{MPI modelling in general-purpose generators}
\label{sec:mcmpi}

We already mentioned in the introduction an unrealistic (but popular!)
simplified model of hadron collisions in which only a pair of asymptotically
free partons from the two incoming protons interact in hard scattering. In this
model, the only theoretical concession to proton structure is that the incoming
parton energies are given by parton density functions (PDFs).

\begin{figure}[tp]
  \centering
  \includegraphics[width=0.6\textwidth]{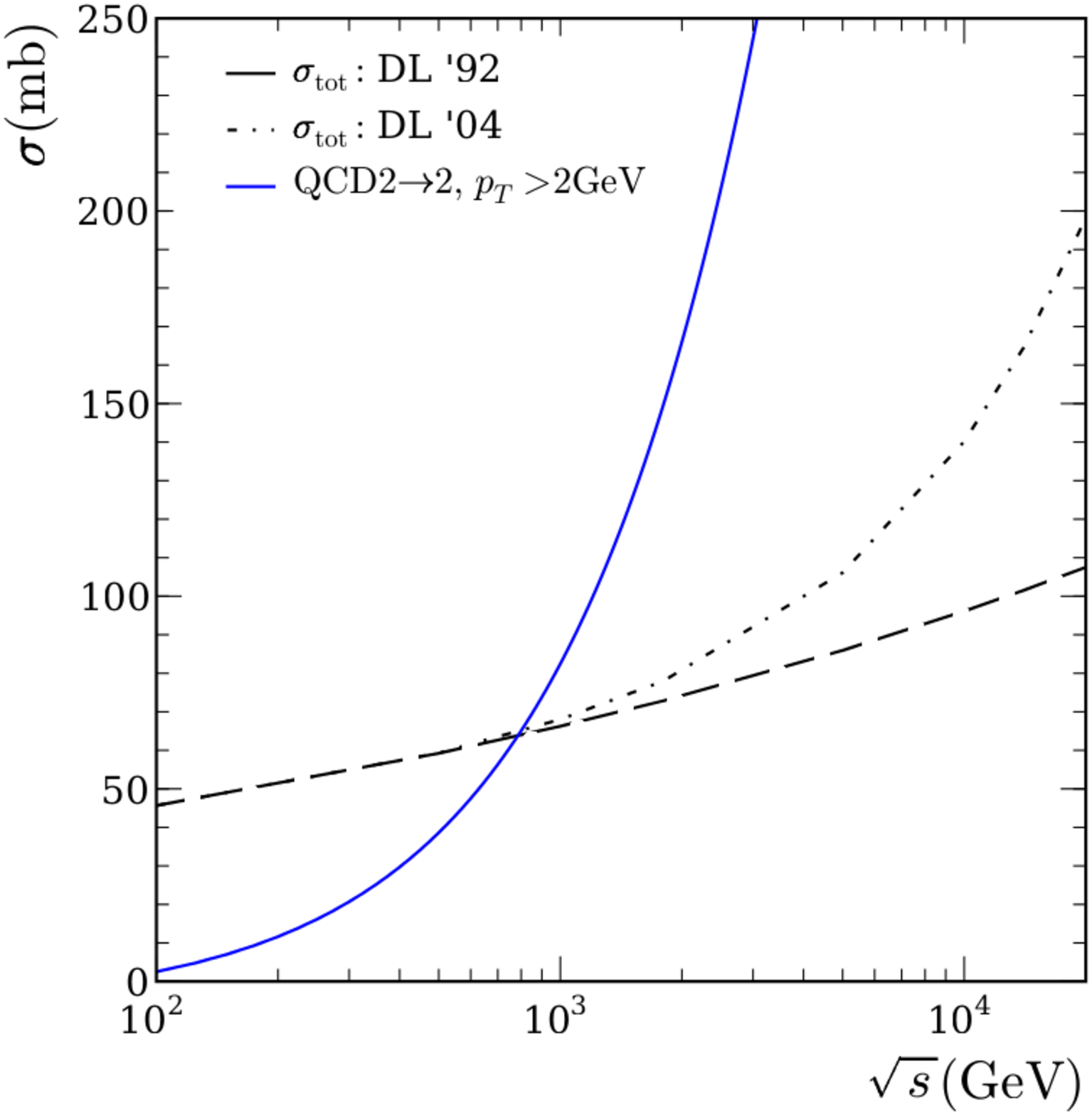}
  \caption{Comparison of cross-section predictions, with the dashed and
    dot-dashed black lines indicating the Donnachie--Landshoff 1992 and 2004
    total $pp$ cross-section parameterisations, constrained by analyticity
    arguments, and the steeper solid blue line showing the partonic dijet cross-section above 2\;GeV. The
    energy at which the partonic cross-section exceeds the total cross-section
    reduces with the partonic \pt cut.}
  \label{fig:mpiunitarity}
\end{figure}

A slightly more realistic model is to treat protons as bags of non-interacting
partons, of which more than one pair can interact in a given collision. In fact,
structure function data from the HERA collider~\cite{Derrick:1993fta} was the
historical catalyst for the adoption of such a model, as the strong rise of the
$F_2$ structure function at low $x$ would without application of unitarization
procedures lead to the the partonic jet cross-section could exceed the measured
total interaction cross-section for sufficiently low minimal jet \pt{s}, as shown
in Figure~\ref{fig:mpiunitarity}. The steep rise in partonic jet cross-section
is driven by interactions at low values of momentum exchange (below a few GeV),
where both the strong coupling and the PDFs diverge.  This apparent
contradiction -- how can the partonic cross-section exceed top-down analyticity
constraints like the Froissart-Martin bound? -- is due to the neglected bulk
interactions of the incoming hadrons. Hence these proton bulk effects must also
be included in any model of hadron collisions which wishes to be infra-red
complete.

Such a model is the Sj\"ostrand--van~Zijl model implemented in the \pythia
generator in 1987~\cite{Sjostrand:1987su}. This model defines the standard
template of MPI modelling adopted by general-purpose MC generators since that
time: the excess of partonic cross-section is interpreted as the mean number of
partonic interactions in a hadron collision at that energy, i.e. $\bar{n} =
\hat{\sigma}_\text{jet}/\sigma_\text{inel}$. An eikonal formalism is then
applied to generate Poisson-distributed numbers of multi-parton interactions
(MPI) from this $\bar{n}$, with use of a $pp$ transverse impact parameter and
nucleon form factor. The effect of this latter feature is to reproduce the
experimental jet pedestal effect%~\cite{}
, where the level of MPI activity
approximately plateaus as a function of the scale of the hardest scattering in
the event: in the eikonal model, this corresponds to an increasing overlap of
proton form factors until the sampled impact parameter $b = 0$ with high
probability and all collisions are fully overlapping (``central'', in the
terminology of double parton scattering and heavy ion physics).

Since the divergence of the partonic cross-section is driven by IR divergences
in the matrix elements and PDFs, and is regularised by the Poisson distribution
of number of interactions, an MPI model requires a mechanism to suppress the
partonic divergence. In the first \pythia model, and in the \jimmy MPI model
developed as a similar extension to the \herwig event
generator~\cite{Butterworth:1996zw}, this is achieved with a simple cutoff on
the scattering \pT, denoted as \ptmin. In later models such as the current
versions of \pythiasix and \pythiaeight, a smoother regularisation of the
divergence is used, with the matrix element $1/\pT^2$ divergence replaced with
\begin{align}
  \label{eq:ptreg}
  \frac{1}{\pT^2} \to \frac{\pT^2}{\left( \pT^2 + (\ptmin)^2 \right)^2}.
\end{align}
This ad hoc form is not theoretically motivated and represents an IR
continuation of perturbative QCD scattering into the regime where the strong
coupling diverges. Not all MC generators use this form: \jimmy retains the sharp
cut-off regularisation, while \herwigpp attempts a more theory-driven
continuation: the optical theorem is used to relate soft inelastic scatterings
to the slope of the elastic scattering cross-section, with experimental input
via CDF data and the Donnachie--Landshoff (DL) total $pp$ cross-section
parameterisations~\cite{Donnachie:1992ny,Donnachie:1998gm,Donnachie:2001xx}. In
all cases, a higher value of \ptmin introduces more screening of the soft
divergence and hence results in less MPI activity than a lower value.

Another major feature of the \pythia MPI model is that \ptmin evolves as a
function of the centre-of-mass collider energy $\sqrt{s}$. The form of this
evolution is again not robustly predicted by QCD theory, but a Regge-inspired
ansatz has long been adopted, in which \ptmin evolves with a power law in $s$,
similarly to the Pomeron term in the DL total cross-section
parameterisation. The specific form used in the \pythia generators is
\begin{align}
  \label{eq:ptminevol}
  \ptmin(\sqrts) = \ptmin(1800~\text{GeV}) \cdot \left(\frac{\sqrts}{1800~\text{GeV}}\right)^{\:e/2}.
\end{align}
Here the tuning parameters are \ptmin(1800~\text{GeV}), and the exponent, $e$,
whose value in the Donnachie--Landshoff fit would be 0.16, but which favours a
higher value of $e \sim 0.25$ in pre-LHC MPI tunes~\cite{Buckley:2009bj}. A
higher value of $e$ means that \ptmin will be higher at LHC energies, and hence
LHC MPI activity will be reduced. The reference energy is set to 1800\;GeV
simply because the first fits were derived as deviations from the Tevatron Run~I
data at that energy: a different energy could be used, e.g. at 7\;TeV for
LHC-driven tunes, but would solely result in an algebraic transformation of the
parameters which would make comparison with old tunes more
difficult.\footnote{Exactly this has occurred with the \pythiasix ``Perugia2011''
  tune set.} Again, not all generators follow the \pythia example: particularly
in the Herwig generator family, the \jimmy MPI model makes no prediction for
energy extrapolation and the more evolved \herwigpp model explicitly attempts to
fit multi-energy data with a single value of \ptmin. The latest tunes of
\herwigpp, however, have decided that the most minimal form of this model is
insufficient to describe all data and have hence introduced an energy evolution
parameterisation of their own.

The remaining common features of Monte Carlo MPI models which are of relevance
to tuning to LHC and other data are the proton form factor and the oft-mentioned
``colour reconnection'' mechanism. The first of these is crucial to description
of the transition between ``minimum bias'' physics (i.e. the bulk of hadron
collider events in which low multiplicity, low \pt inelastic scattering
dominates) and underlying event physics (in which the MPI interactions are in
addition to a hard partonic scattering). The \pythia family, always keen to
provide a wide range of phenomenological handles, offer a range of form factor
parameterisation options, from single- and double-Gaussian parameterisations of
the form factor itself (with tweakable relative widths and populations in the
double-Gaussian case) to a general $O(b) \propto \exp{\left( -b^n \right)}$ form for the
\emph{overlap} function $O(b)$. The \herwig family, by comparison, fixes the form
factor shape to the Fourier transform of the proton electromagnetic form factor,
\begin{align}
  G(\vec{b}) =
  \int \frac{\mathrm{d}^2\vec{k}}{(2\pi)^2}
  \frac{e^{i\vec{k}\cdot\vec{b}}}{(1 + k^2/\mu^2)^2},
\end{align}
where $\mu^2$ is an inverse radius scale-factor introduced to account for the
possible difference in distribution of electric and colour charge: as in \pythia
this width parameter is considered free. In recent versions of \herwigpp and
\pythiaeight, a refinement of this scheme has been introduced in which the
density of the matter distribution is dependent on the momentum fraction $x$ of
the hardest scattering: this ``hot spot'' model is both supported by
data~\cite{Bahr:2008dy,Bahr:2009ek} and theoretically
motivated~\cite{Corsetti:1998mr,Godbole:2004kx}.

The consistency requirements of explicit event generator implementation force
the introduction of additional complexities, since the colour charges of the
resulting beam remnants must be resolved into colour singlet final-state
hadrons. The initially simple form of this connection between MPI scatterings
has been refined in recent years by the work of Skands, Sj\"ostrand, and Corke
in implementing colour string reconnection, MPI rescattering, and $x$-dependent
proton size models in \pythiasix and latterly \pythiaeight.  Colour reconnection
is the final aspect of MPI modelling that we will discuss here. The motivation
for this is that with many colour strings/dipoles being created by the multiple
scattering, some form of annealing may take place on the timescale of
hadronisation to form more energy/action-efficient topological
configurations. This model was originally introduced as part of \pythia
hadronisation, with an addition refinement to re-suppress the effect of such
annealing for high-\pt colour strings (motivated by the idea that such strings
will have less time to participate in annealing)~\cite{pythia6}, and has recently
been introduced into \herwigpp, although differently formulated for cluster
rather than string hadronisation~\cite{Gieseke:2011xy}. \pythiaeight has additionally
introduced a related form of topological reconfiguration called
``rescattering'', whereby MPI interactions may interact at a diagrammatic level:
we will not consider this further in this note. Colour reconnection introduces
one or more parameters related to the strength of the reconnection probability
in the annealing process.

This concludes the summary of MC MPI modelling most prevalent in the
general-purpose MC generators in use at the LHC. Alternative, although in most
cases strikingly similar, models have also been developed, notably with
inspiration from Regge models such as in PHOJET~\cite{phojet}, nuclear
collective excitations as in EPOS~\cite{Werner:2008zza}, the use of CCFM parton
shower evolution as in CASCADE~\cite{Jung:2000hk}, and dipoles as in the
DIPSY~\cite{Flensburg:2011kk} model. However, the eikonal partonic scattering
model pioneered in \pythia remains the mainstay of MPI simulation in general
purpose event generators such as \jimmy~\cite{Butterworth:1996zw},
\herwigpp~\cite{Bahr:2008pv,Gieseke:2011na}, \pythia{}s 6
and~8~\cite{pythia6,Sjostrand:2007gs}, and Sherpa~\cite{Gleisberg:2008ta}. As a
result, this is the model with most current influence on LHC signal and
background MPI simulation, and is the one on which the phenomenological aspects
of the remainder of this note will concentrate.

\section{ATLAS measurements of soft QCD observables}

The crucial inputs to improvements in the quality of soft QCD modelling at the
LHC are of course measurements of observables at the LHC which are particularly
sensitive to MPI model features. There is no observable which is a purely MPI
phenomenon -- quantum mechanics tells us that we must consider all compatible
processes as contributing towards any observable, and in the case of soft QCD
observables competing effects such as initial state radiation (ISR), diffractive
process matrix elements, hadronisation, etc. are all potential contributors to
nominally ``MPI'' observables. The issue of decoupling these modelling aspects
in the process of model tuning will be addressed in Section~\ref{sec:tuning}.

In the ATLAS experiment, the analyses of most importance to constraining MPI
models are as follows:
\begin{itemize}
\item Diffractive and inelastic cross-section measurements;
\item Minimum bias measurements at 900\;GeV and 7\;TeV;
\item Underlying event measurements with leading cluster and leading track at 900\;GeV and 7\;TeV;
\end{itemize}
We will now briefly summarise each of these analyses, as well as in-progress
analyses expected to contribute to future soft QCD phenomenology studies.

\subsection{Diffractive and inelastic cross-section measurements}
\label{sec:xsec}

As mentioned in Section~\ref{sec:mcmpi}, MPI models make extensive use of
parameterisations of total $pp$ cross-section, in particular one of the
Donnachie--Landshoff parameterisations. All MPI models choose the mean number of
partonic scatters based on the ratio of jet cross-section to
$\sigma_\text{inel}$, the \pythia MPI models evolve their regularisation scale
\ptmin with an ansatz inspired by the DL Pomeron slope, and \herwigpp's MPI
model makes an explicit analytic connection to the elastic slope determined from
the DL parameterisation via the eikonal formalism. It is hence important to
constrain the (components of the) $pp$ total cross-section at 7\;TeV based on
experimental data.

\begin{figure}[tp]
  \centering
  \includegraphics[width=0.7\textwidth]{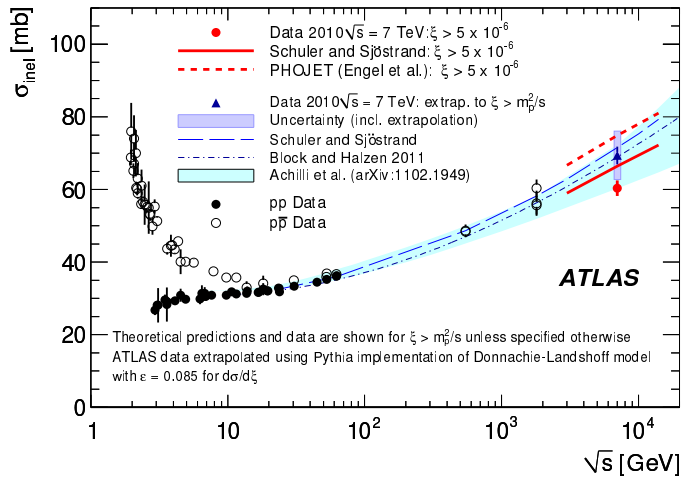}
  \caption{Inelastic $pp$ cross-section as a function of centre-of-mass energy
    $\sqrt{s}$, with ATLAS' measurement indicated with the solid red dot at
    7\;TeV, by comparison with the parameterisation predictions shown with thick
    red lines. The blue triangular point and associated blue vertical bar is the
    ATLAS measurement extrapolated to full elastic acceptance, for comparison
    with the long-dashed and dot-dashed thin blue lines and shaded areas.}
  \label{fig:inelxsec}
\end{figure}

\begin{figure}[tp]
  \centering
  \includegraphics[width=0.7\textwidth]{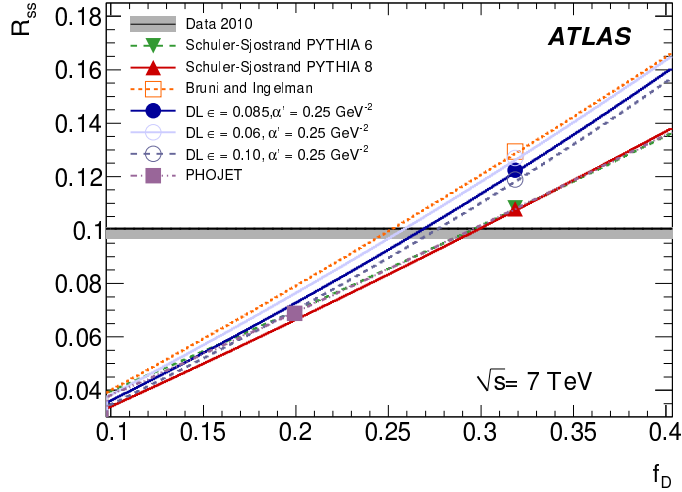}
  \caption{Comparison of the measured same-side-to-inclusive inelastic event
    rate ratio, measured as $R_\text{ss} = [10.02 \pm 0.03 {}^{+0.01}_{-0.04}]\%$ by
    ATLAS, to various model predictions as a function of the diffractive
    fraction of the inelastic cross-section $f_\text{D}$ in the models. The
    markers indicate the default values of $f_\text{D}$ for each model: the
    indication is that at 7\;TeV the fraction of diffractive contribution to the
    inelastic cross-section should be reduced from $\sim 32\%$ to $\sim 27\%$ in
    most models, except \phojet in which an increase of $f_\text{D}$ from $\sim
    20\%$ is required.}
  \label{fig:rss-vs-fd}
\end{figure}

The ATLAS measurement of the inelastic $pp$ cross-section at 7\;TeV is based on
use of the forward minimum bias trigger scintillator (MBTS) detectors -- a pair
of 16-element scintillators located at the calorimeter endcaps on both sides of
the interaction point at $z = \pm 3.56\;\text{m}$, covering the range $2.09 <
|\eta| < 3.84$ -- and a luminosity measurement to a precision of 3.4\% with the
LUCID Cerenkov detector at $z = \pm 17\;\text{m}$. The experimental definition
of an inelastic event is that at least two of the 32 MBTS segments has a charge
above the noise threshold, i.e. that there is measurable proton dissociation on
at least one side of the detector. The measured inelastic cross-section at
7\;TeV was measured within acceptance to be $60.33 \pm 2.10\;\text{mb}$,
slightly in tension with previous fits as shown in Figure~\ref{fig:inelxsec},
with an extrapolation to the elastic limit indicating better agreement but with
a much-increased systematic error due to the extrapolation. Additionally, a
subset of the events were identified as ``single-sided'' when at least two MBTS
segments were activated on one side, and none on the other side: single-sided
events are expected to be dominated by single diffractive $pp$ scattering. The
measured fraction of single-sided to inclusive inelastic events $R_\text{ss}$
was measured as just over 10\%, which is shown by comparison to various MC
models as a function of their diffractive cross-section fraction $f_\text{D}$ in
Figure~\ref{fig:rss-vs-fd}. A slight reduction in model diffractive
cross-sections is favoured.

\subsection{Minimum bias measurements}
\label{sec:mb}

``Minimum bias'' is a much misused term in soft QCD physics: depending on
whether one is speaking from an experimental or theoretical perspective, it
respectively indicates a class of observables constructed on events selected
using minimally strict conditions (either trigger or offline), or it is used to
classify an event type in which there is generally no very hard interaction and
where soft multiple scattering is the dominant physical process. These two
concepts are closely related: if one minimally biases experimental selection
criteria, then the majority of events will be dominated by such an interaction
mode -- but the distinction is still useful to draw, not least because an
experimental minimum bias selection will also select all kinds of ``hard''
processes, and because the phrase is also sometimes used to indicate only
``non-single-diffractive'' (NSD) processes: another case of misleading leakage
from the calculational division of process types into the classification of
real-world collider events.

All the measurements discussed here are measured using the ATLAS ``minimum
bias'' trigger stream, but the specific collection of observables usually
regarded as ``minimum bias'' are simple observables such as the $\eta$ and $\pt$
distributions of tracks and calorimeter clusters, and the distribution of the
number of (charged) particles in an event or the correlation of other mean
properties with such an event-level property. Observables which further
explicitly ``bias'' the event selection, such as the ``underlying event''
observables discussed in the next section, are considered distinct.

ATLAS measurements of these minimum bias observables have been made at 900\;GeV
and 2.36 \& 7\;TeV, using the low pile-up 2010 dataset to obtain clean
measurements. Again, the MBTS trigger scintillators were key to the analysis: at
least one MBTS hit was required on each side of the detector, in addition to a
number $n_\text{trk}$ charged particles above a track \pt cut within the tracker
acceptance of $|\eta| < 2.5$. Various values of the track \pt cut and
$n_\text{trk}$ were used to change the phase space within which the observables
mentioned above are computed: the more particles in the tracker acceptance, and
the higher their \pt cut, the more the events are expected to be dominated by
perturbative QCD. Jet structure is expected to start emerging with the
restriction of phase space, but these observables do not highlight that
transition. The use of the two-sided MBTS requirement is a purely experimental
version of the model-dependent NSD definition used at previous colliders, and is
expected to suppress (but not eliminate) single-diffractive and elastic
scattering events.

\begin{figure}[pb]
  \centering
  \includegraphics[width=0.3\textwidth]{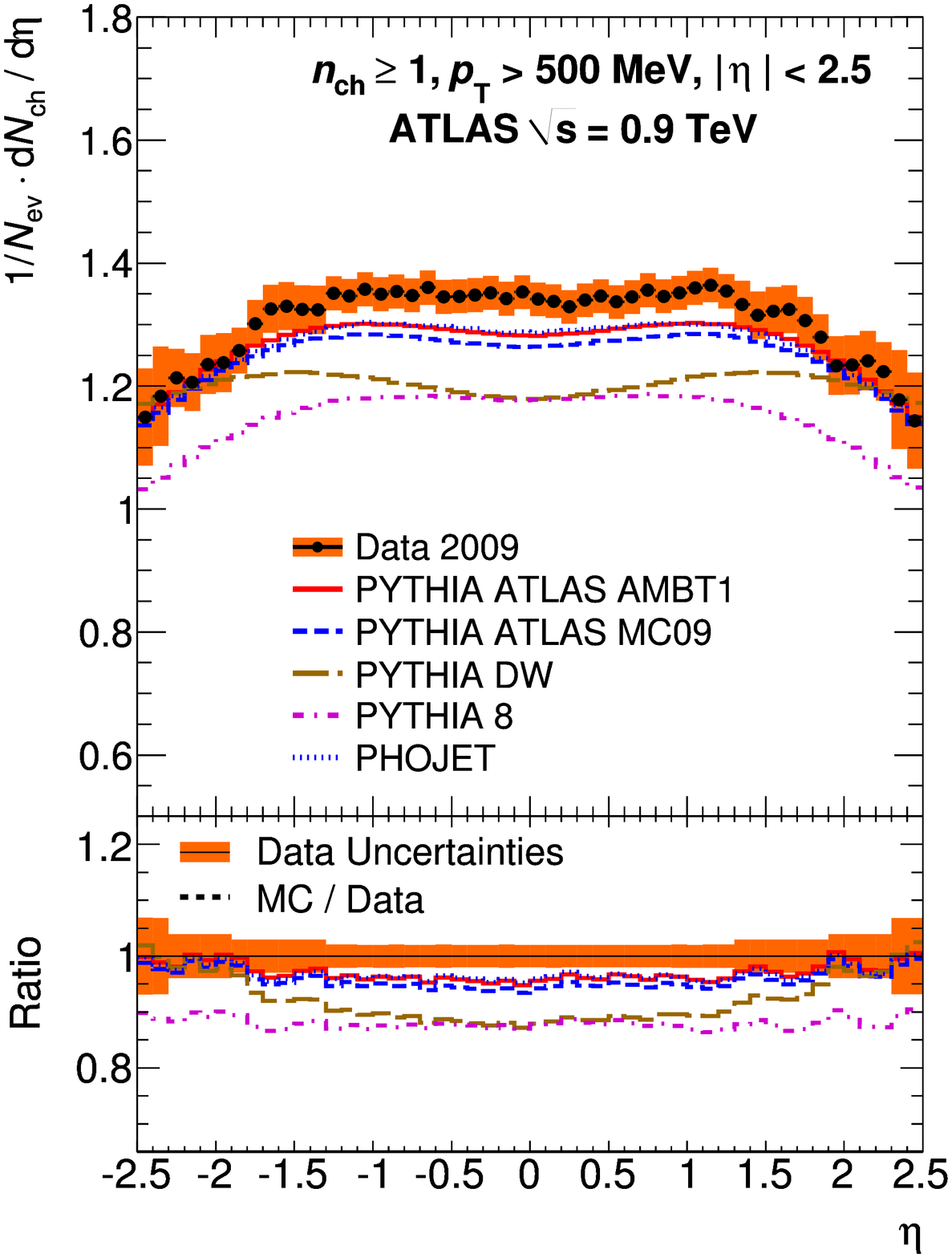}
  \includegraphics[width=0.3\textwidth]{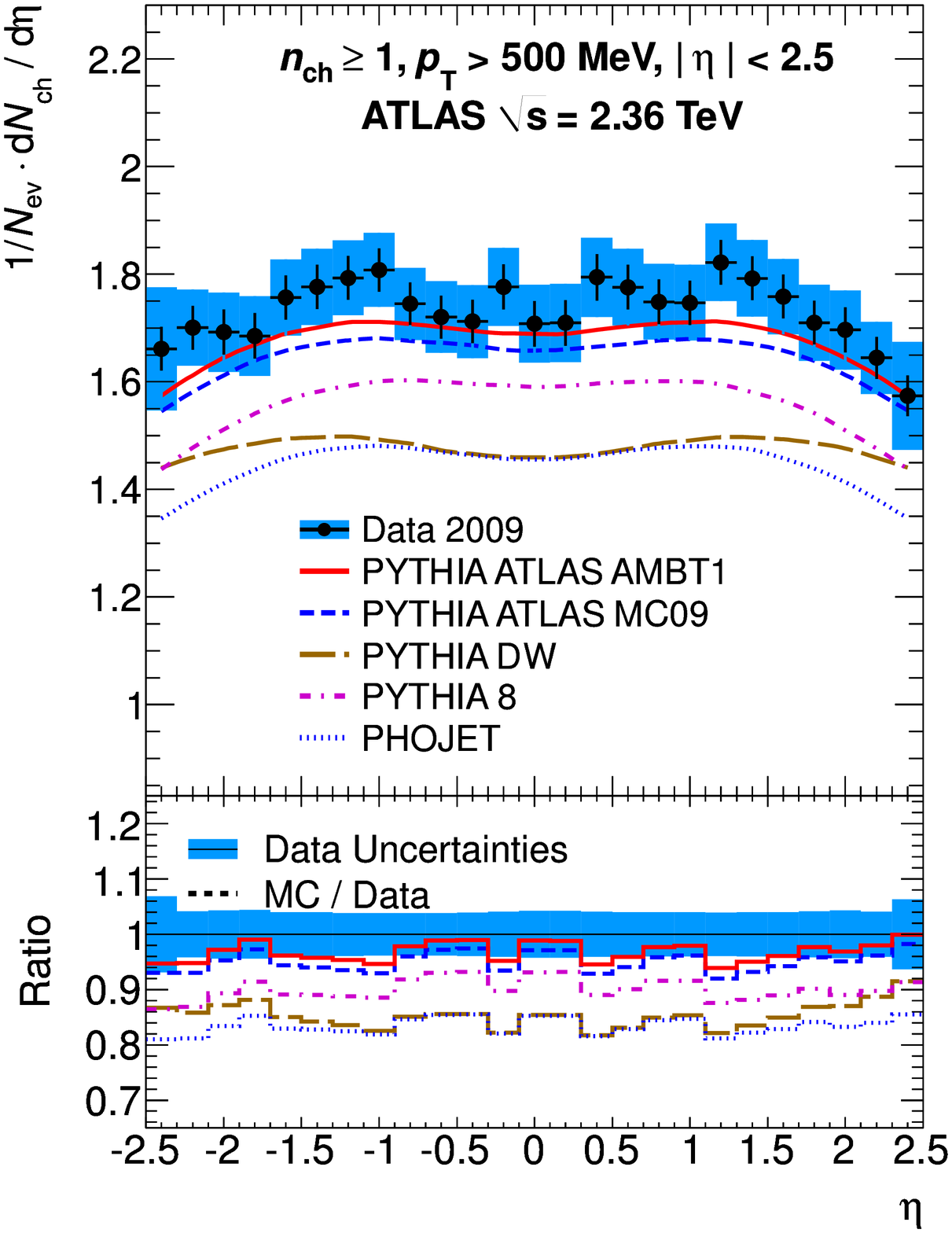}
  \includegraphics[width=0.3\textwidth]{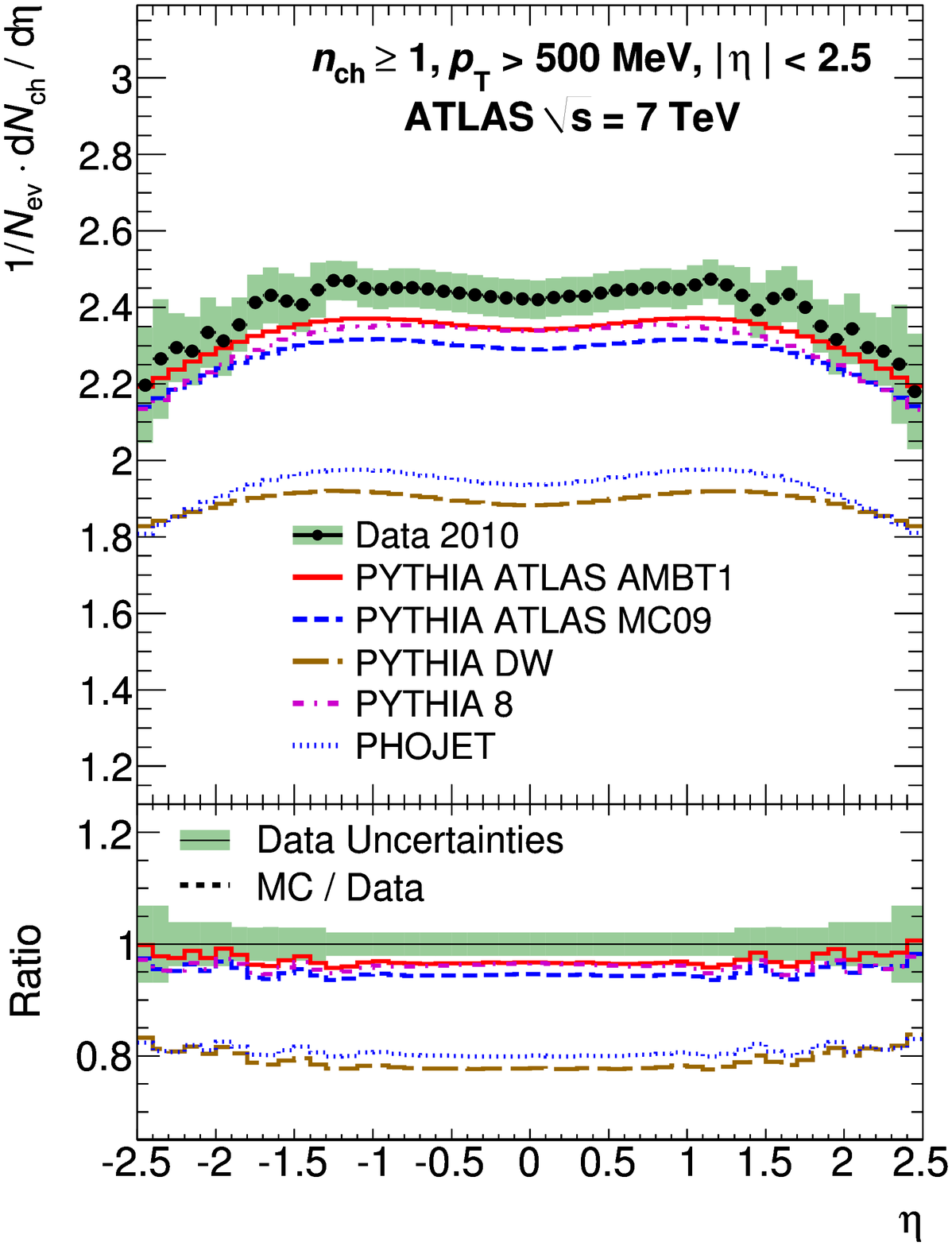}
  \includegraphics[width=0.3\textwidth]{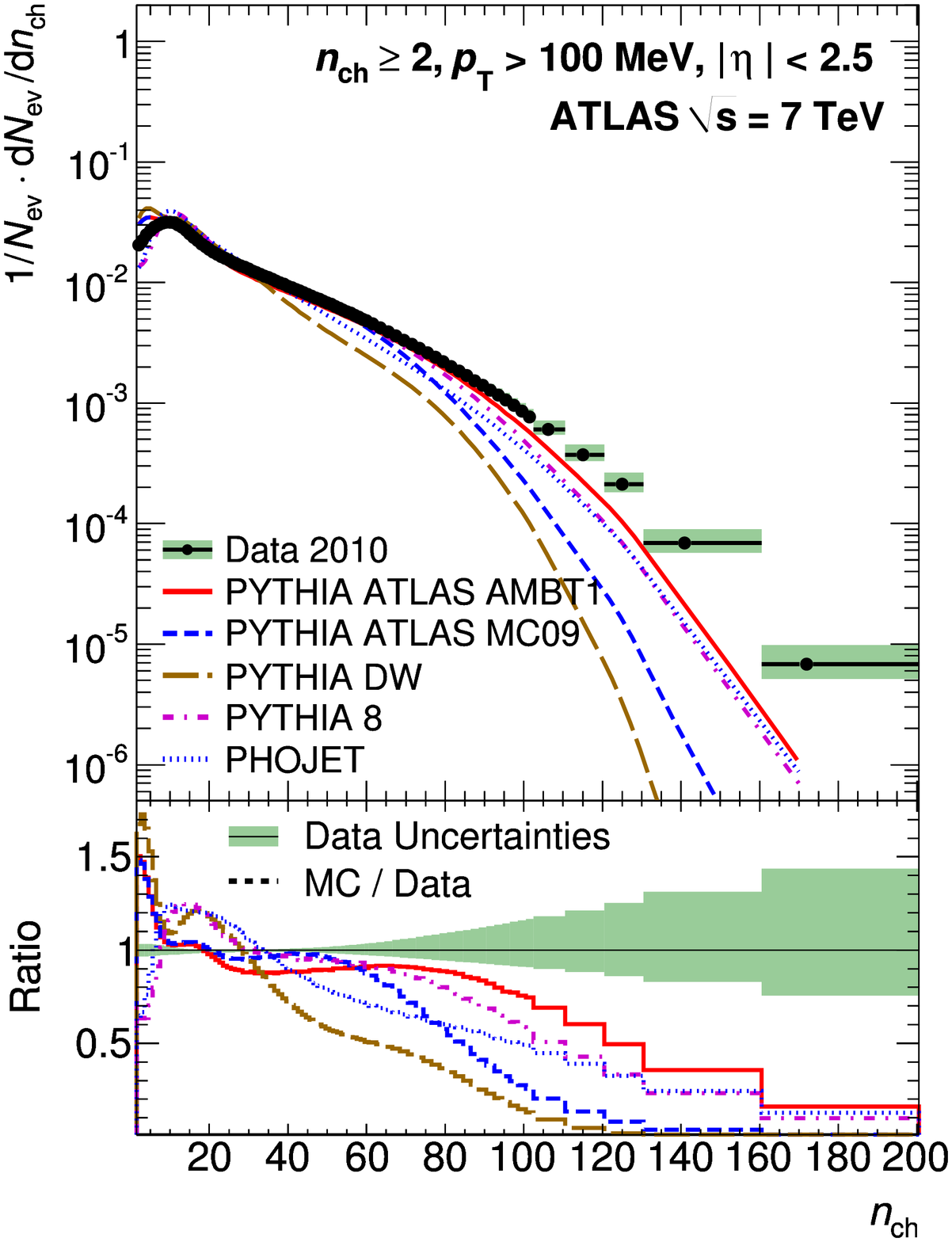}
  \includegraphics[width=0.3\textwidth]{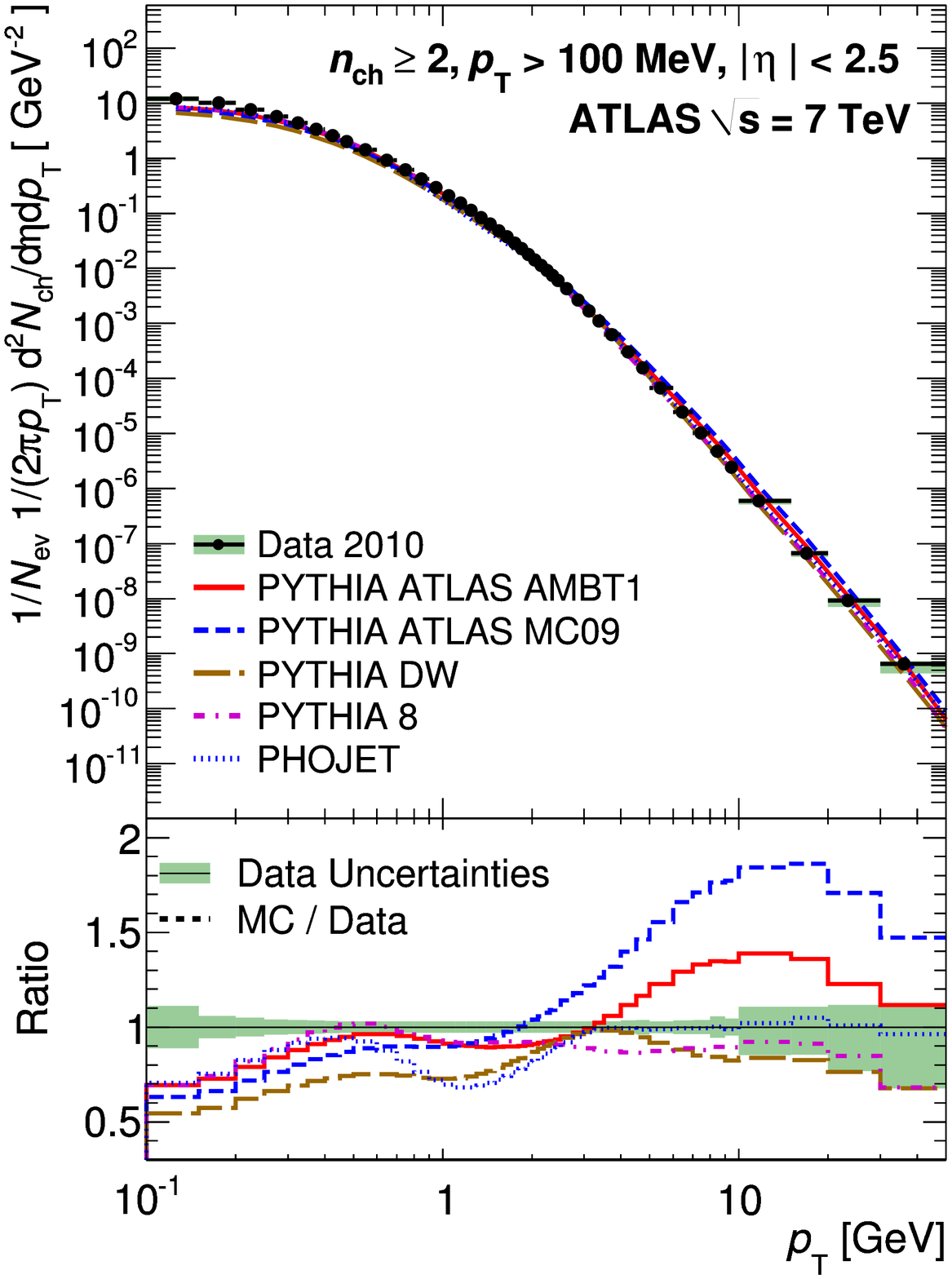}
  \includegraphics[width=0.3\textwidth]{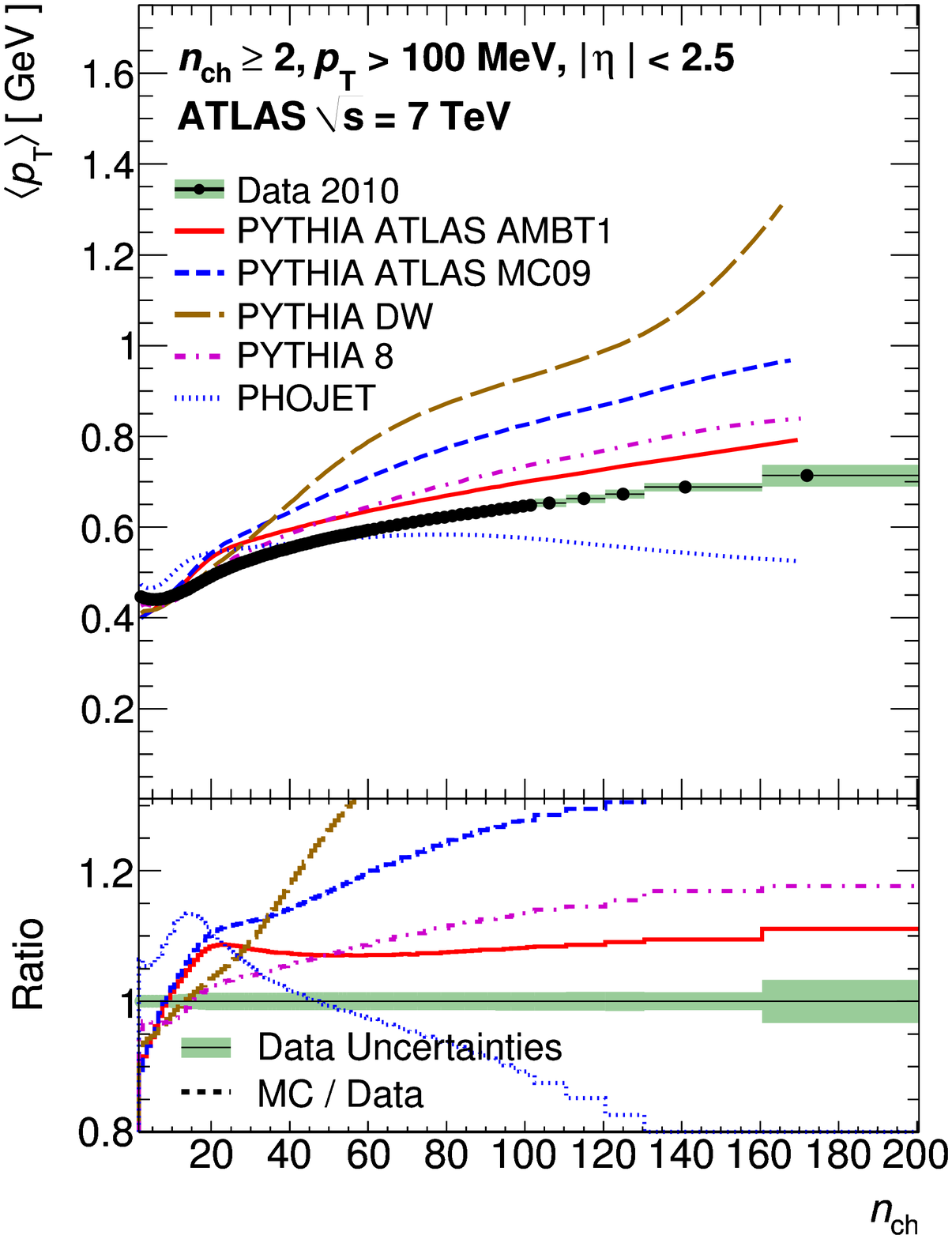}
  \includegraphics[width=0.3\textwidth]{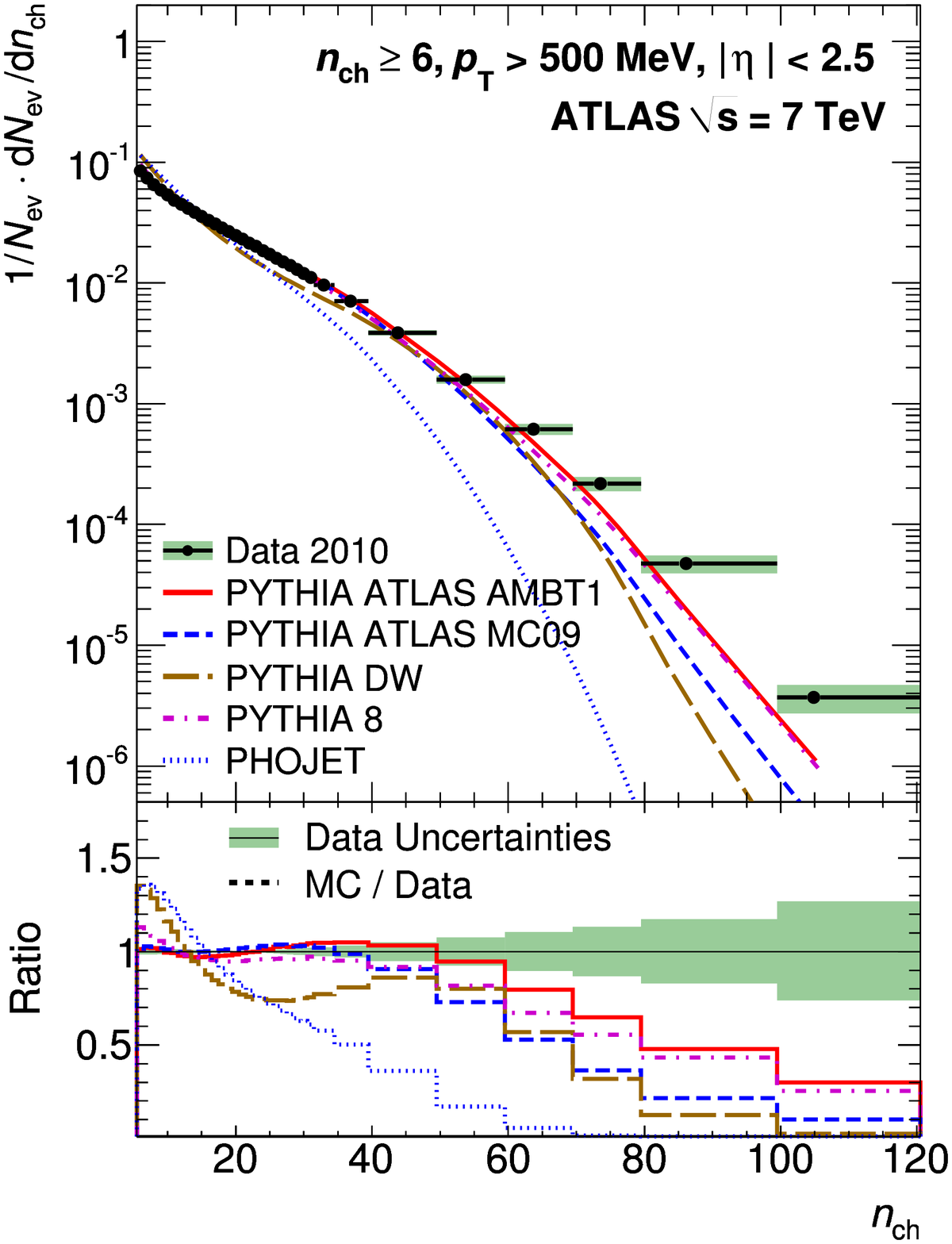}
  \includegraphics[width=0.3\textwidth]{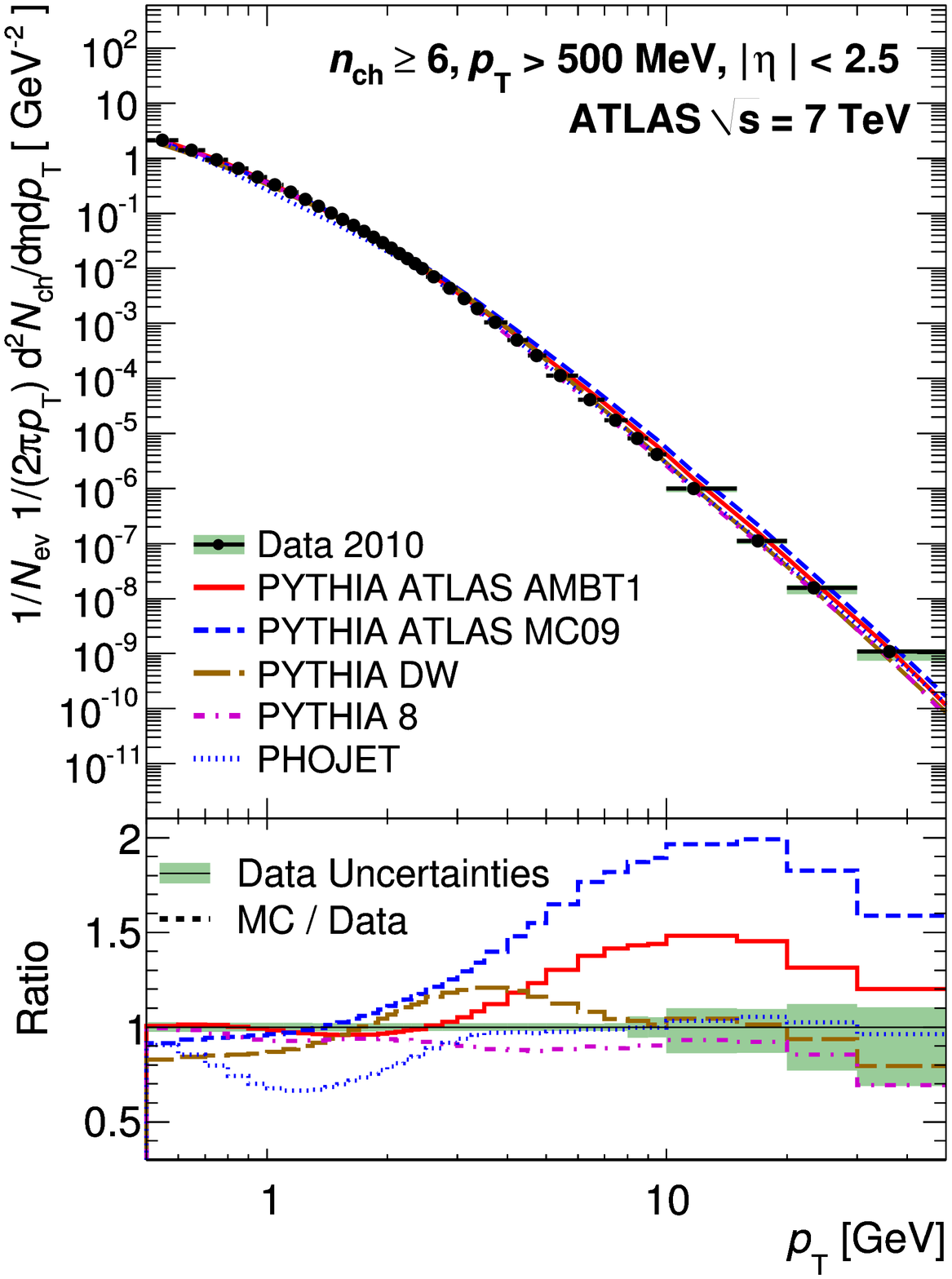}
  \includegraphics[width=0.3\textwidth]{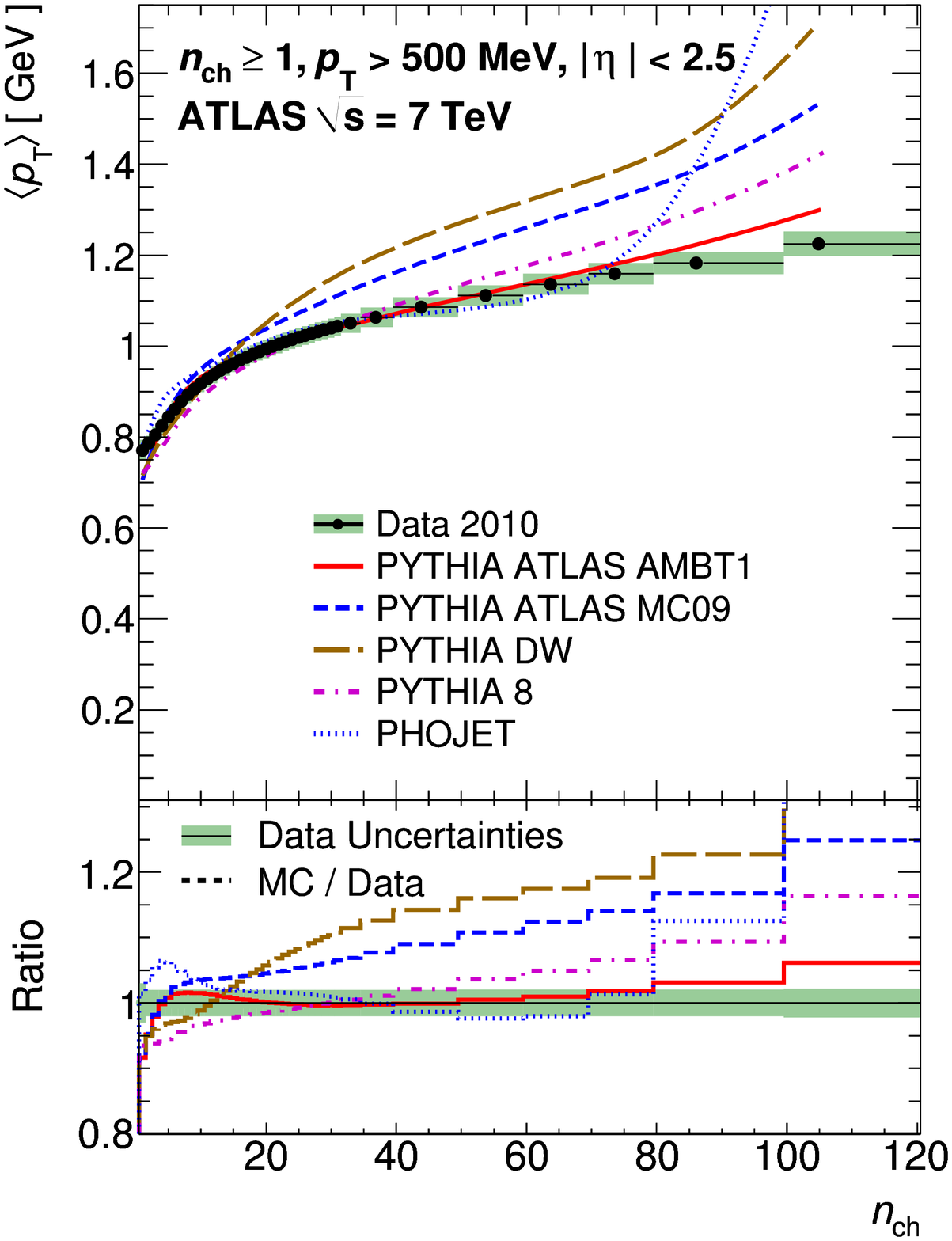}
  \caption{Minimum bias measurements at 900\;GeV, 2360\;GeV, and 7\;TeV, with a
    variety of different phase spaces defined by requirements on the charged particle
    \pt cut, and the number of charged particles which pass that cut. The top row
    of plots shows the $1/N_\text{ch} \mathrm{d}N_\text{ch}/\mathrm{d}\eta$ distribution
    at all three energies, while the second and third rows show comparisons of
    charged multiplicity, \pt spectra, and $\langle \pt \rangle$ vs. $N_\text{ch}$
    between track \pt cuts of 100 and 500\;MeV.}
  \label{fig:mbexamples}
\end{figure}

The $(\pt^\text{trk}, n_\text{trk})$ phase spaces used in this measurement are
as follows: (100\;MeV,~1), (100\;MeV,~2), (100\;MeV,~20), (500\;MeV,~1),
(500\;MeV,~6), (2500\;MeV,~1). Examples are shown in
Figure~\ref{fig:mbexamples}.

\FloatBarrier
\subsection{Underlying event measurements}
\label{sec:ue}

The ``underlying event'' (UE) is the name that we give to all elements of a
hadron collision which cannot be directly identified with the hard scattering
process. This is a rather na\"ive view, and it is completely correct to say that
``there is no underlying event; there is only event''\footnote{\copyright~Rick
  Field, MPI@LHC 2008} -- however, it reflects the perspective that must be
taken to make progress towards new physics discoveries: that there are hard
interactions of interest whose clear experimental signatures are complicated and
diluted by extra contributions related to MPI and ISR. In terms of soft QCD
measurements however, UE is almost always taken to mean observables which have
been constructed to focus on non-hard aspects of event structure, and in
particular to study the evolution of such aspects as a function of the hard
scale. It should be said right away that the UE is not necessarily ``soft'' --
fluctuations in MPI and ISR may produce new semi-hard jets, particularly if the
hardest scattering in the event is very hard, e.g. a TeV-scale QCD or EW
interaction.

\begin{figure}[t]
  \centering
  \includegraphics[width=0.4\textwidth]{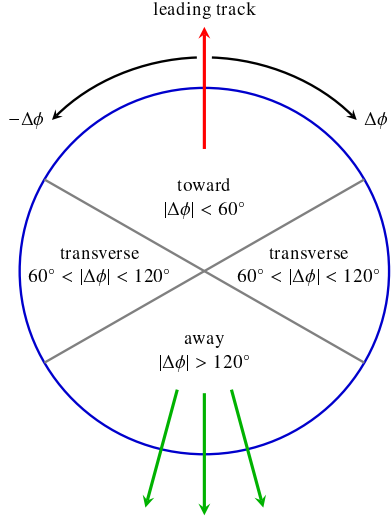}
  \caption{Topological decomposition of dijet/Drell-Yan events for underlying
    event measurements. }
  \label{fig:ue-topo}
\end{figure}

ATLAS has published two UE measurements based on the 2010 dataset (again to
avoid the pile-up contamination of the 2011 runs). Both use the topological
decomposition illustrated in Figure~\ref{fig:ue-topo}, which was first
established by the pioneering UE measurements of the CDF experiment. In this
construction, events are azimuthally oriented along an axis which represents the
flow of energy in the hardest scattering in the event, so that aspects of the UE
may be seen most clearly (with minimal contamination from the hard process) in
the transverse directions. This leading axis could be determined using e.g. a
tensorial diagonalisation of some kind, but is more usually taken to simply be
the direction of the leading jet or reconstructed boson. Most UE observables are
constructed to show the dependence of the \pt and multiplicity observed in each
region as a function of the \pt of the hard process. If the \pt of the hard
process may be safely used down to the lowest scales, UE observables hence show
the transition of MPI from ``minimum bias'' physics into the hard scattering
regime.

The ATLAS measurements use two different detector elements to make their
measurements of UE quantities at 900\;GeV and 7\;TeV: the first follows the lead
of the CDF measurements by using tracking information in the azimuthal regions,
whereas the second is the first measurement of UE properties using calorimeter
clusters. In both cases, to avoid systematics issues with calorimeter jets in
the ATLAS commissioning phase the leading object in the event is taken to be a
``single particle'' -- a charged track or calorimeter cluster respectively --
rather than a jet. This limits the range of validity of the measurement, since
at some scale the leading particle will not necessarily be in the leading jet,
and so both measurements are made using the minimum bias trigger stream, with a
scale reach only up to $\sim 20\;\text{GeV}$. Future ATLAS UE measurements will
extend this programme to use leading track jets, calorimeter jets, and $Z$ and
$W$ events.

\begin{figure}[p]
  \centering
  \includegraphics[width=0.45\textwidth]{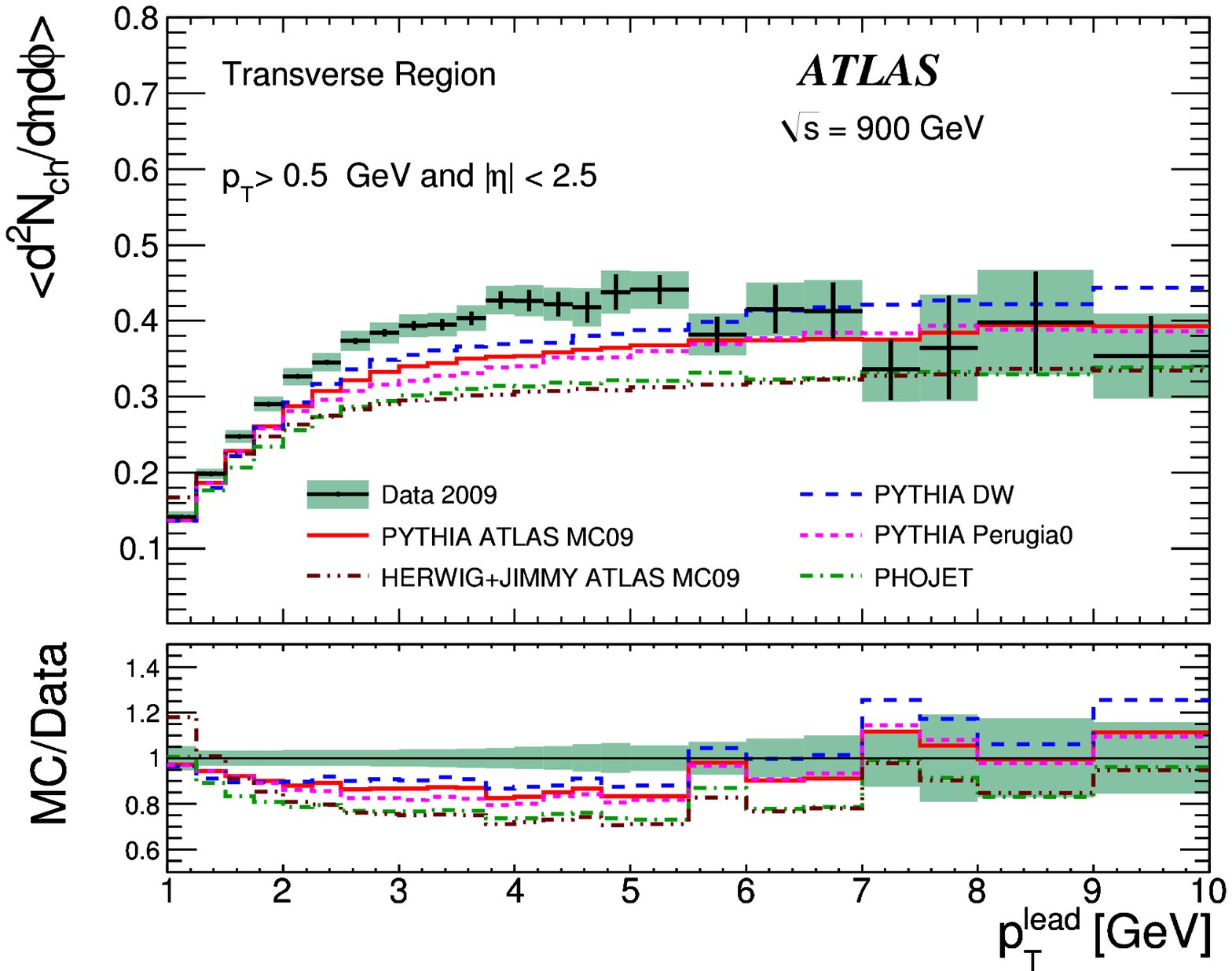}
  \includegraphics[width=0.45\textwidth]{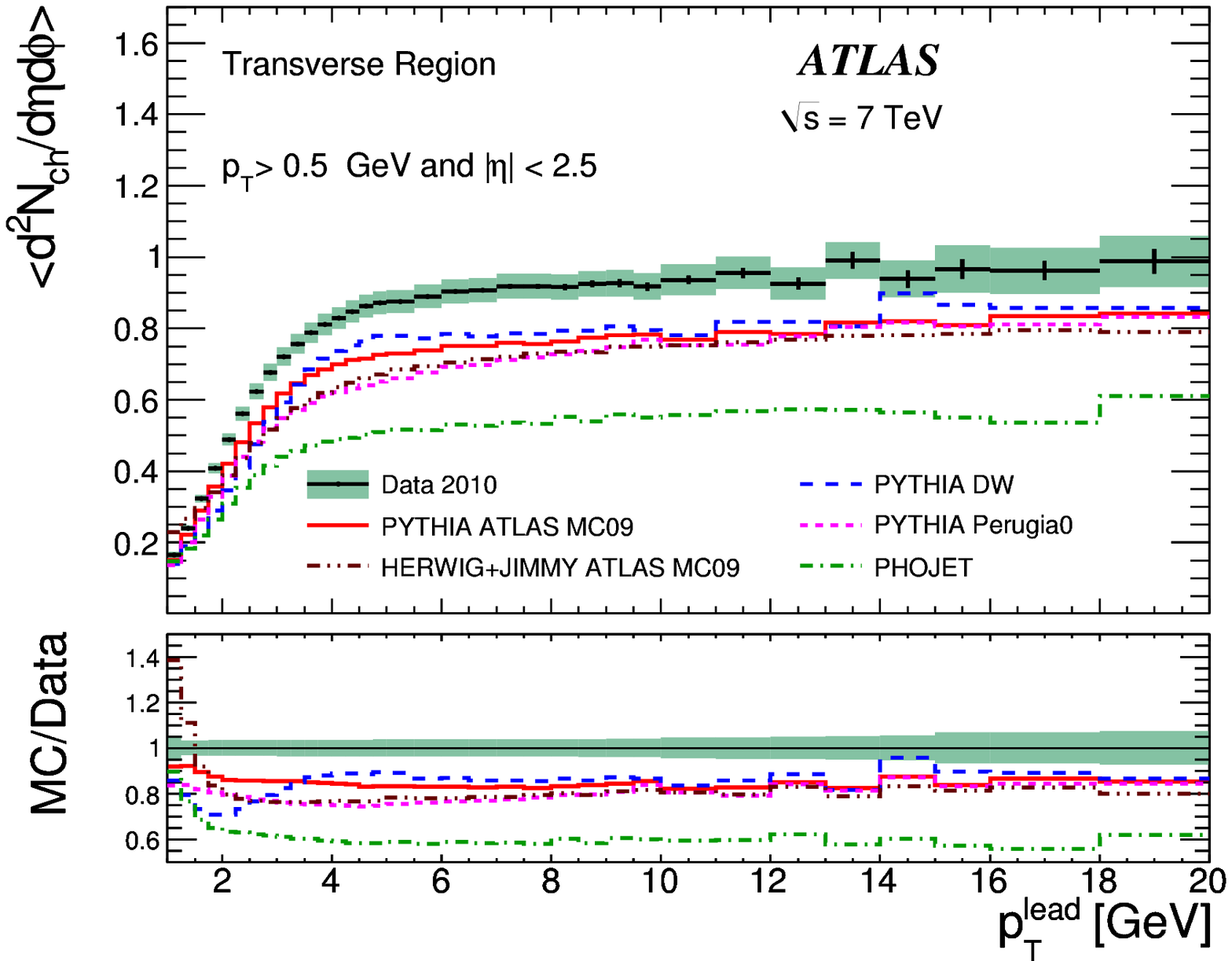}
  \includegraphics[width=0.45\textwidth]{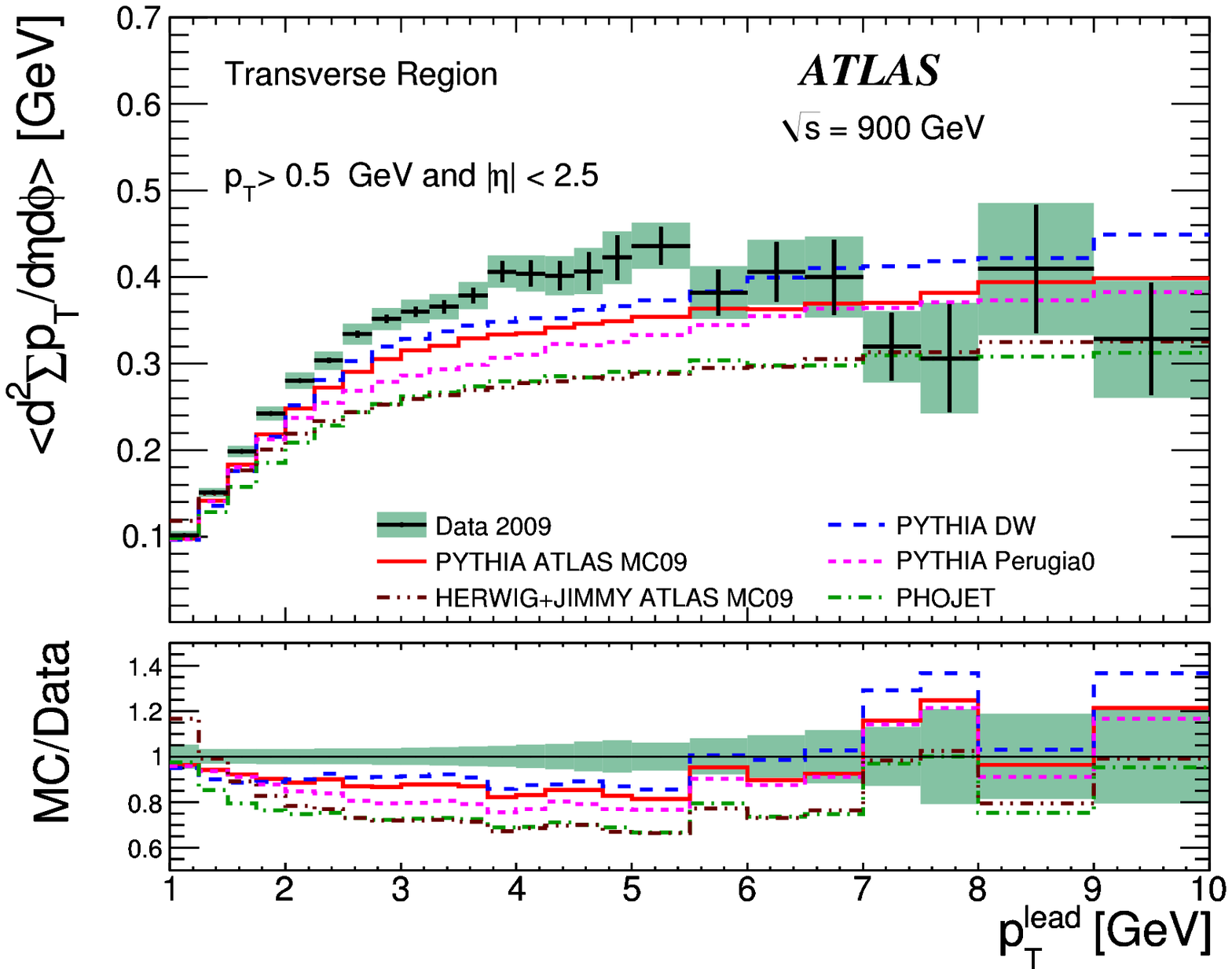}
  \includegraphics[width=0.45\textwidth]{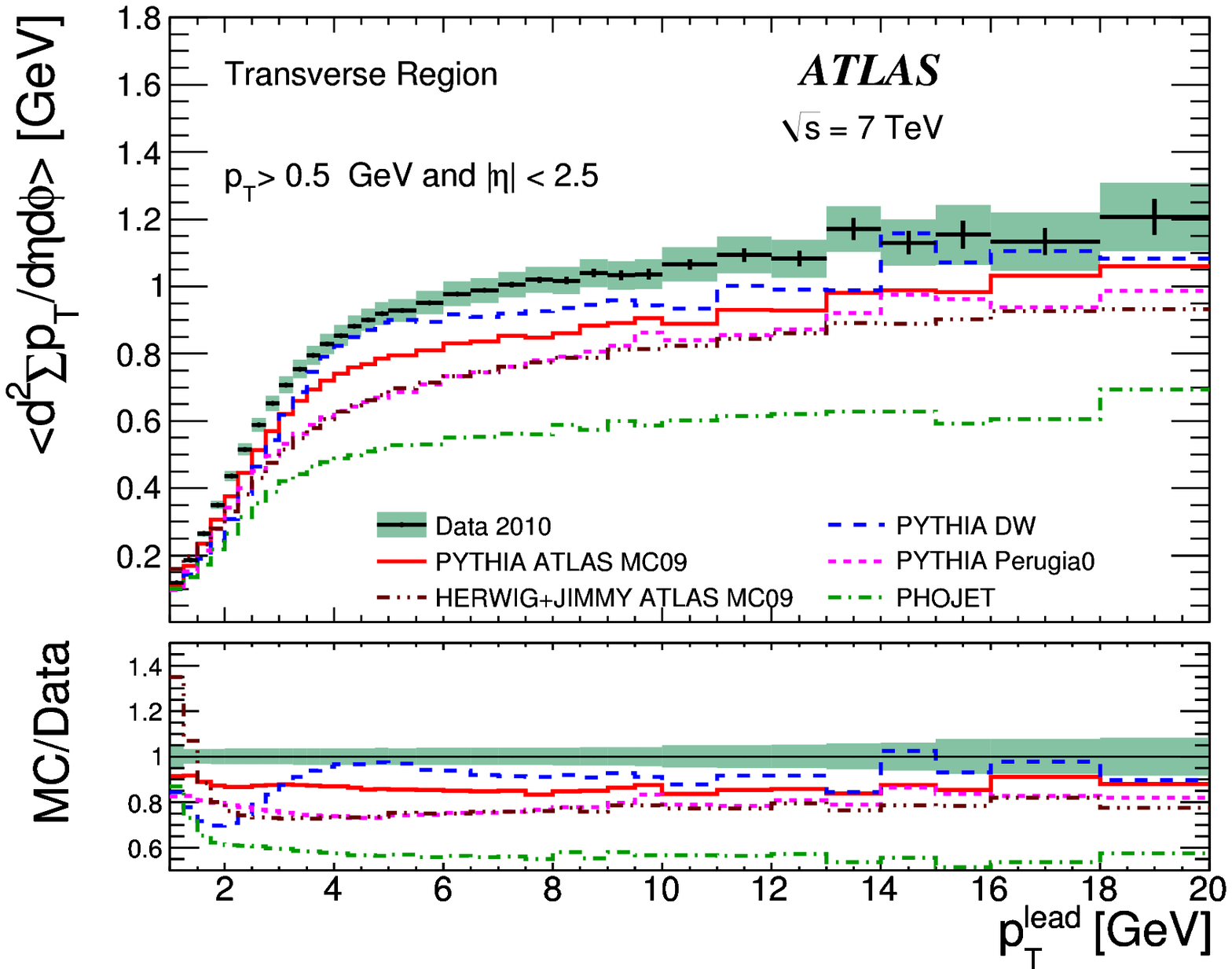}
  \includegraphics[width=0.45\textwidth]{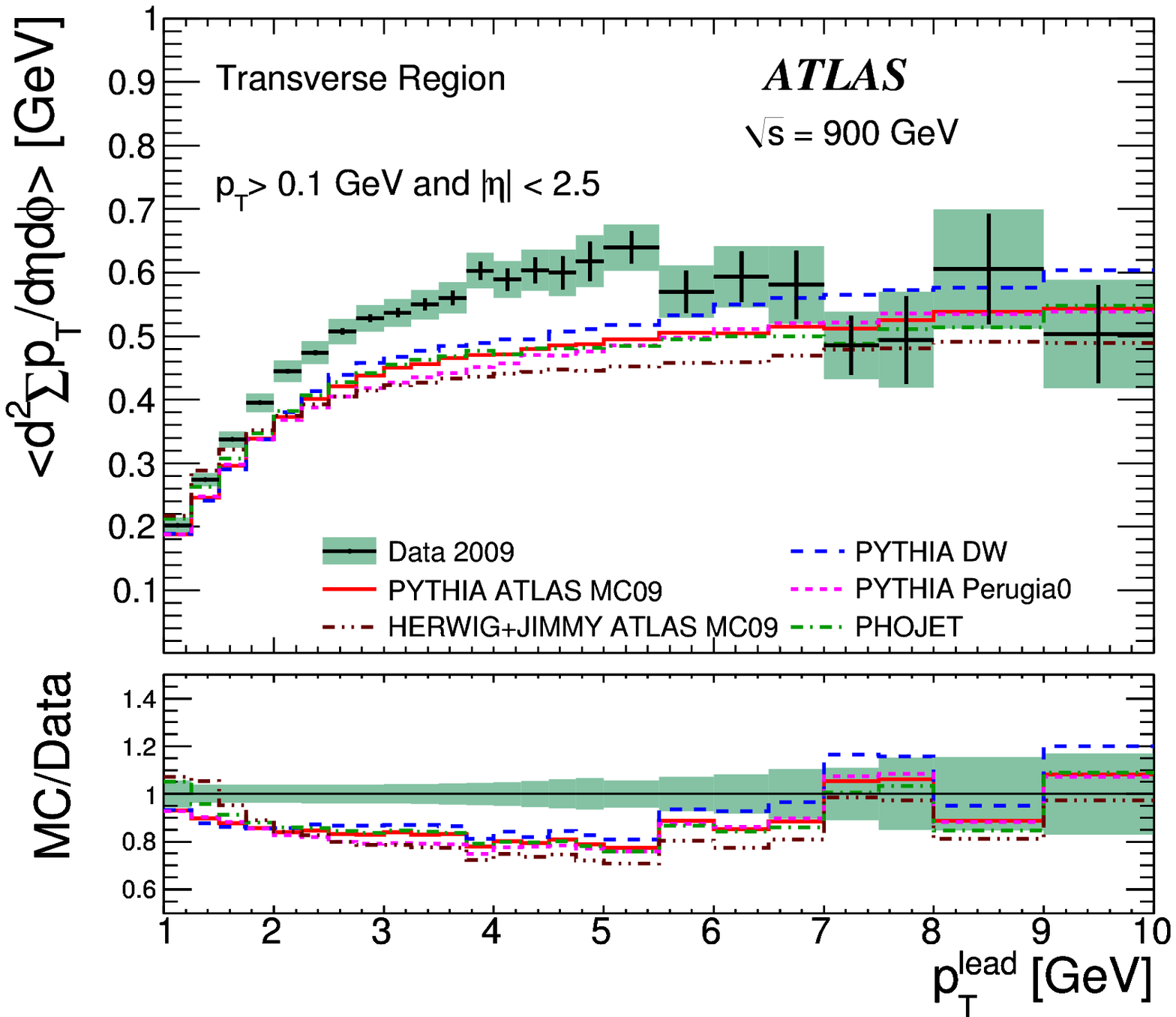}
  \includegraphics[width=0.45\textwidth]{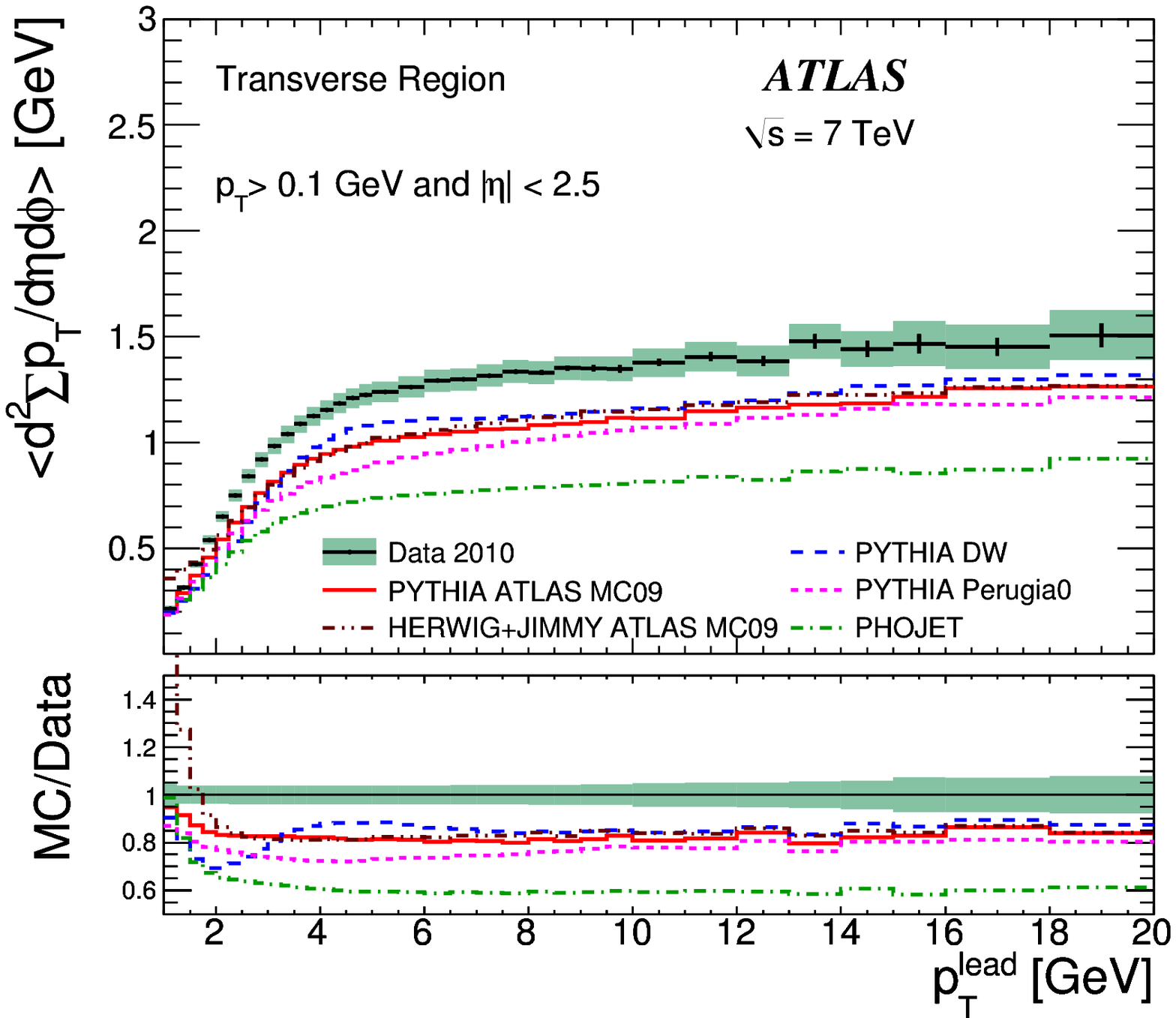}
  \caption{Leading-track-based underlying event measurements at 900\;GeV (left column)
    and 7\;TeV (right column). The first row is the dependence of the mean number of
    particles with $\pt > 500\;\text{MeV}$ in the transverse regions as a function of
    the \pt of the leading track; the second row shows the same evolution for the mean
    \pt of those particles; and the final row shows the scale evolution of mean \pt for
    the more inclusive track \pt requirement of 100\;MeV.}
  \label{fig:ueexamples}
\end{figure}

\begin{figure}[p]
  \centering
  \includegraphics[width=0.45\textwidth]{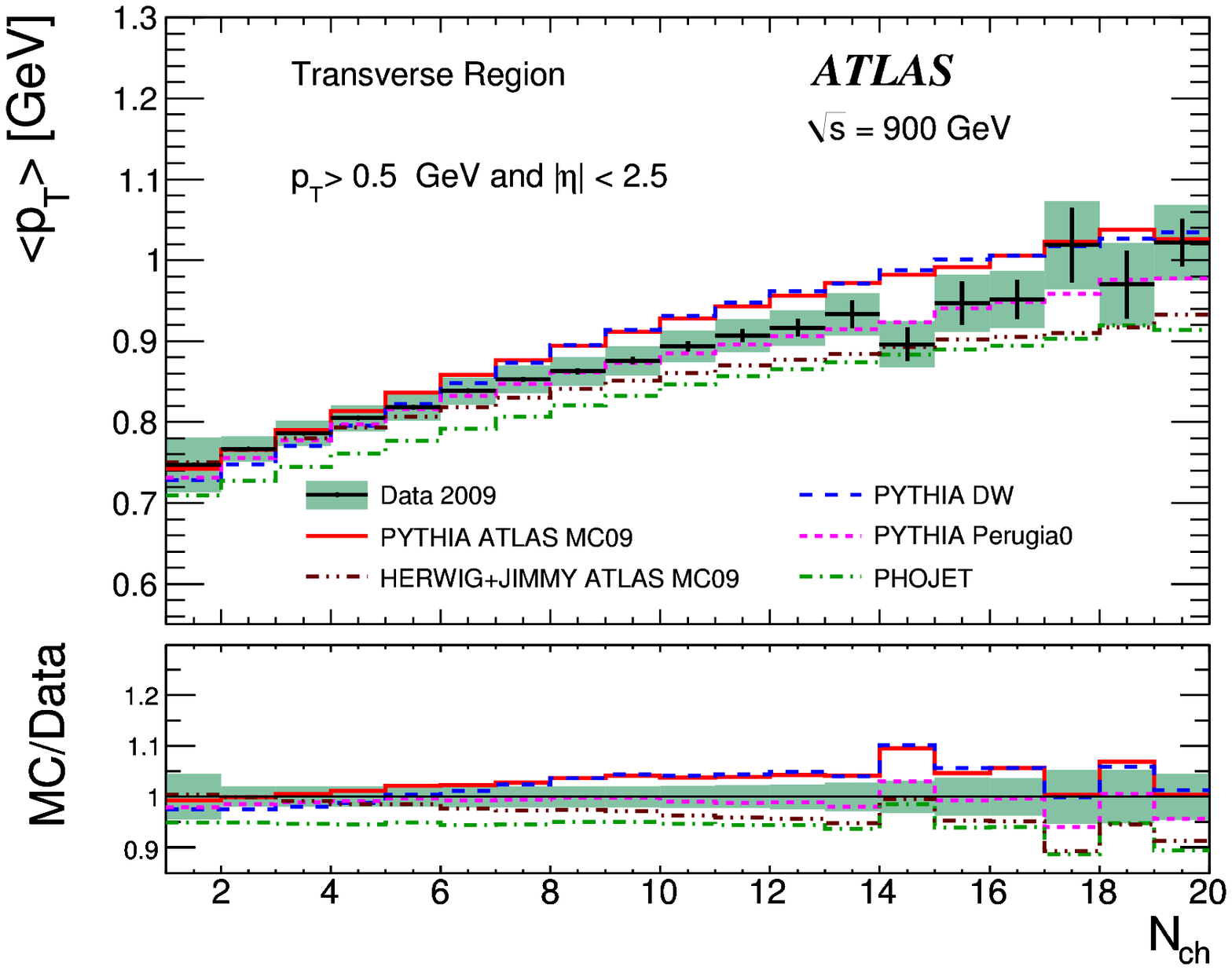}
  \includegraphics[width=0.45\textwidth]{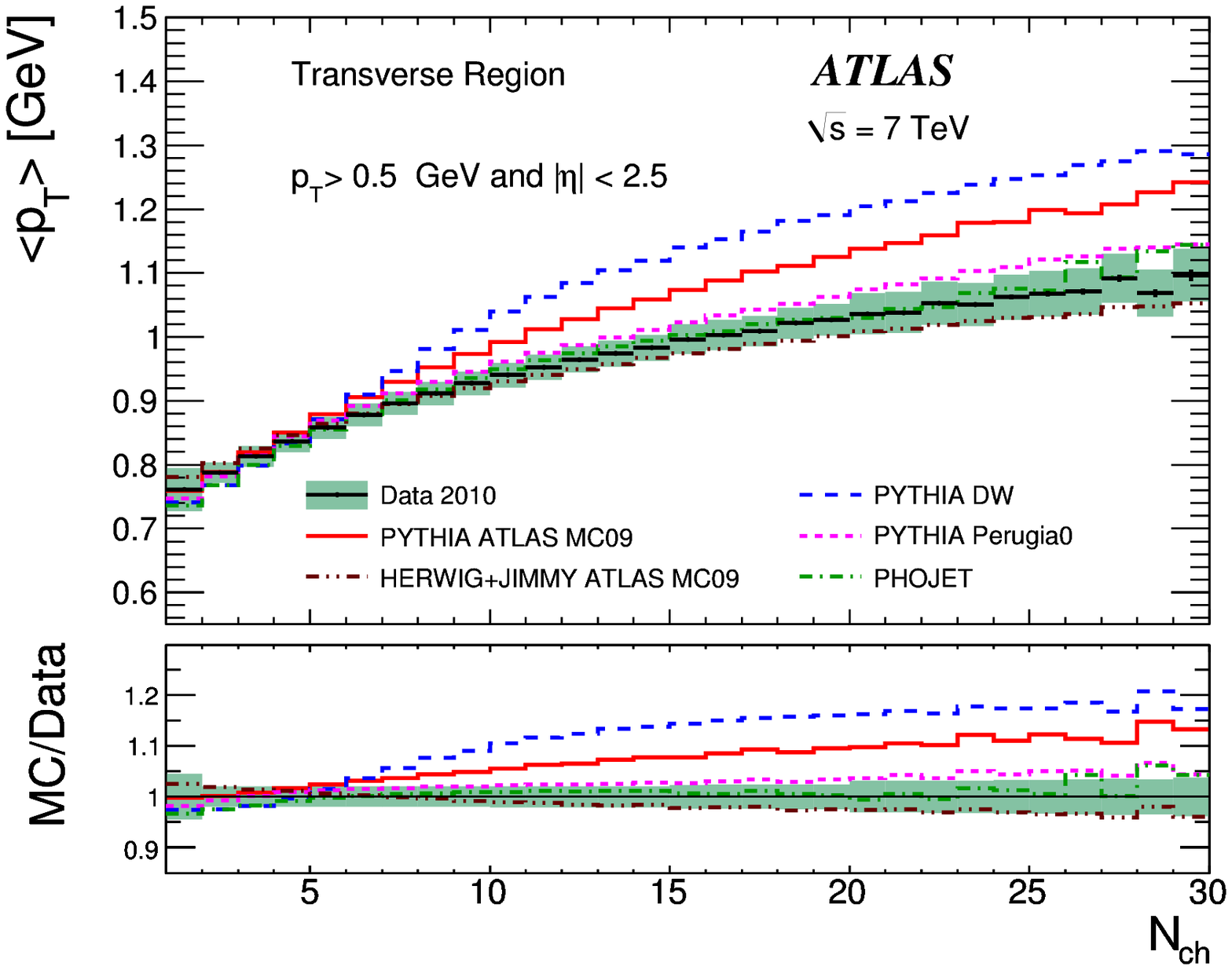}\\
  \includegraphics[width=0.85\textwidth]{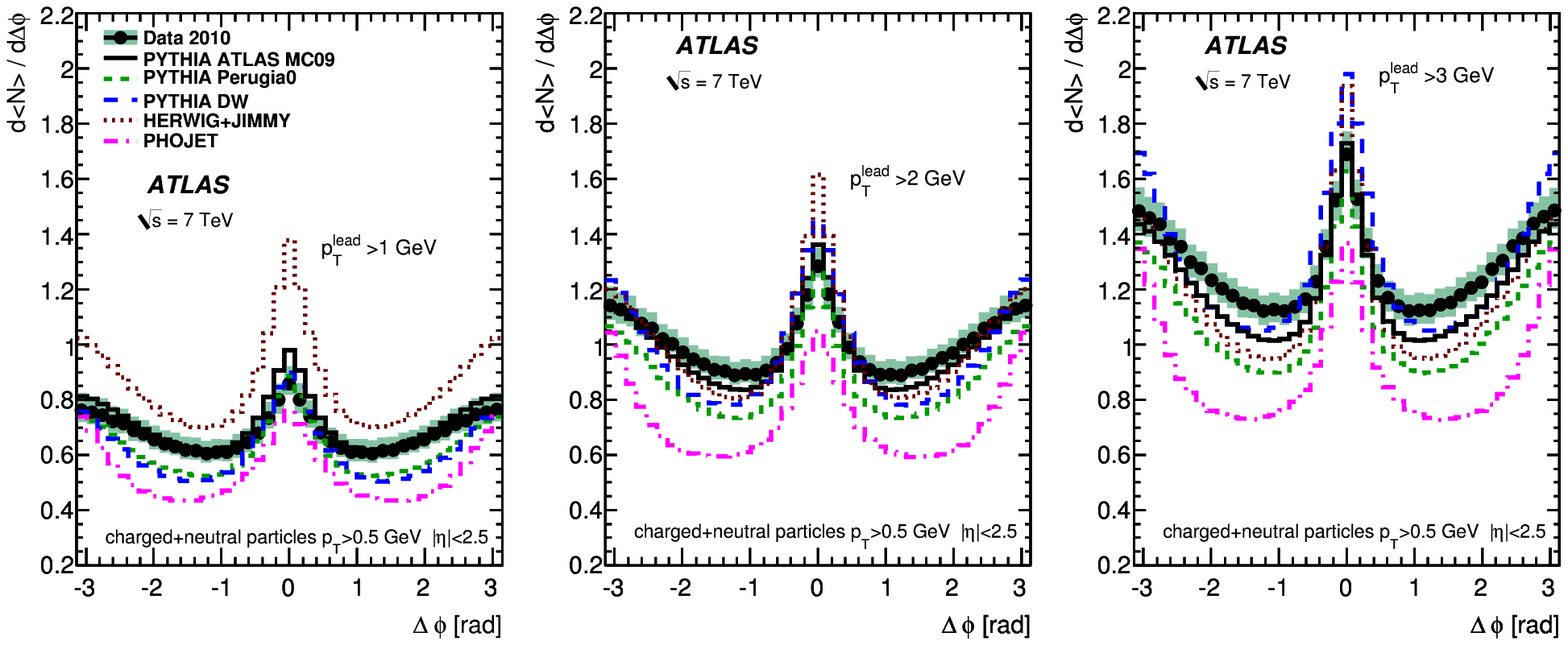}\\
  \includegraphics[width=0.4\textwidth]{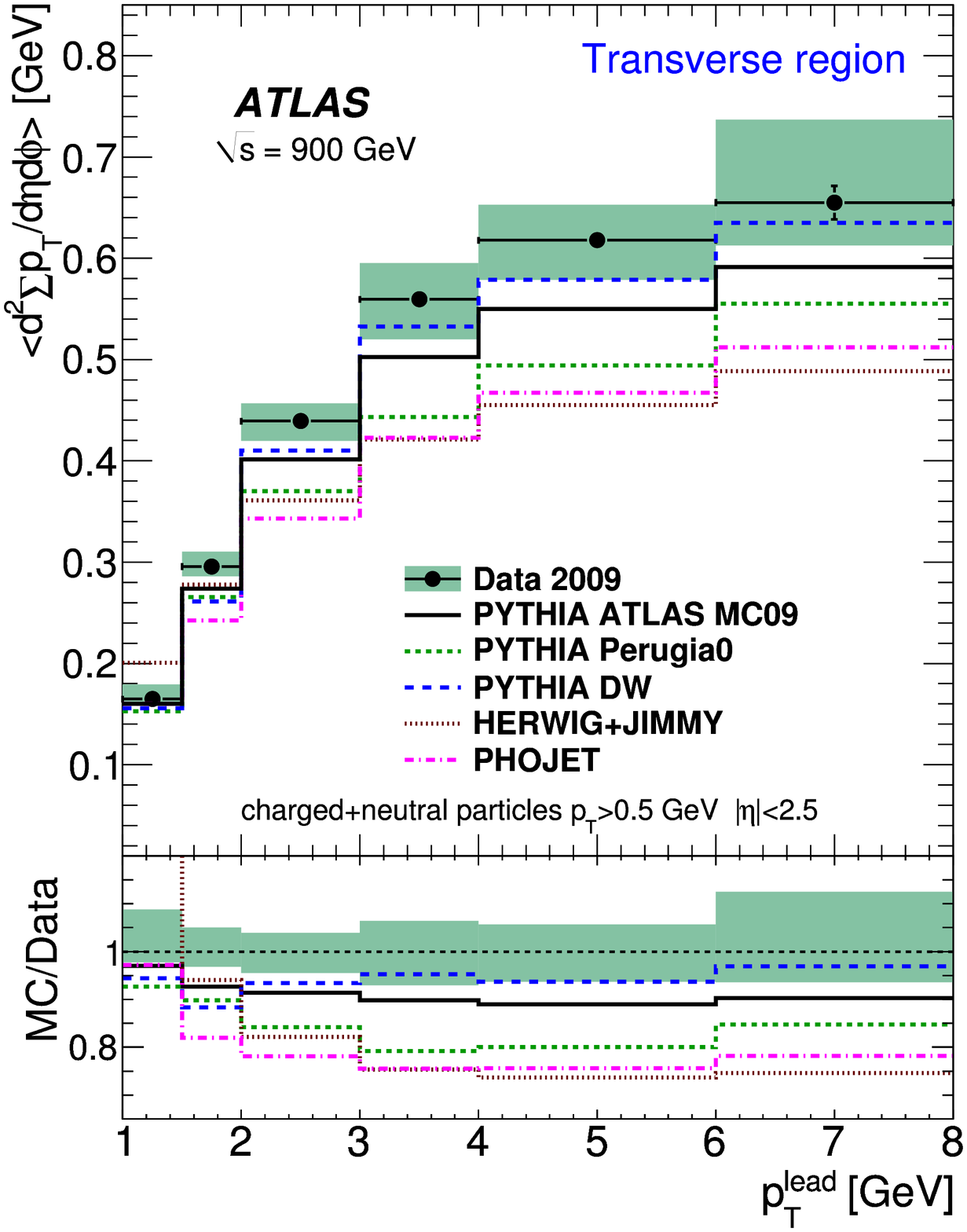}
  \includegraphics[width=0.4\textwidth]{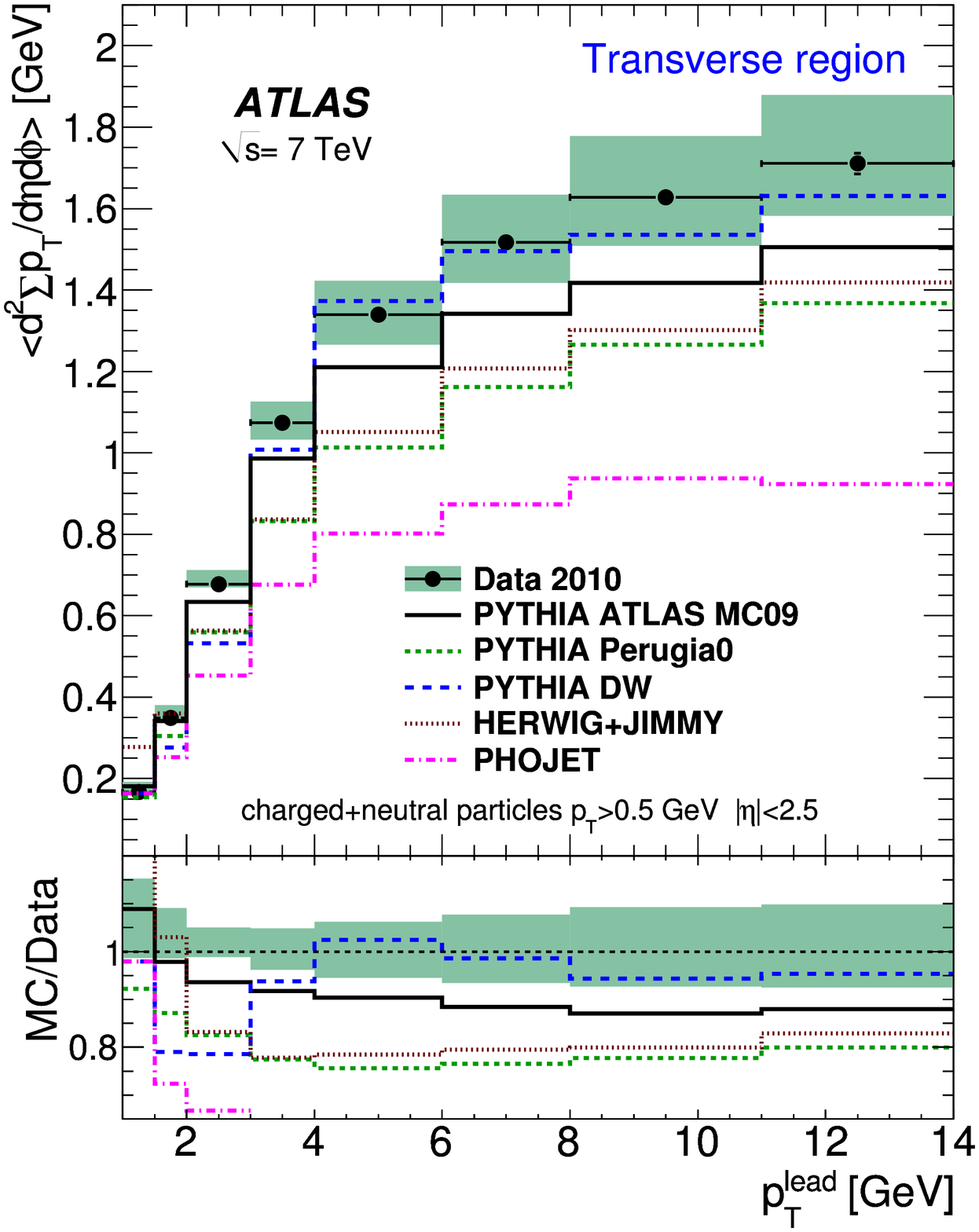}
  \caption{Underlying event measurements at 900\;GeV and 7\;TeV. The first row
    shows the correlation of mean track \pt and region multiplicity in the
    transverse region of the leading track UE analysis. The second row shows the
    emergence of azimuthal jet structure in the leading cluster analysis as the
    leading cluster scale increases; and the final row shows the sum of transverse
    region cluster \pt density as a function of leading cluster \pt.}
  \label{fig:ueexamples2}
\end{figure}

Similarly to the minimum bias analysis, the leading track analysis uses two
different track \pt cuts, 100 and 500\;MeV. The leading cluster analysis uses
all clusters. Examples of observables from these analyses are shown in
Figures~\ref{fig:ueexamples} and~\ref{fig:ueexamples2}. The dominant features are the ``ramp and plateau''
structure in the UE plots against leading object scale: this is the ``pedestal''
structure driven by the increase and saturation of hadron form factor overlap as
the hard event scale increases. Several connections between the plots are worth noting:
\begin{itemize}
\item the plateau heights represent roughly twice the charged particle number
  and \pt density as seen in the minimum bias analysis with the same cuts;
\item the level of UE plateau activity increases by a factor of two between
  900\;GeV and 7\;TeV;
\item a factor of 1.5 increase in $\sum \pt$ is seen in reducing the track \pt
  cut from 500 to 100\;MeV;
\item and the cluster-based analysis (which also measures neutral particles) has
  a higher \pt density plateau value than the most inclusive track-based
  observable, by a factor of roughly 30\%.
\end{itemize}

\subsection{Other soft QCD analyses}

Finally we summarise further analysis efforts which are either in progress or
which simply cannot be given due weight in this contribution.

Several other ATLAS analyses are in progress with direct relevance to soft QCD
and MPI. The most obvious are extensions of the leading track/cluster underlying
event analyses to use composite leading objects such as track and calorimeter
jets: both are ongoing, as is a UE measurement in $Z \to e^+ e^- / \mu^+ \mu^-$
events.

There are also a series of analyses probing the flavour and correlation
structure of minimum bias events: specifically the measurement of production
rates and \pt{s} of $\Lambda$ and $K_\text{s}$ hadrons, and correlations between
both pairs of charged particles and between matched forward/backward
pseudorapidity intervals.

A final set of analyses is focused on looking for specific soft QCD processes:
jet gap analyses provide a direct study of diffractive topologies at 7\;TeV,
while a set of explicit searches for hard double parton scattering (DPS) in
various event types will investigate the validity and universality of the
``$\sigma_\text{eff}$'' model used so far to formalise DPS calculations.

While not an explicit physics analysis, commissioning of pile-up modelling which
provides a good description of LHC data is also a major driver of MPI model
exploration and tune iteration.

\FloatBarrier
\section{Testing and tuning of MPI models}
\label{sec:tuning}

As already alluded to in Section~\ref{sec:mcmpi}, the various families of
general-purpose event generators have distinct philosophies of which modelling
aspects may be tuned, and which should be robustly predicted from theoretical
inputs.

A particularly controversial area is that of tuning the parton showers in the
\pythia generator family: parton showers are an iterative process whereby
partons recursively radiate by means of \emph{splitting functions} which are
collinear expansions of full LO QCD matrix elements. As the evolution scales and
splitting functions are definitively perturbative, generator authors disagree
about whether fudge factors can be justified on the scales and couplings in the
various types of parton shower -- although the rapid rise of shower-matched
higher-order matrix element generators such as POWHEG, AlpGen, and MC@NLO does
appear to be driving the field in a more theoretically constrained direction.

Much less controversial is tuning of processes which involve the divergent
strong coupling, such as hadronisation and MPI. As there is no IR-complete model
of QCD, any process which has to approach or transcend $\Lambda_\text{QCD}$ must
be phenomenologically constructed and hence possess free parameters to be tuned
to experimental data. The only disagreement is in the number of parameters
available: again, the \pythia family favours configurability, while the \herwig
and Sherpa generators attempt to be more minimal. Each has their place in
collider physics research, and much of the ubiquity of \pythiasix in the LHC
experiments is due to the ability to make its output look very much like data --
the price for this is reduced predictivity.

Typically hadronisation models introduce between 10 and 30 parameters: string
hadronisation models tend to require more parameters for flavour structure, as
the string breaking is not predictive about this, while cluster hadronisation
requires more parameters to fix the kinematics of cluster splitting. MPI models
add an extra 5 or more parameters, the number depending strongly on the degree
of available refinement in parameterising the proton form factor and the colour
reconnection mechanism. Tuning 30+ parameters all at once is not a
computationally feasible approach, even with modern semi-automated tools and
most certainly not if the tuning is done by hand. Hence some factorisation of
parameters is required: hadronisation, for example, if assumed to be universal
between lepton and hadron colliders, can be cleanly tuned to LEP, JADE, and SLD
data without any need for a functioning MPI configuration. A more specific
example is that of the $b$ and $c$ quark fragmentation functions, which usually
receive special treatment and can be tuned in isolation (if necessary) once a
base tune of the light quark fragmentation has been established. The observables
most sensitive to MPI (and shower) configuration may then be tuned using the
final state setup constructed from the $e^+ e^-$ data.

This approach is the one taken by ATLAS' tuning group, with the constraint to
data being made with the \rivet~\cite{Buckley:2010ar} analysis toolkit and
\professor~\cite{Buckley:2009bj} tune optimisation program. This toolchain has
been key to systematising the process of event generator tuning for the LHC, as
\rivet provides standardised and validated Monte Carlo versions of all the
relevant experimental analyses, and \professor is a numerically efficient and
scalable system for numerical optimisation of the event generator parameters
with respect to the reference data. The specific method used by \professor is as follows:
\begin{enumerate}
\item randomly sample points in the (possibly factorised) parameter space;
\item at each point run a full set of high-statistics MC runs, for every
  collider configuration and process type that should contribute to the
  tuning. This part may take several days for each run -- hence serial
  optimisation with a standard gradient descent minimiser is impractical -- and
  hence the scalability of the \professor system relies on the ability to
  batch-parallelise this generation step;
\item for each bin of each distribution taken independently, use the
  pseudoinverse method via a singular value decomposition to algebraically
  determine optimal coefficients for an arbitrary-order polynomial
  parameterisation of the bin value as a function of the MC parameters. Special
  treatments are also made for the statistical and theoretical errors;
\item the many bin parameterisations are aggregated and compared to the
  reference data to compute a goodness of fit measure. This is then numerically
  optimised, since evaluating the generator observable predictions is now
  exceedingly fast: the result is an optimal generator tune.
\end{enumerate}

The same speed of evaluation of the parameterised MC generator means that
\professor can also:
\begin{itemize}
\item provide an interactive GUI explorer for generator configurations;
\item use multiple equivalent parameterisations to obtain an error estimate on
  the accuracy of the parameterisation;
\item and compute objective error tunes (or ``eigentunes'') similar to the
  error sets produced in PDF fitting using a Hessian formalism.
\end{itemize}

\subsection{ATLAS tunes of \pythiasix, \pythiaeight, and \herwigjimmy}

ATLAS has made several iterations of MC tunes, particularly for the \pythiasix
generator which until recently has been responsible for the vast bulk of ATLAS
simulation production. ATLAS' involvement in generator tuning began with the
\pythiasix and \herwigjimmy ``MC08'' tunes in 2008, and began using the
automated \rivet and \professor tools with elements of the MC09 pre-LHC tunes.
The advent of early LHC data provided the first constraints on MPI models at
7~TeV (as well as an extra low-energy $pp$ point at 900~GeV), and the ATLAS
minimum bias observables of Section~\ref{sec:mb}, as well as the UE data at
900~GeV, were used to drive the \pythiasix AMBT1~\cite{ATLAS:1266235}, and the
\herwigjimmy AUET1~\cite{ATLAS:1303025} tunes in 2010.

The most recent tuning series from ATLAS has for the first time incorporated
\pythiaeight into the tuning, as part of the general migration of MC simulation
to use the C++ era replacements for the venerable Fortran generators. The second
round of ATLAS tuning to its own data tunes both the initial state shower and
the MPI model in \pythiasix, in an attempt to describe both hard and soft QCD
modelling with a single configuration. The tunings of \pythiaeight and
\herwigjimmy were restricted to the MPI modelling only, in the case of
\pythiaeight because its description of jet structure was already very good,
unlike \pythiasix AMBT1. In all cases, the tunings were performed for a range of
leading order and MC-adapted LO (or ``mLO'') PDFs, with NLO PDFs and tuning of hybrid
generators such as AlpGen and POWHEG being reserved for a later study.

The shower tuning stage for \pythiasix used CDF and ATLAS jet shape and
track-jet fragmentation data~\cite{Acosta:2005ix,Aad:2011kq,Aad:2011sc}, D\O{}
and ATLAS dijet azimuthal decorrelation data~\cite{d0dijets,daCosta:2011ni}, and
CDF $Z$ \pt data~\cite{drellyan_zeept}. The latter study turns out to
unfortunately bias the shower tune to the detriment of the ATLAS data $Z$ \pt,
which will be addressed in the \emph{next} round of ATLAS shower tuning. The
hadronisation, including specific $b$-fragmentation behaviour, had previously
been tuned to LEP data, and the fitting weights were chosen to bias the fit
toward a good description of the ATLAS observables. The results were largely
successful, with particular improvement of jet structure observable description
although at the cost of multiple $\Lambda_\text{QCD}$ values in the code -- this
latter point has much relevance to the interaction of showers with higher-order
matrix element generators and is under active investigation. At the close of the
AUET2 tune construction (which also includes and MPI tune), it was observed that
the Perugia2010 tune series which had inspired the shower treatment resulted in
some perverse behaviours of \pythiasix-derived non-perturbative corrections in
QCD jet studies: as a result of this, a second tuning round for the ATLAS MC11
production was started, which used the more conventional shower configuration of
AMBT1 but tuned the three parameters required to optimise jet structure and
near-$\pi$ dijet decorrelation description. This tune series was also successful
at describing jet data and, denoted as AMBT2B/AUET2B, was extended to include
the \lostar, \lostst, CT09MC2, CTEQ6L1, and MSTW2008LO PDFs~\cite{mrstlostar},
all of which were then used as a base for MPI tuning.

For the MPI tuning of all three generators, the ATLAS data described in
Sections~\ref{sec:mb} and~\ref{sec:ue} was used, in addition to the full range
of minimum bias and underlying event measurements from CDF: minimum
bias~\cite{Acosta:2001rm,Aaltonen:2009ne}, leading track UE at
1800~GeV~\cite{Affolder:2001xt}, leading jet UE in jet events at 630, 1800, and
1960~GeV~\cite{Acosta:2004wqa,Aaltonen:2010rm}, and the UE in $Z \to e^+e^-$ Drell-Yan events
at 1960~GeV~\cite{Aaltonen:2010rm}. The weights in the fit were again chosen to bias the fit
toward a good description of the ATLAS observables at the potential expense of
the Tevatron ones, and for the 7~TeV data more than the 900~GeV data: this is
designed to optimise the description of the observables that ATLAS most needs to
simulate for the next two years. As the \jimmy MPI model specifically has no
treatment of purely soft scattering, i.e. it is only formulated in the presence
of a hard scattering and makes no attempt to IR-complete the MPI scatterings
below \ptmin, it was only tuned to UE data. The Pythia-like energy evolution
ansatz from eq.~\eqref{eq:ptminevol} was manually applied to \jimmy's \ptmin
value by means of sampled meta-parameters.

\begin{figure}[p]
  \centering
  \includegraphics[width=0.4\textwidth]{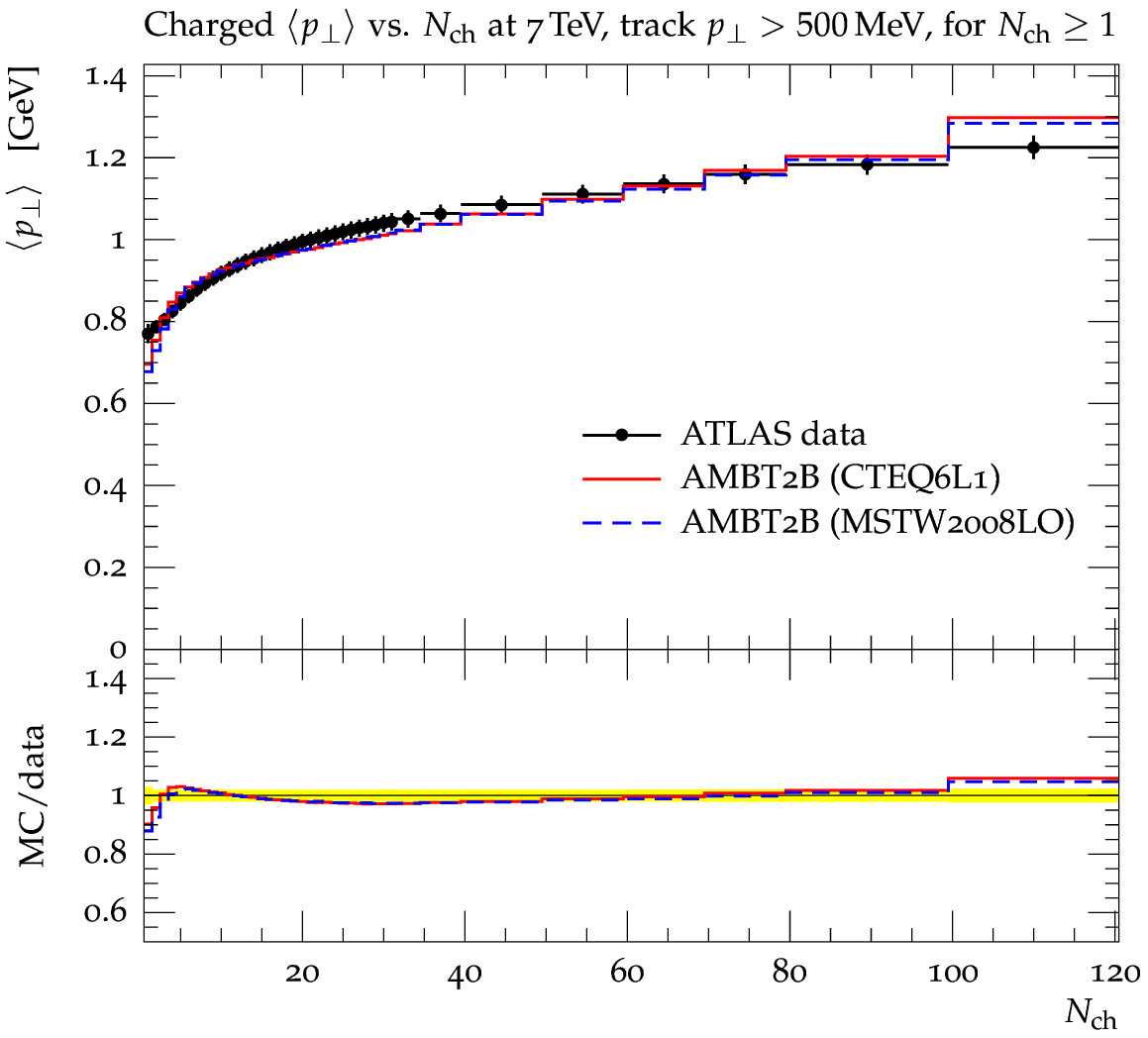}
  \includegraphics[width=0.4\textwidth]{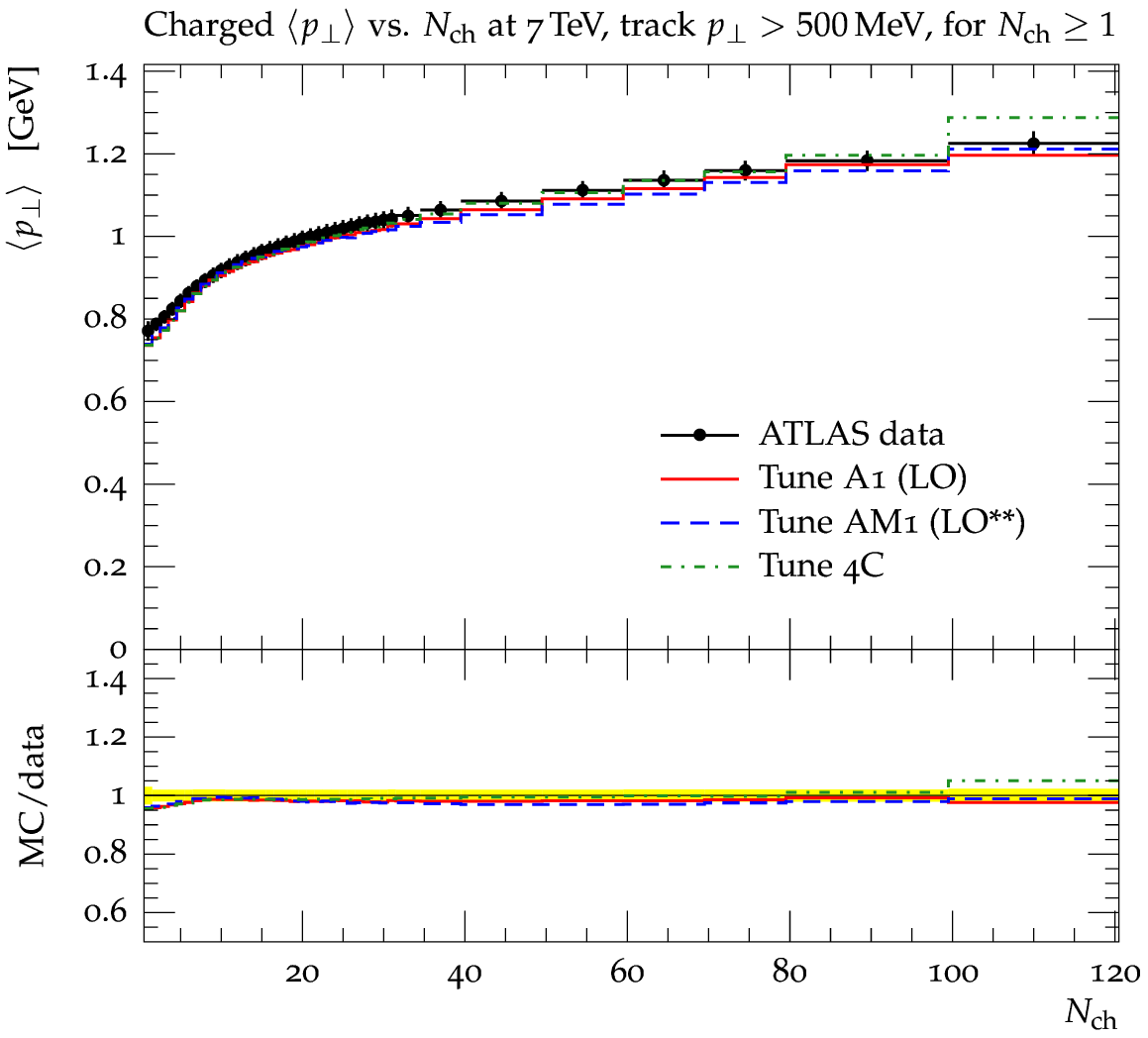}\\
  \includegraphics[width=0.4\textwidth]{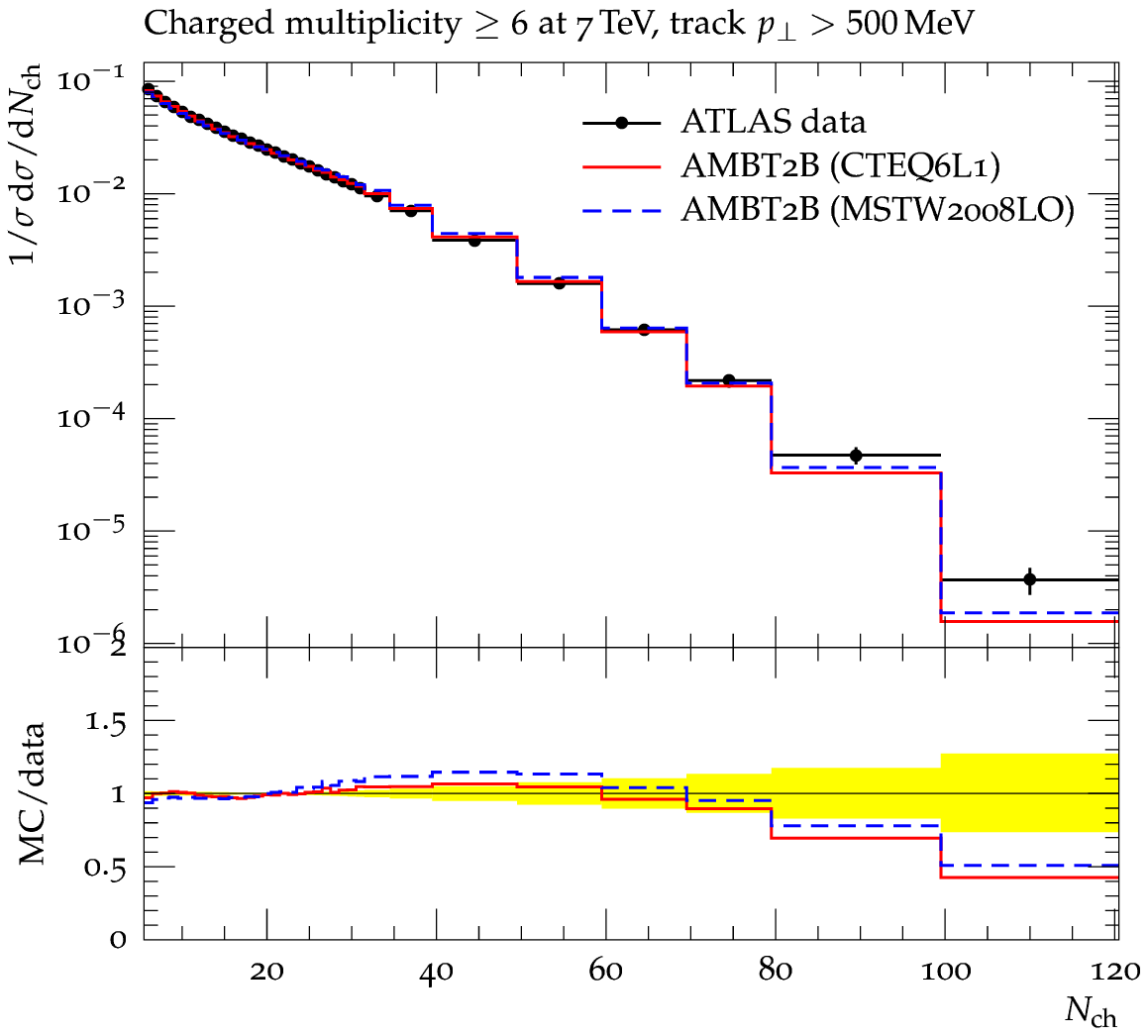}
  \includegraphics[width=0.4\textwidth]{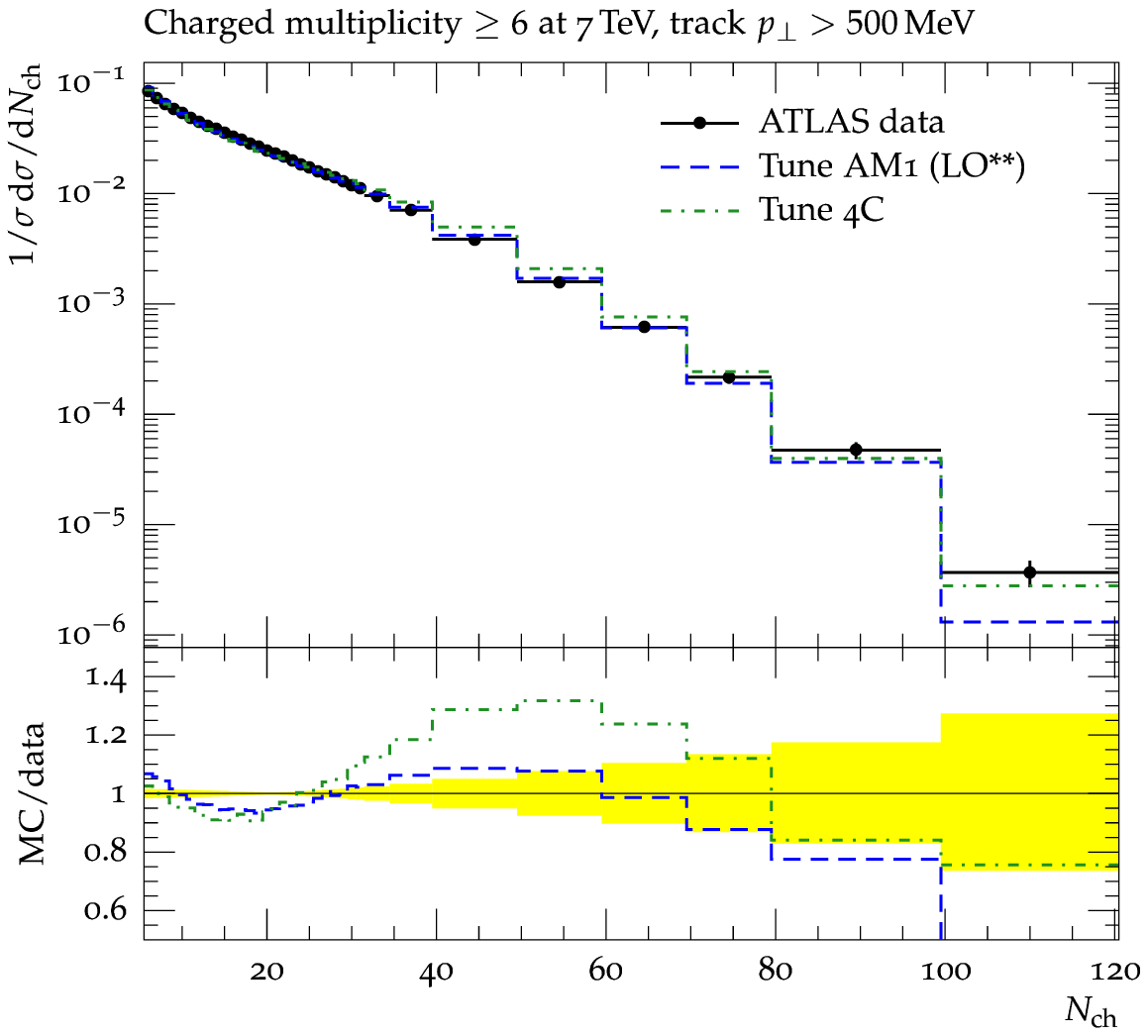}\\
  \includegraphics[width=0.4\textwidth]{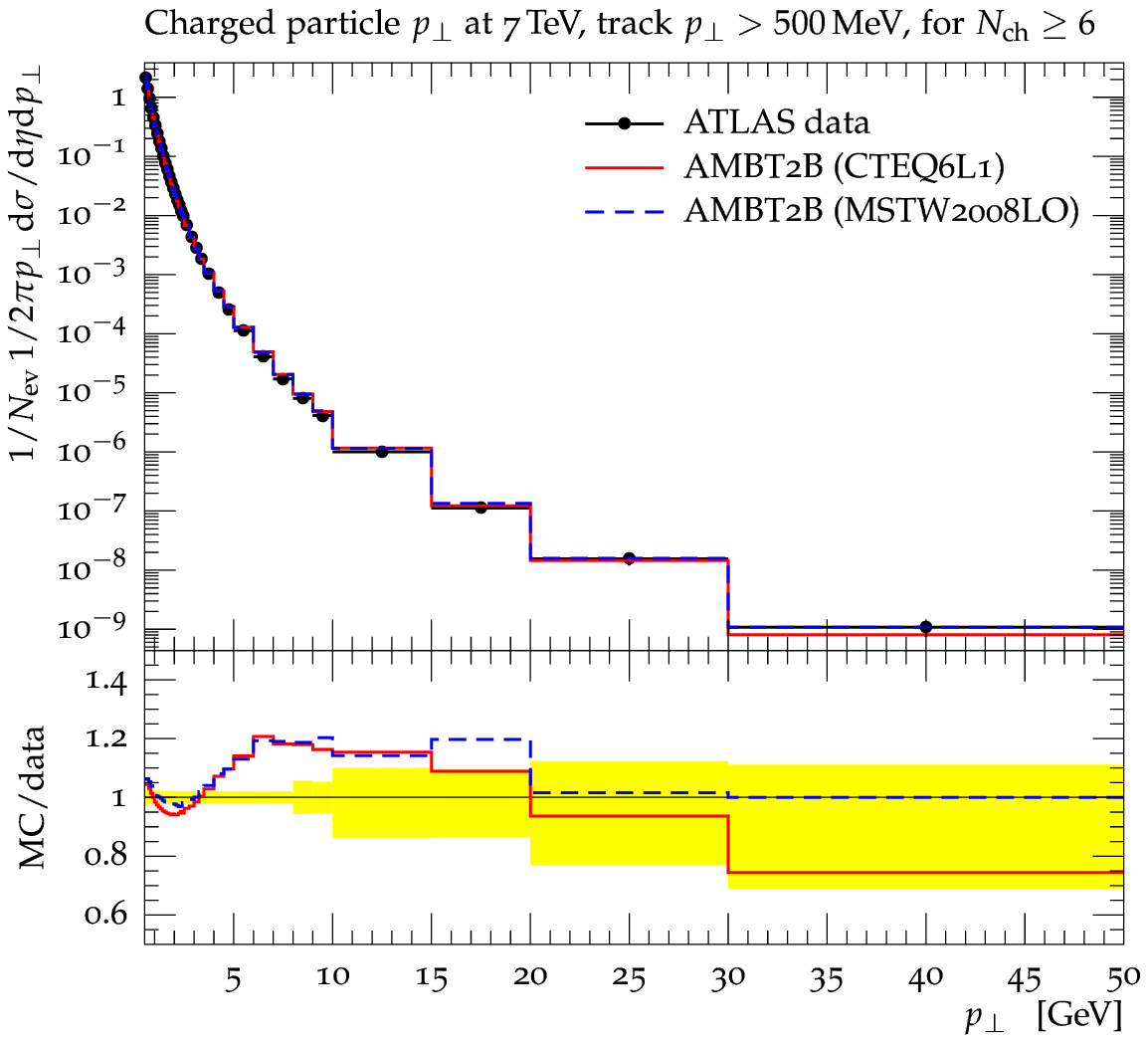}
  \includegraphics[width=0.4\textwidth]{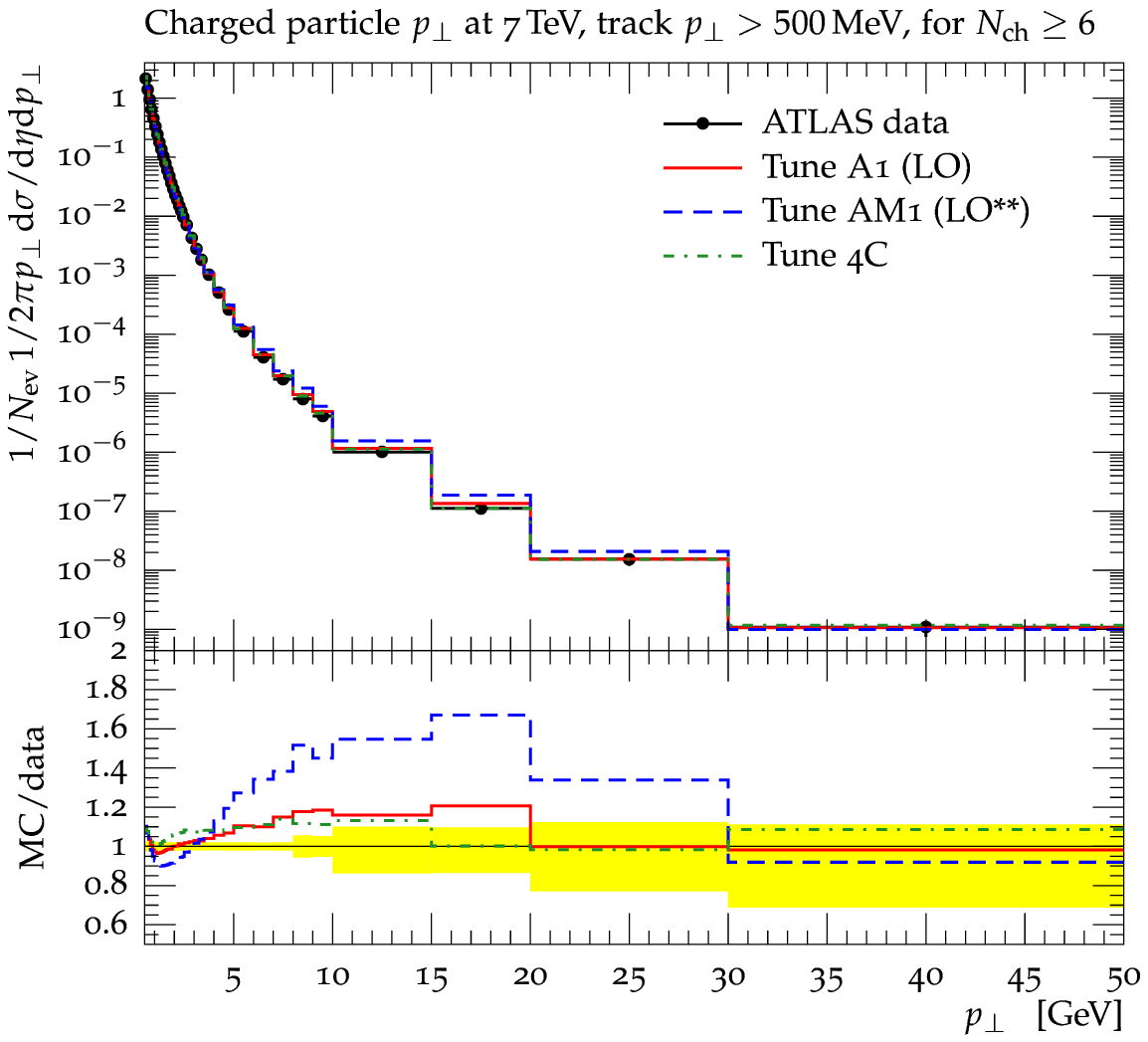}
  \caption{ATLAS minimum bias tunes of \pythiasix (left column) for LO PDFs and
    \pythiaeight (right column) for an LO and an mLO PDF, compared to ATLAS
    minimum bias observables (and the 4C author tune of \pythiaeight). The data
    description is generally good, with most difficulty being observed in the
    description of the \pt spectrum. This is particularly the case for the
    \lostst PDF tune of \pythiaeight -- a similar effect is seen for \pythiasix
    in the next figure.}
  \label{fig:mbtuning}
\end{figure}

\begin{figure}[tp]
  \centering
  \includegraphics[width=0.4\textwidth]{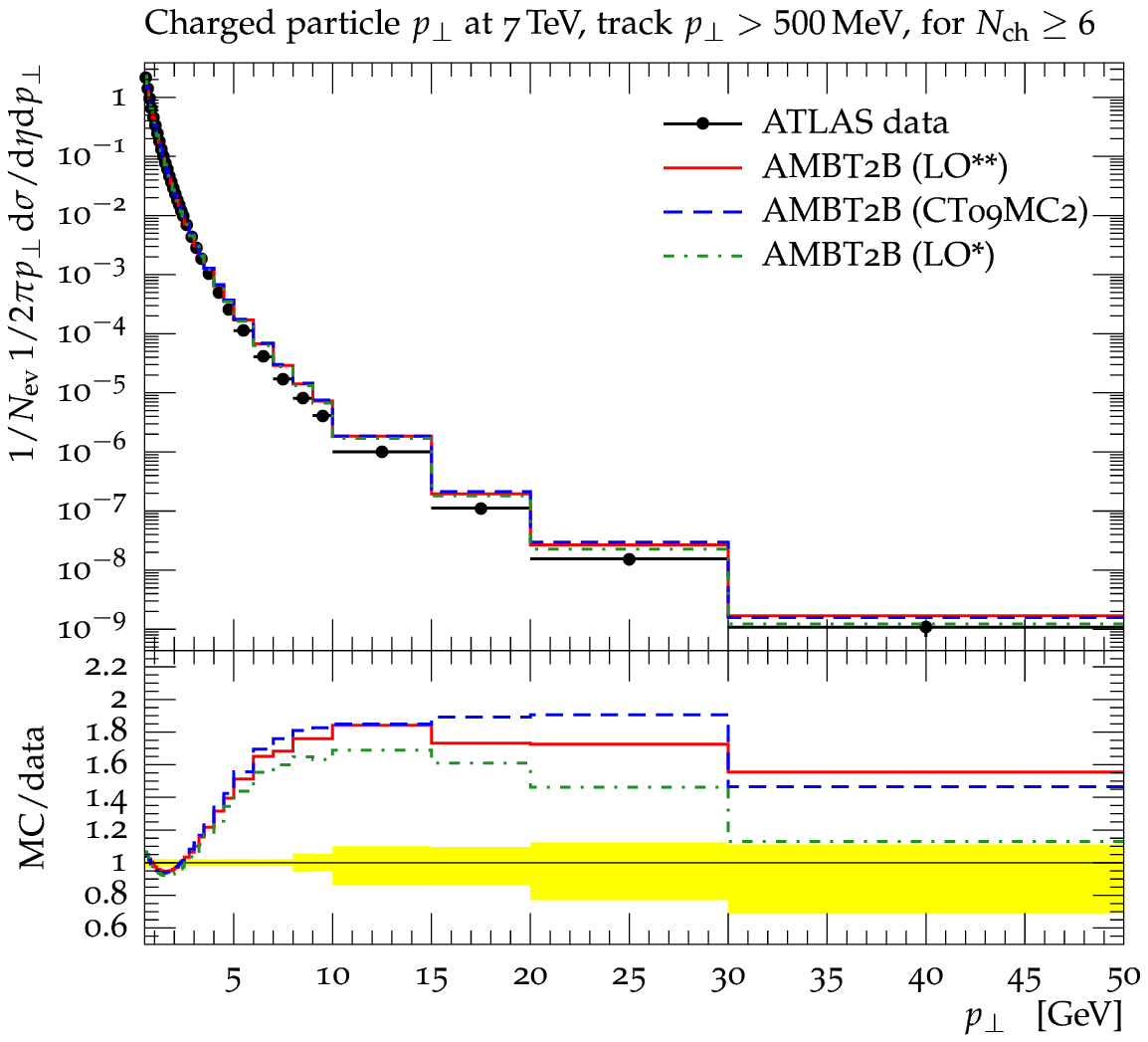}
  \includegraphics[width=0.4\textwidth]{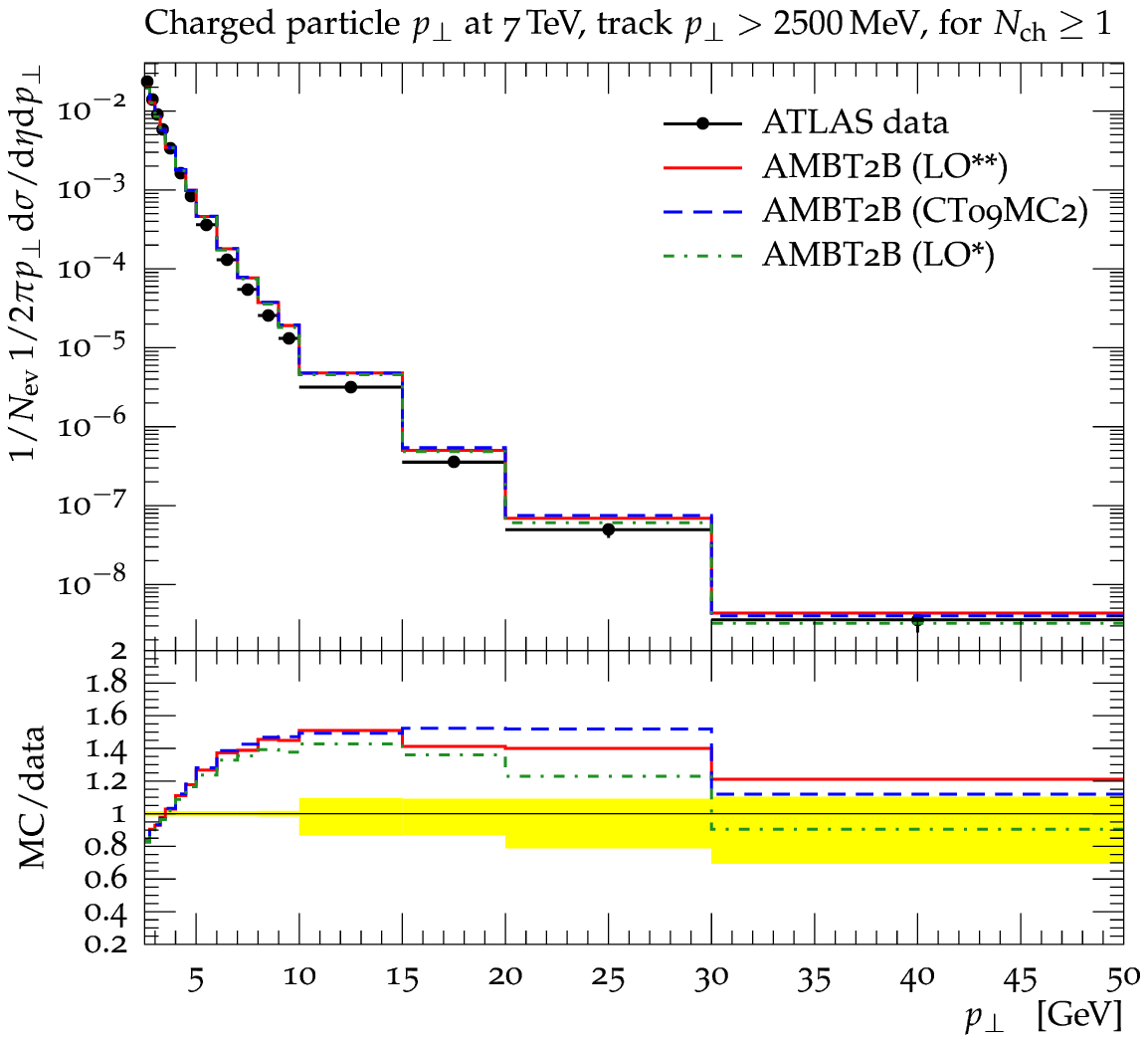}
  \caption{ATLAS tunes of \pythiasix to mLO PDFs, compared to ATLAS minimum bias
    \pt spectrum data, for two different track \pt cuts. In both cases large
    excesses above data are seen from 5--30~GeV, up to a factor of 1.8 excess
    for the lower track \pt cut. This feature was found to be a regular result
    of using mLO PDFs, and could not be significantly altered by any use of MPI
    model parameters.}
  \label{fig:mbtuningpy6mlo}
\end{figure}

\begin{figure}[p]
  \centering
  \includegraphics[width=0.33\textwidth]{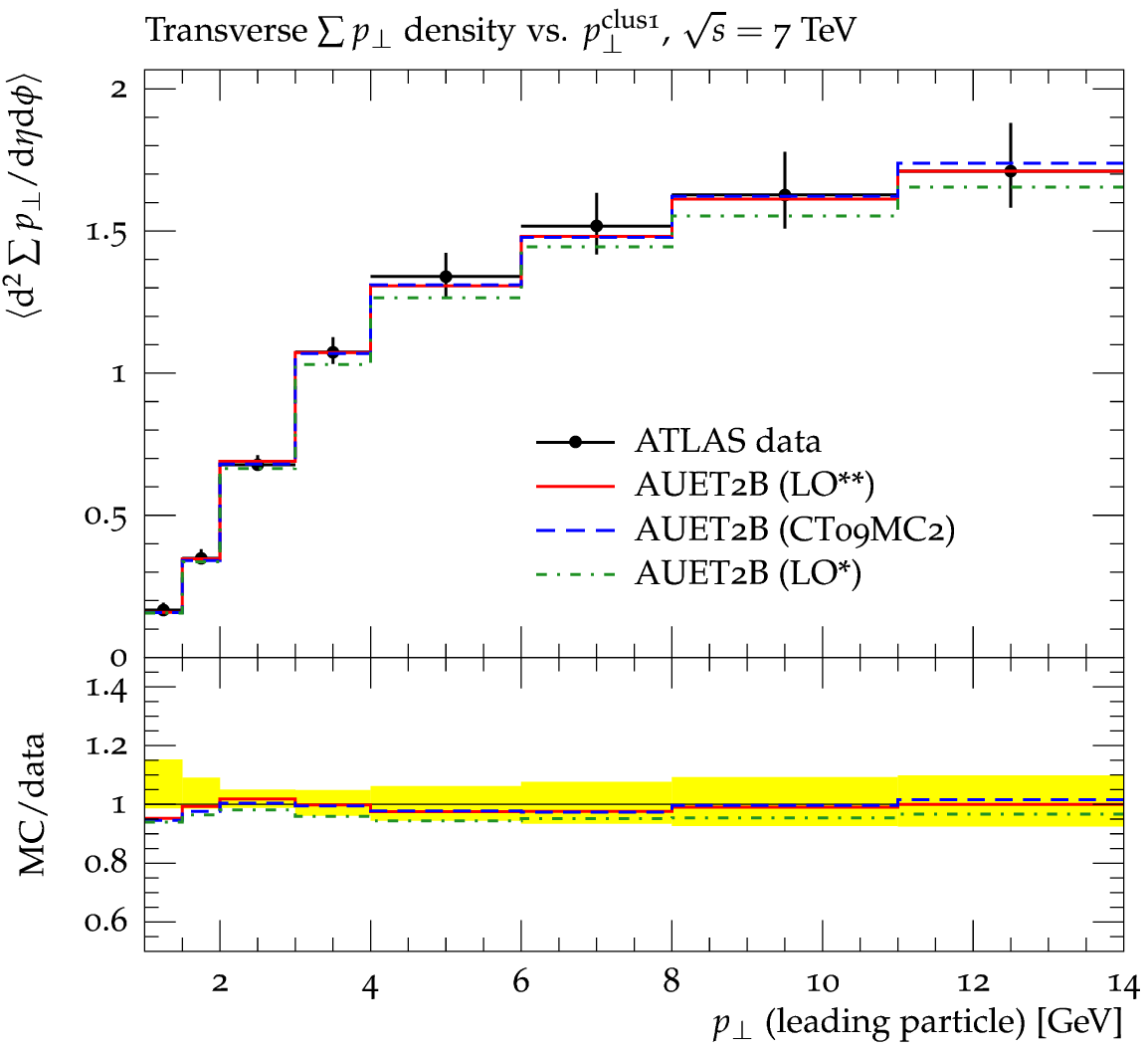}
  \includegraphics[width=0.33\textwidth]{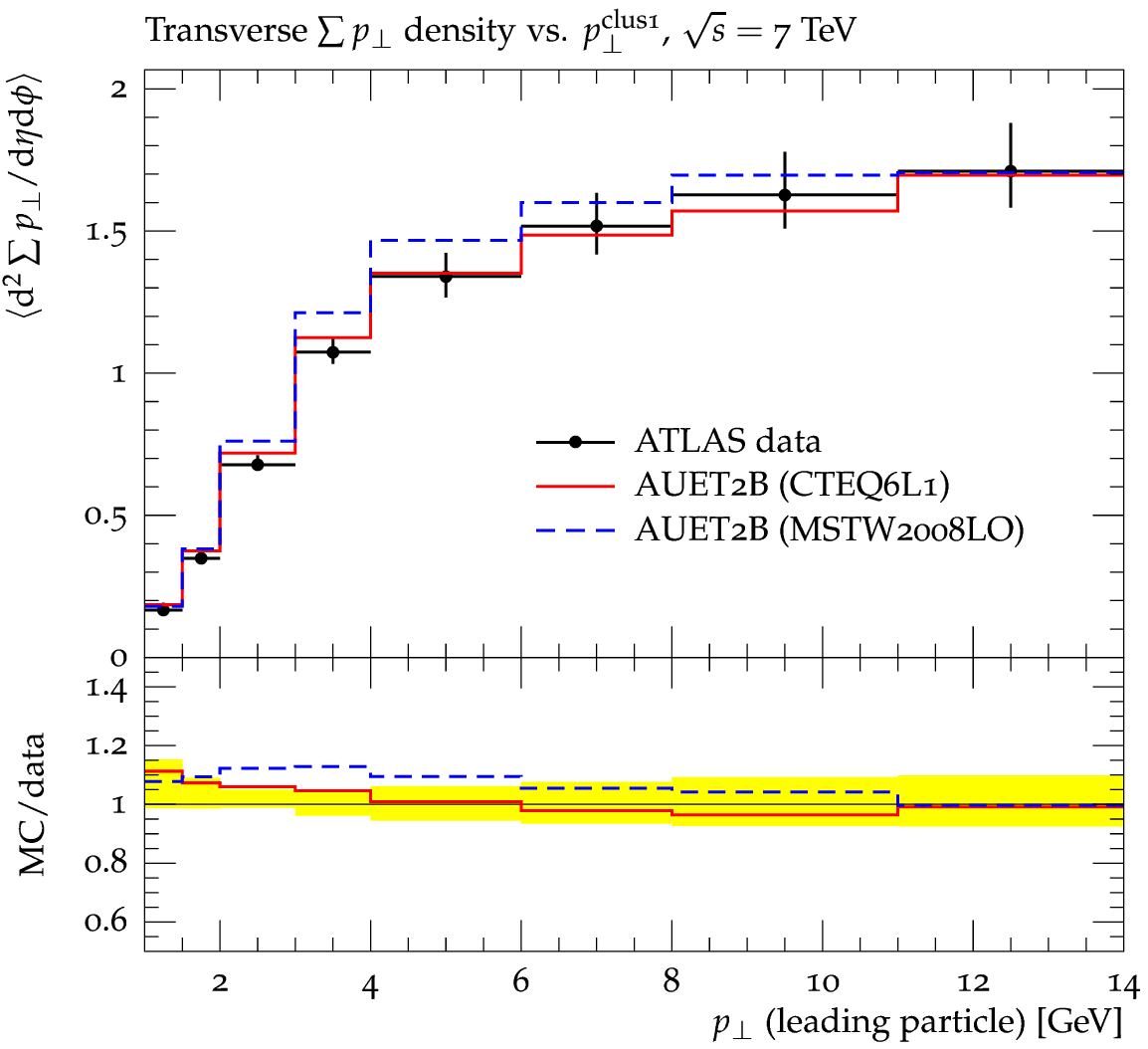}\\
  \includegraphics[width=0.33\textwidth]{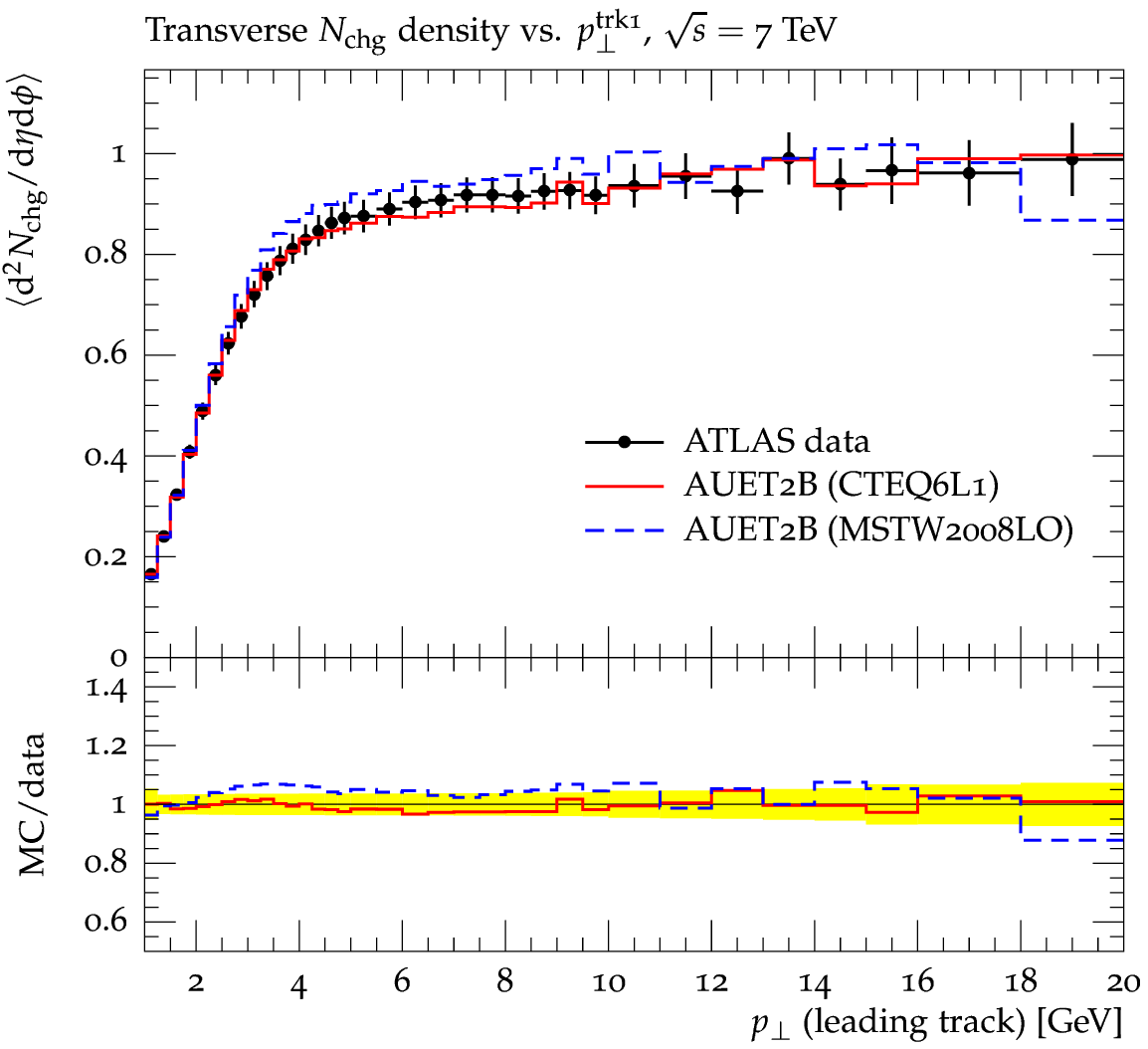}
  \includegraphics[width=0.33\textwidth]{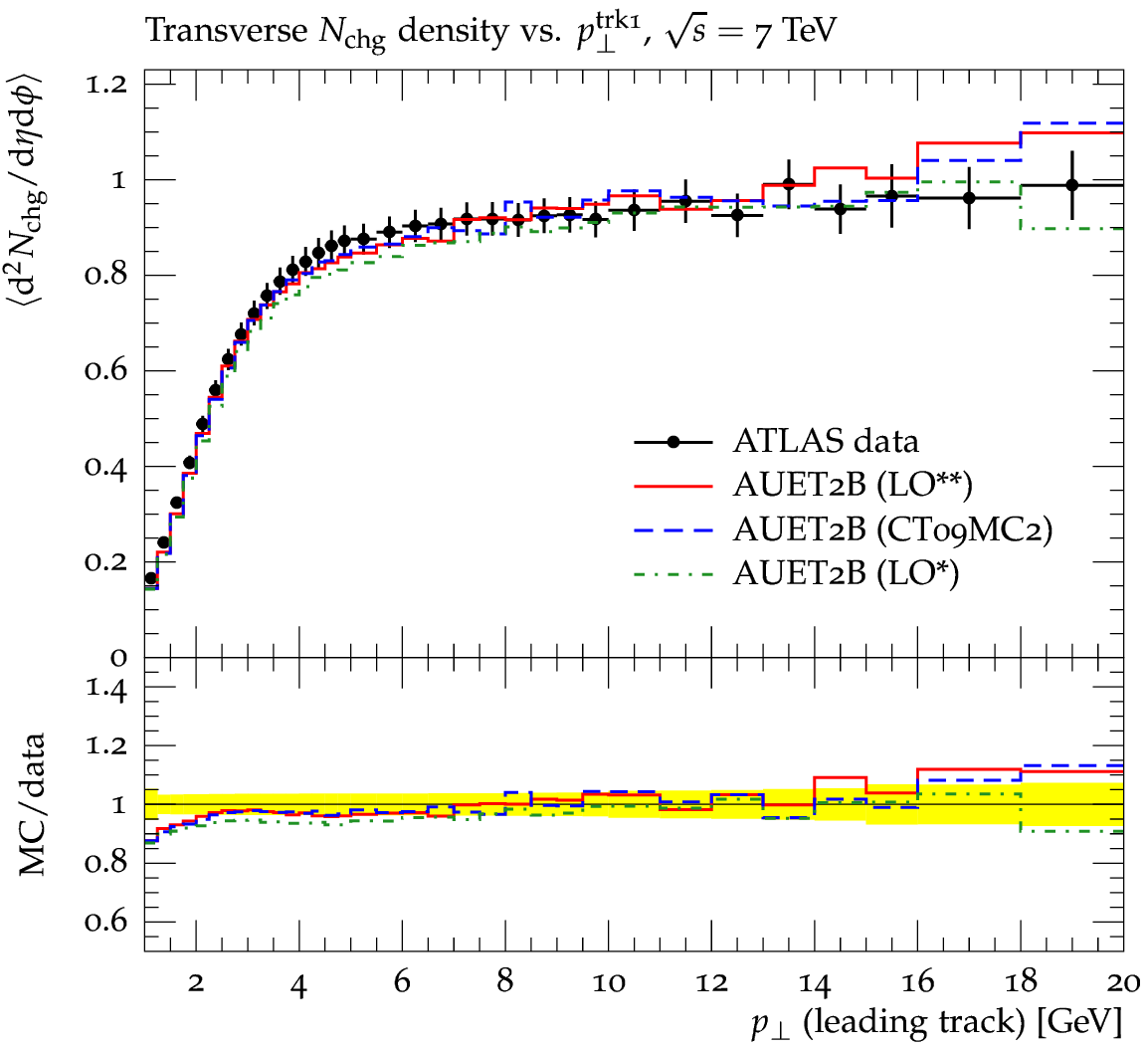}\\
  \includegraphics[width=0.33\textwidth]{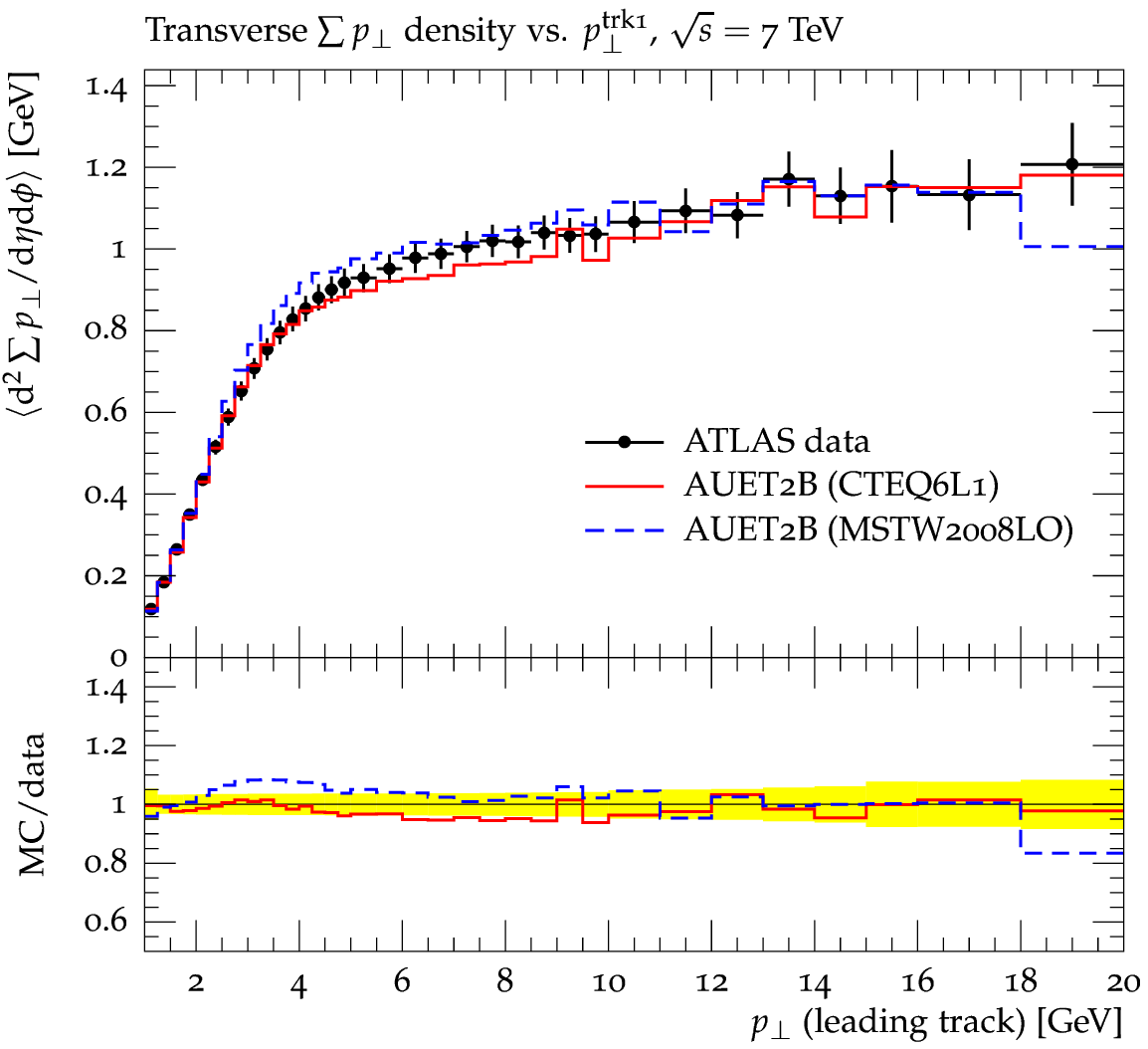}
  \includegraphics[width=0.33\textwidth]{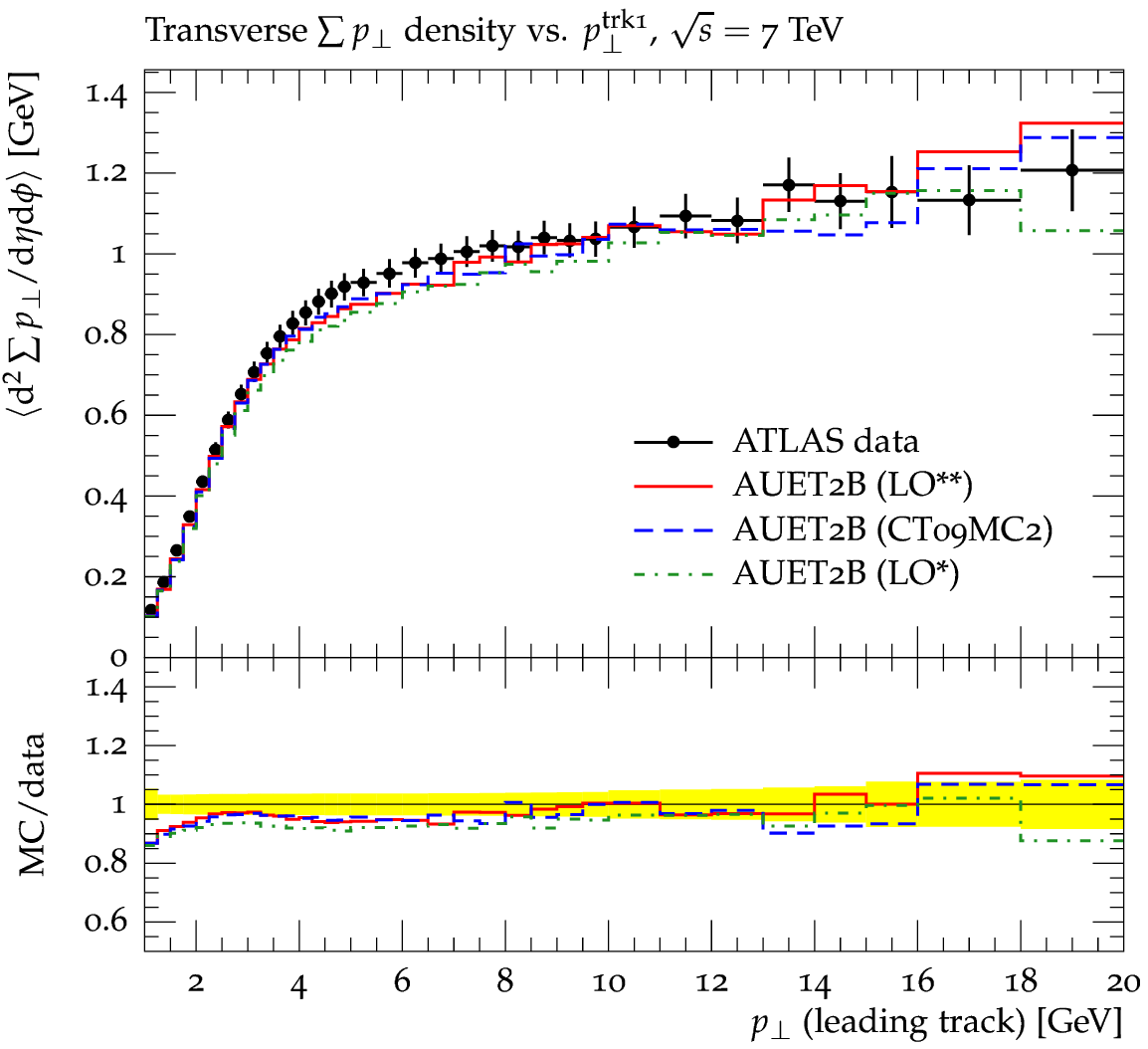}
  \includegraphics[width=0.33\textwidth]{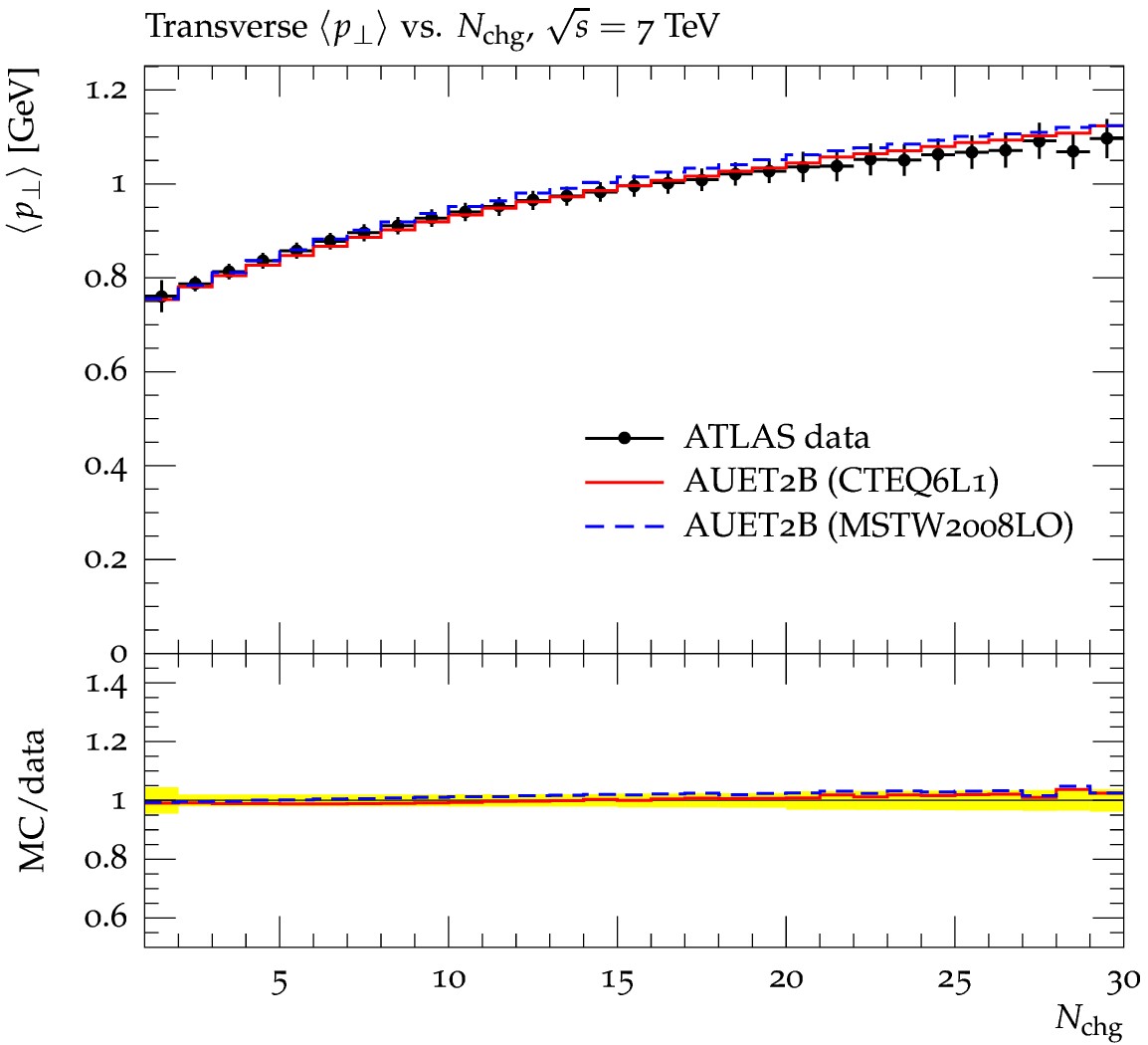}
  \includegraphics[width=0.33\textwidth]{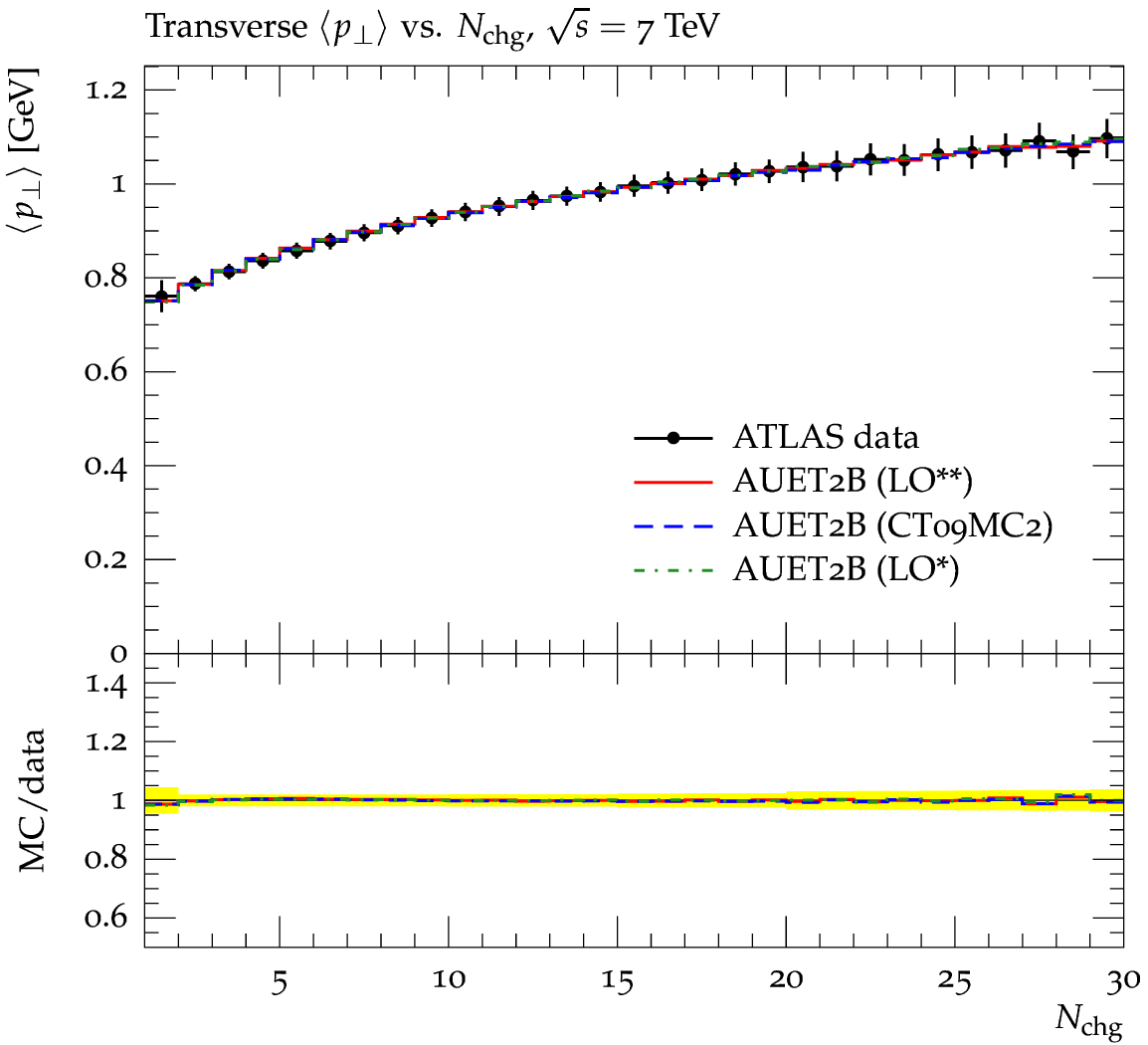}\\
  \caption{ATLAS tunes of \pythiasix compared to ATLAS underlying event data,
    for LO (left column) and mLO (right column) PDF tunes. Both types of PDF
    describe UE data well, and so there is no significant MPI-oriented problem
    to using mLO PDFs for simulation of the UE in hard-scale event simulation,
    unlike the case for minimum bias generation seen in the previous
    figure. There does appear to be a slight effect of mLO PDFs in that all
    tunes using them slightly undershoot the ``turn-over'' region of the
    $N_\text{ch}$ and $\sum \pt$ profiles.}
  \label{fig:uetuningpy6}
\end{figure}

\begin{figure}[tp]
  \centering
  \includegraphics[width=0.4\textwidth]{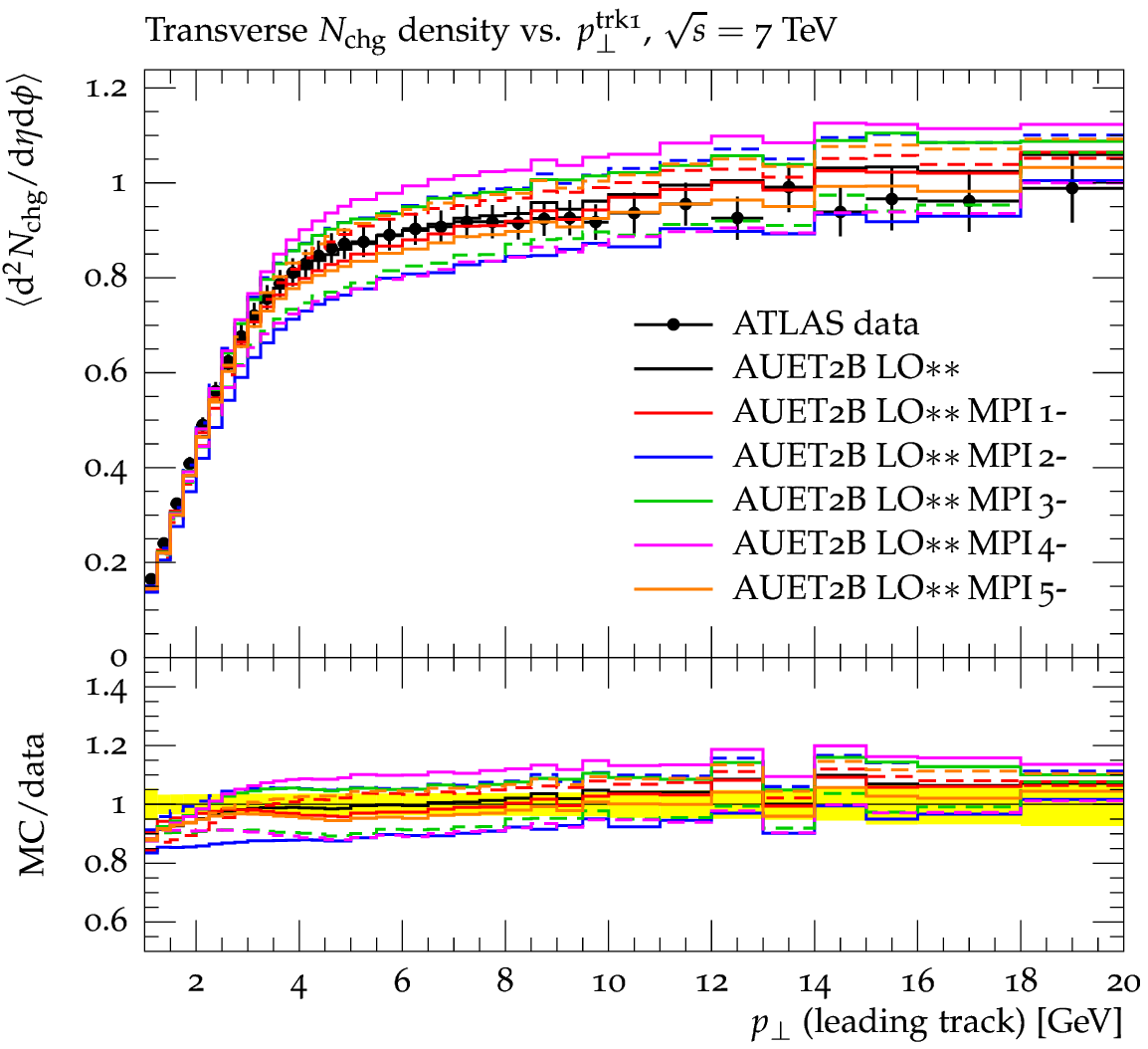}
  \includegraphics[width=0.4\textwidth]{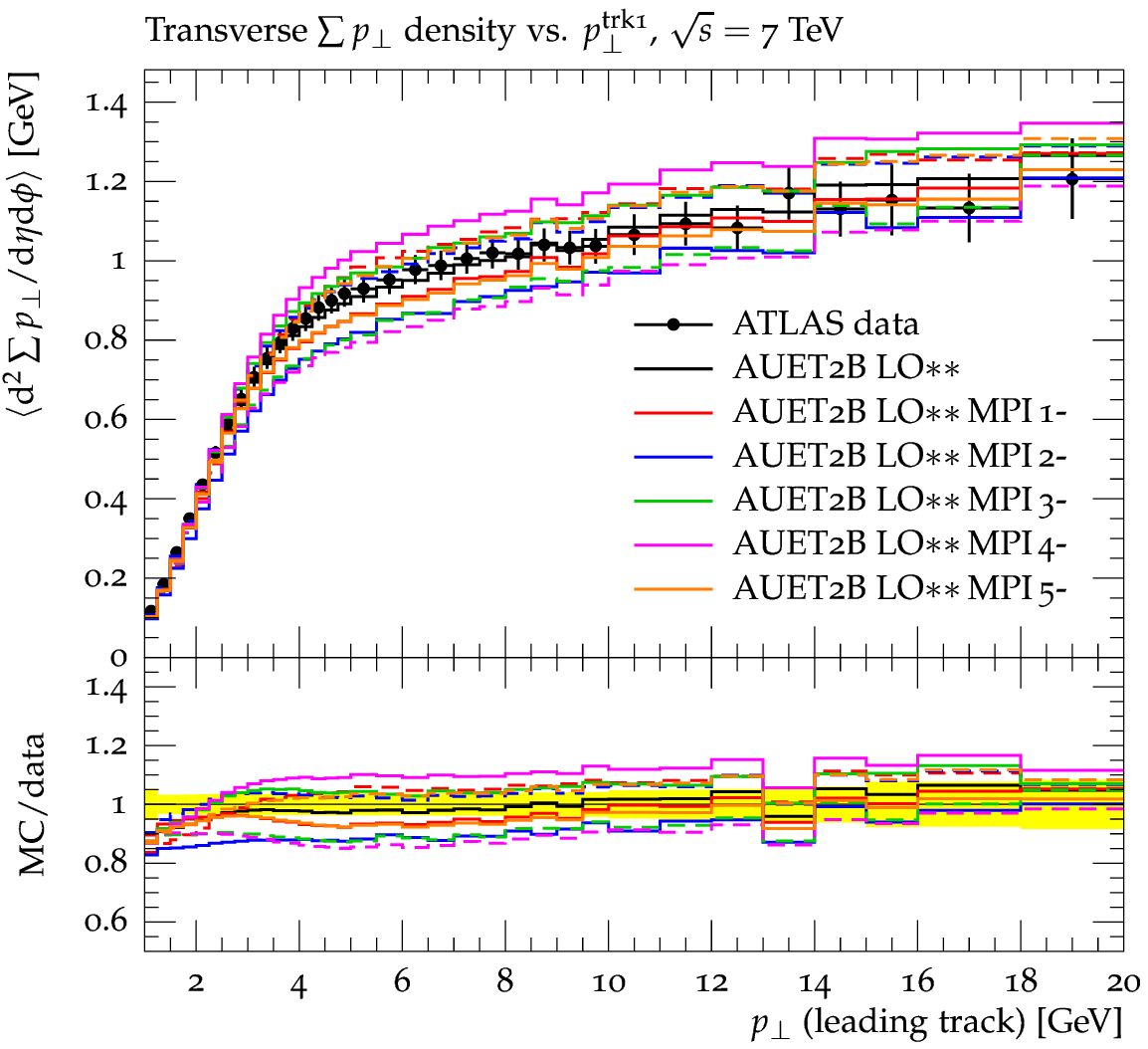}
  \caption{ATLAS systematic MPI error tunes of \pythiasix compared to ATLAS UE
    data at 7~TeV. These tunes have been constructed similarly to PDF Hessian
    error sets, by requiring fixed deviations in goodness of fit from the
    optimised tunes along diagonalised principle directions in the parameter
    space, and hence provide a set of tunes which quantitatively represent
    the uncertainties in \pythiasix MPI modelling and ATLAS UE data.}
  \label{fig:uetuningpy6eigen}
\end{figure}

\begin{figure}[p]
  \centering
  \includegraphics[width=0.4\textwidth]{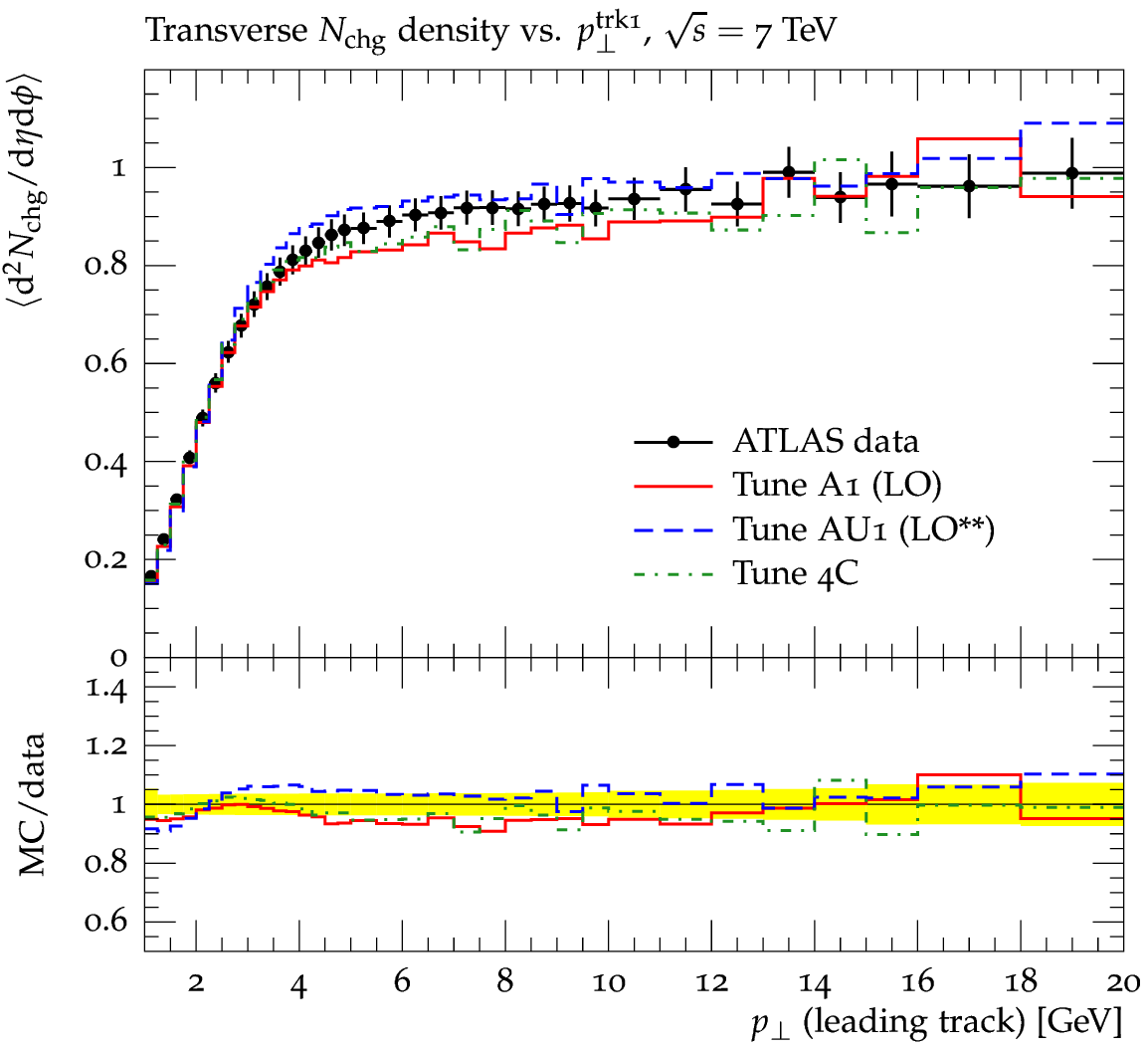}
  \includegraphics[width=0.4\textwidth]{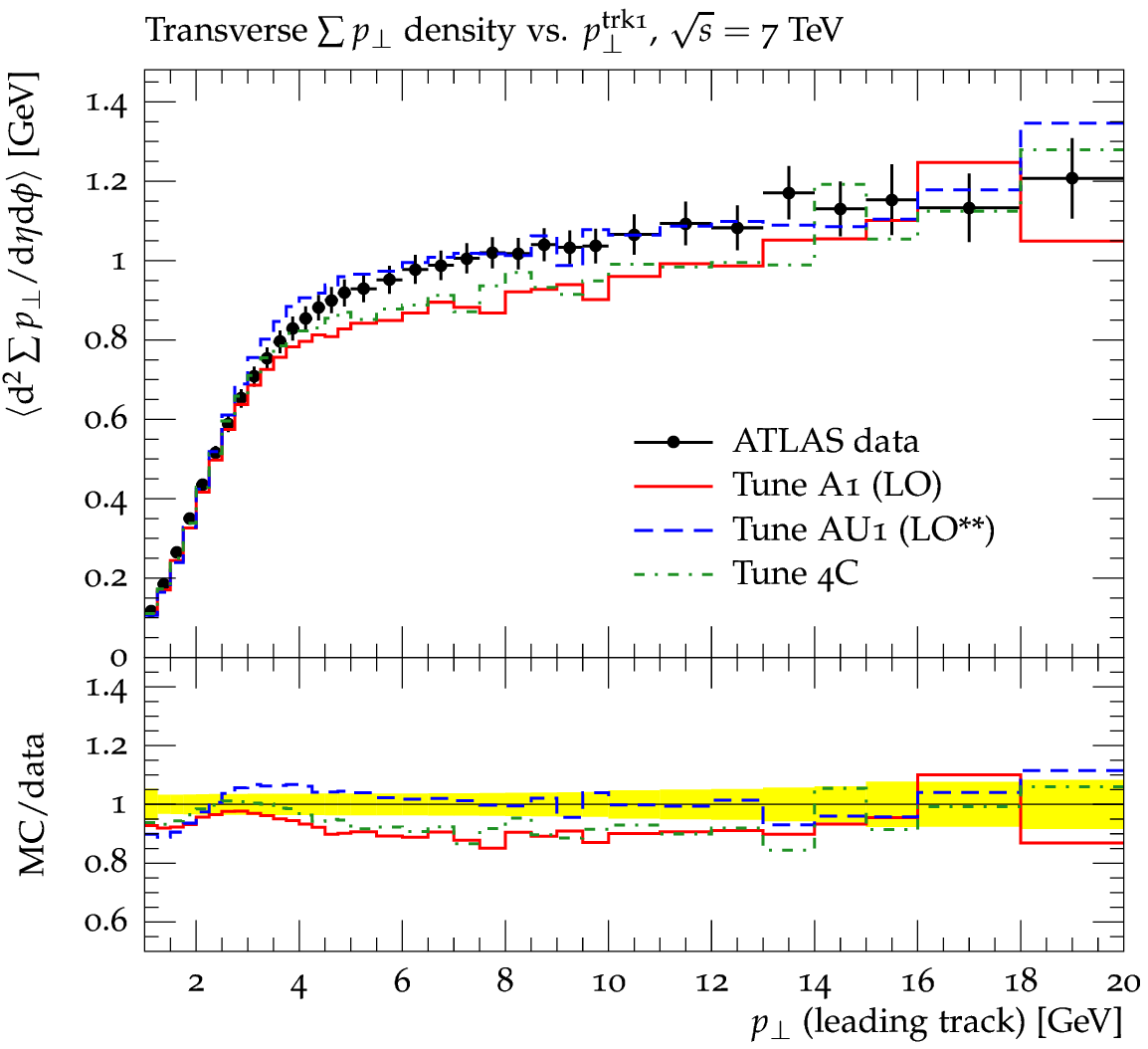}\\
  \includegraphics[width=0.4\textwidth]{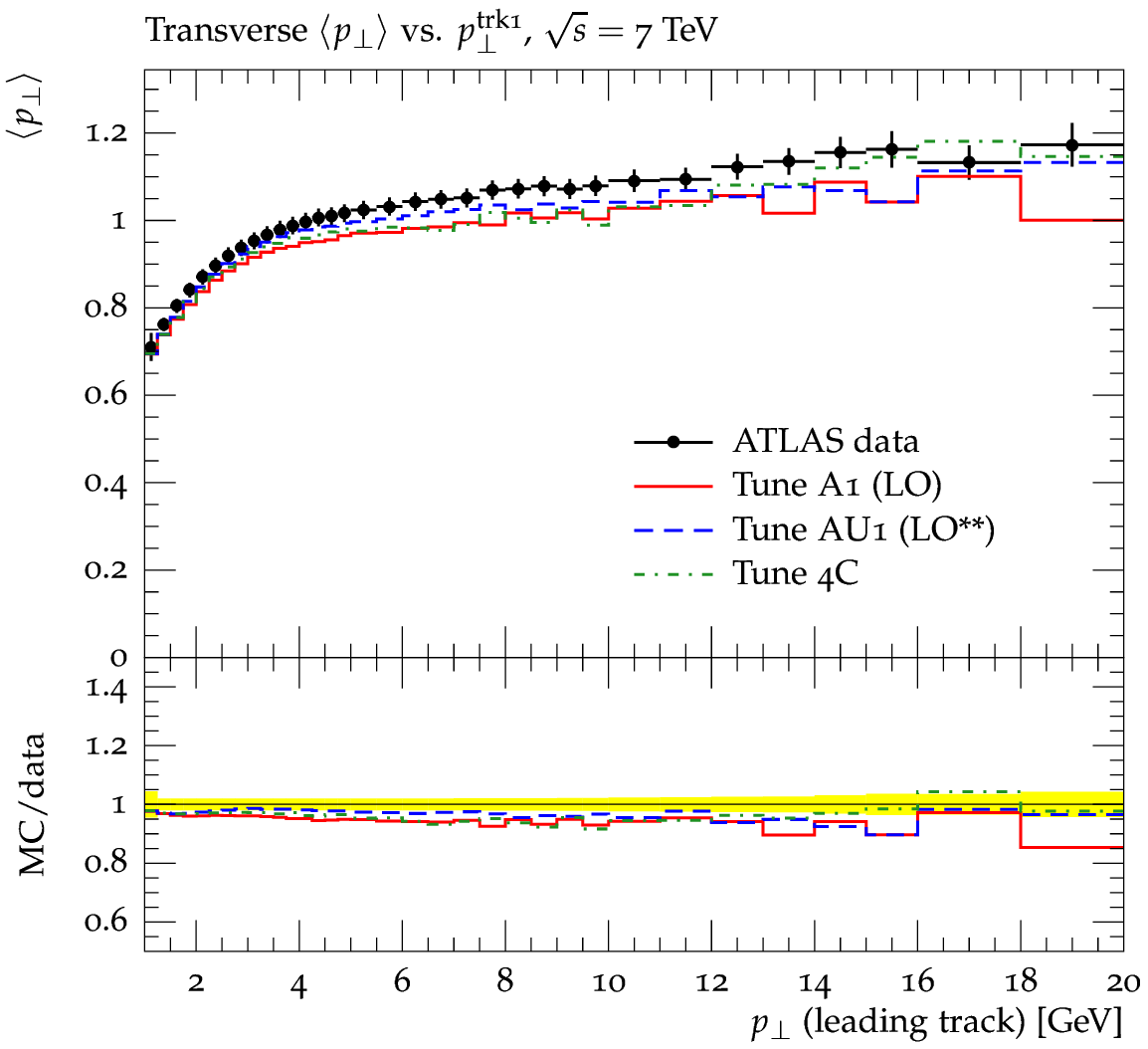}
  \includegraphics[width=0.4\textwidth]{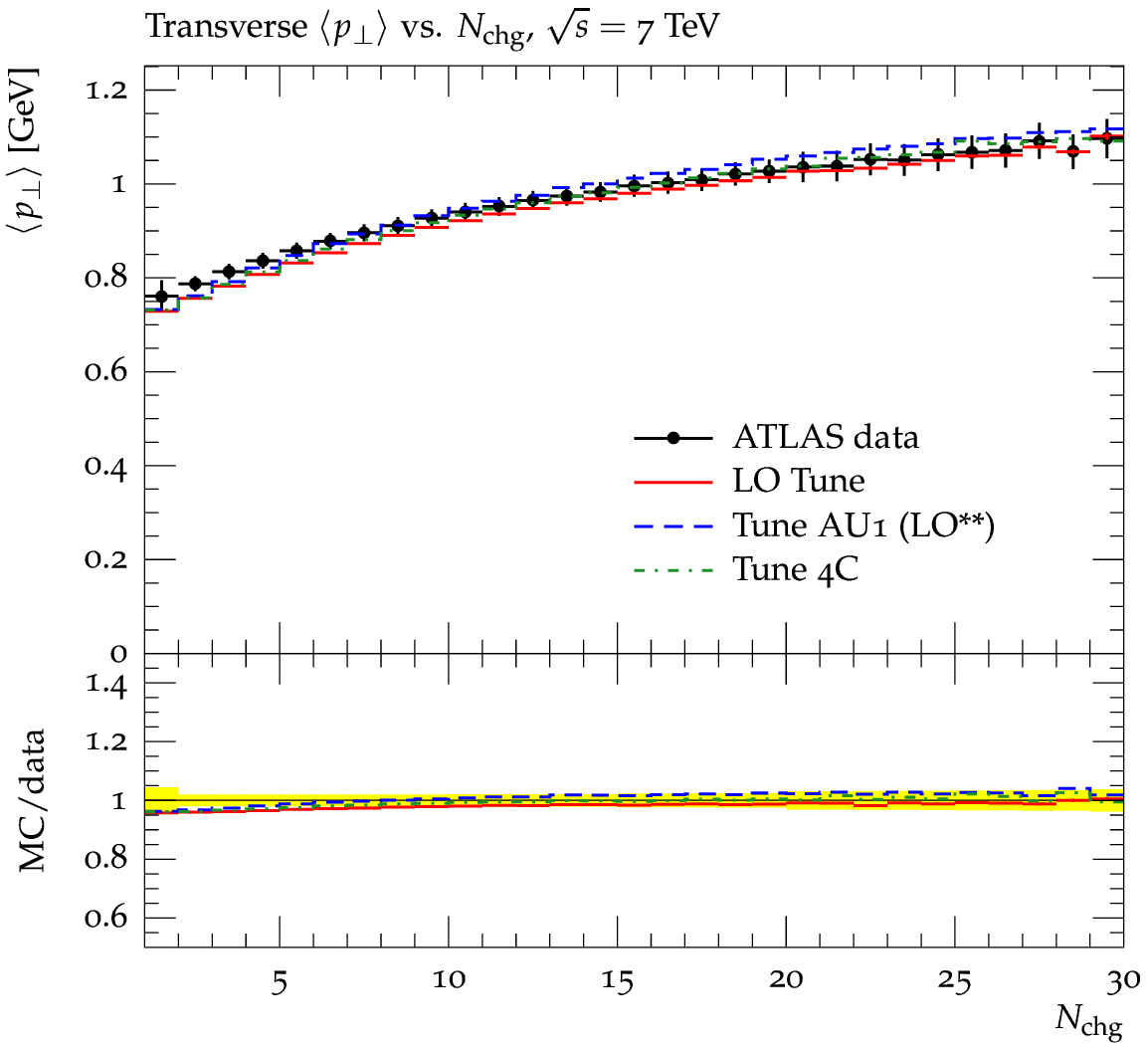}\\
  \includegraphics[width=0.4\textwidth]{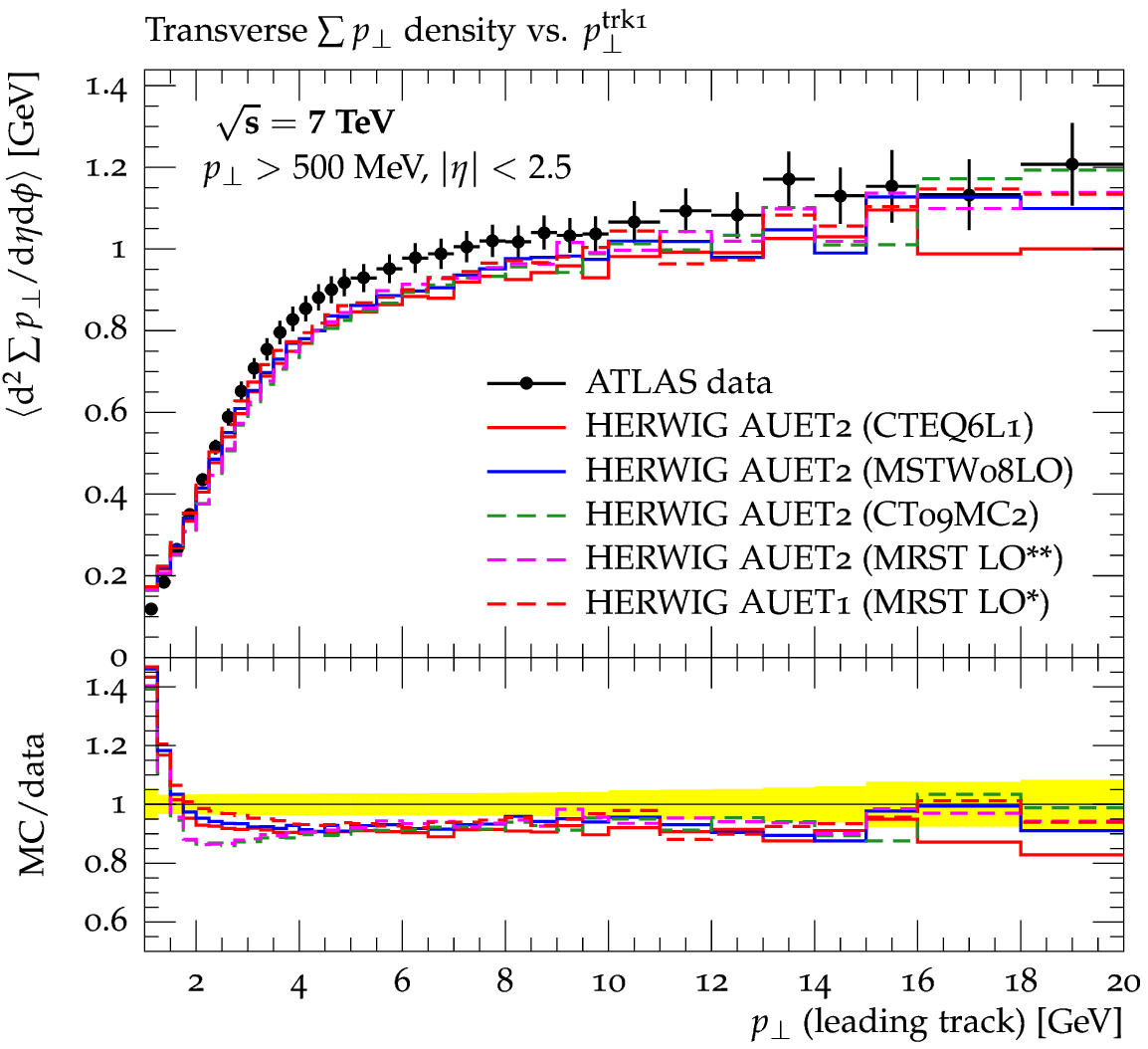}
  \includegraphics[width=0.4\textwidth]{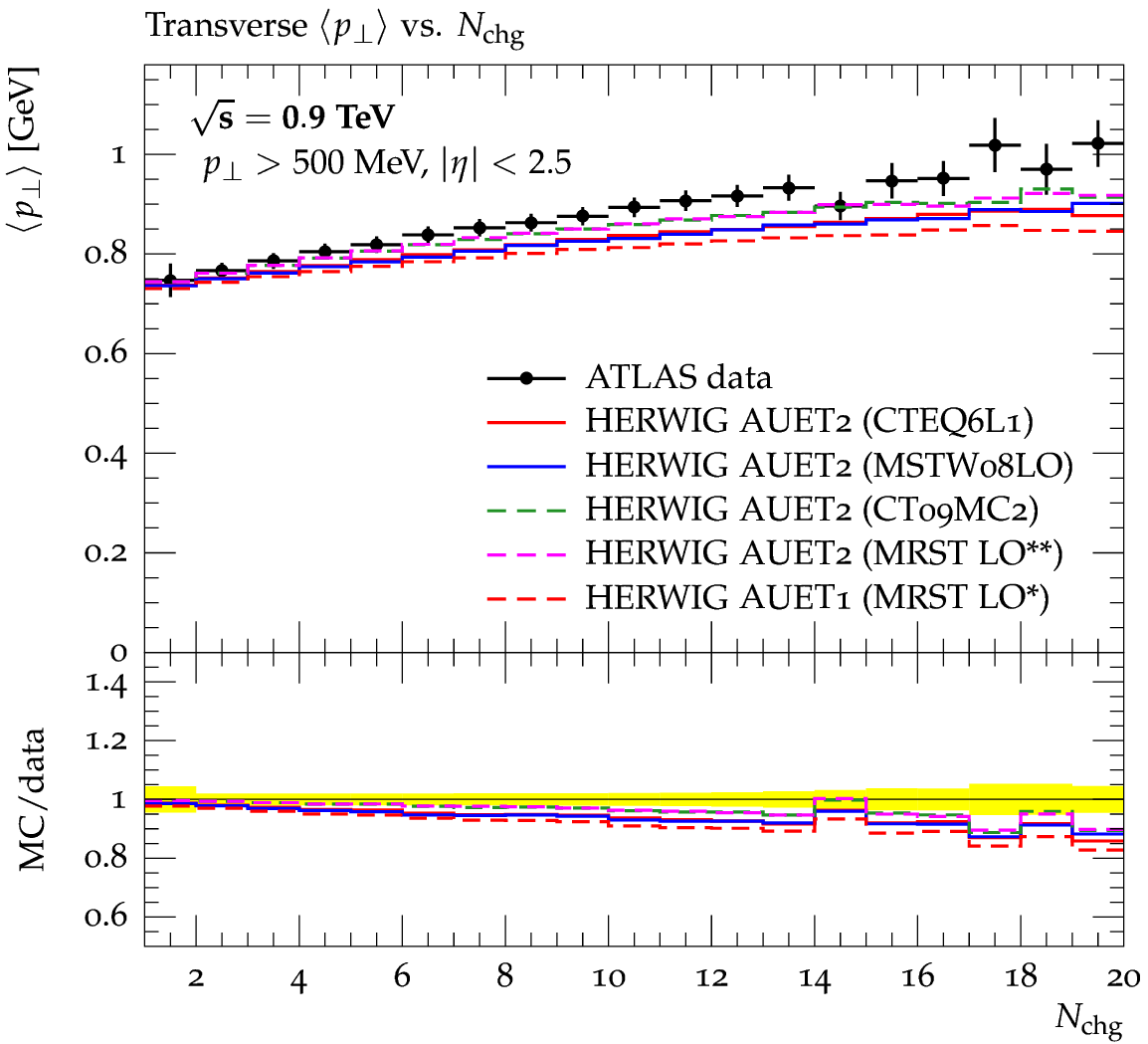}
  \caption{ATLAS underlying event tunes of \pythiaeight and \herwigjimmy,
    compared to the same ATLAS UE data at 7~TeV as shown in earlier figures for
    \pythiasix. The quality of description is again good for \pythiaeight, but
    for \herwigjimmy while the overall level of plateau activity is seen to be
    correct, its modelling both undershoots the turn-over region (bottom left)
    and does not quite capture the necessary balance between charged particle
    multiplicity and \pt (bottom right). These limitations could not be
    addressed by tuning of the \jimmy MPI model, and no further \jimmy tuning
    will be pursued by ATLAS.}
  \label{fig:uetuningpy8jimmy}
\end{figure}

The results of the MPI fitting for all three generators are very strongly
dependent on the PDF being used, since the MPI secondary scatterings in the
models are mostly driven by the nature of the low-$x$ gluon PDF, for $Q^2 \sim
10~\text{GeV}^2$ and $x \sim 10^{-4}$. The modified leading order PDFs, intended
specifically to make LO event generator simulation of hard processes more
``NLO-like'', typically have larger low-$x$ gluon fractions than the standard LO
and NLO PDFs, and hence the \ptmin screening parameter naturally increases to
produce the same level of activity as for a less MPI-active PDF. Plots of the
minimum bias tunes for both Pythia-family generators are shown in
Figures~\ref{fig:mbtuning}, \ref{fig:mbtuningpy6mlo}, and underlying event tunes
for all three generators in Figures~\ref{fig:uetuningpy6},
\ref{fig:uetuningpy6eigen}, and \ref{fig:uetuningpy8jimmy}.

The major results from the latest ATLAS MPI tuning are as follows:
\begin{itemize}
\item A fully consistent description of MB and UE observables could not be
  obtained with either \pythiasix or \pythiaeight. The ATLAS UE observables
  favour slightly more ``active'' MPI parameter configurations than the MB ones
  do. This appears to require a modelling extension, and it is possible that the
  addition of an $x$-dependent proton matter distribution to the latest
  \pythiaeight versions may help to describe this data. The result was that
  separate MB and UE tunes were constructed for each generator, hence the
  distinction between \pythiasix's AUET2B and AMBT2B tunes.
\item For all three generators, distinct groupings of MPI cutoff values were
  discovered, with all mLO PDFs preferring higher values of \ptmin than the
  groups of LO PDFs. This is as expected, but is the first explicit and
  quantitative demonstration of this model behaviour.
\item Surprising features were observed for some minimum bias observables in
  both \pythiasix and \pythiaeight when using mLO PDFs. In particular, an excess
  of MC above data by a factor of 1.8 was seen in the MB track \pt spectrum for
  both \pythiasix and \pythiaeight when using the \lostar, \lostst, and CT09MC2
  PDFs. This effect could not be changed by use of any MPI model parameters, and
  attempts to identify which PDF features were responsible were unable to
  pinpoint any single PDF aspect which was reliably correlated to the anomalous
  behaviours. The decision was made to not use mLO PDFs for minimum bias
  simulation -- while some artefacts were also observed in UE observables, they
  were much less extreme than in MB, to the extent that they should be a very
  minor issue compared to the benefits to the hard process simulation of using
  an mLO PDF.
\item For all generators, leading order PDFs were capable of the best
  description of MPI observables. The CTEQ6L1 PDF was seen to be particularly
  good at describing MPI data with both Pythia generators. This is consistent
  with theory, as there is no motivation for mLO PDFs to improve the description
  of the dijet matrix elements used for MPI scattering simulation, but is again
  the first observation that not all PDF effects can be ``tuned away'' by MPI
  models.
\item No model is currently capable of describing either MB or UE data for
  tracks with \pt below 500~MeV.
\end{itemize}

The AUET2 tunes of \herwigjimmy are the last such tunes which will be
constructed by ATLAS: they provide generally good descriptions of the UE for
signal processes, and cover sufficient PDFs for use in PDF systematics studies
or for combination with NLO generators such as MC@NLO. There is, however, not
enough flexibility in the \jimmy model to describe the detailed structure of UE
observables: essentially the 3 available parameters (\ptmin, $\sqrt{s}$
evolution exponent, and matter radius) all move the UE plateaus up and down but
without any more nuanced changes in shape. The emphasis for MPI models other
than \jimmy will now move to \herwigpp and Sherpa, and \herwig dependence is
itself being phased out of ATLAS production in favour of the newer and more
capable C++ generators. It is notable, however, that despite the simplicity of
the \jimmy model -- no colour reconnection, a very minimal matter distribution
parameterisation, etc. -- it does describe most UE data very well! In
particular, the indication appears to be that the detail of how \ptmin is used
to regularise MPI scattering cross-sections is not particularly crucial, at
least above a track \pt of $\sim 500~\text{MeV}$.

\begin{figure}[t]
  \centering
  \includegraphics[width=0.4\textwidth]{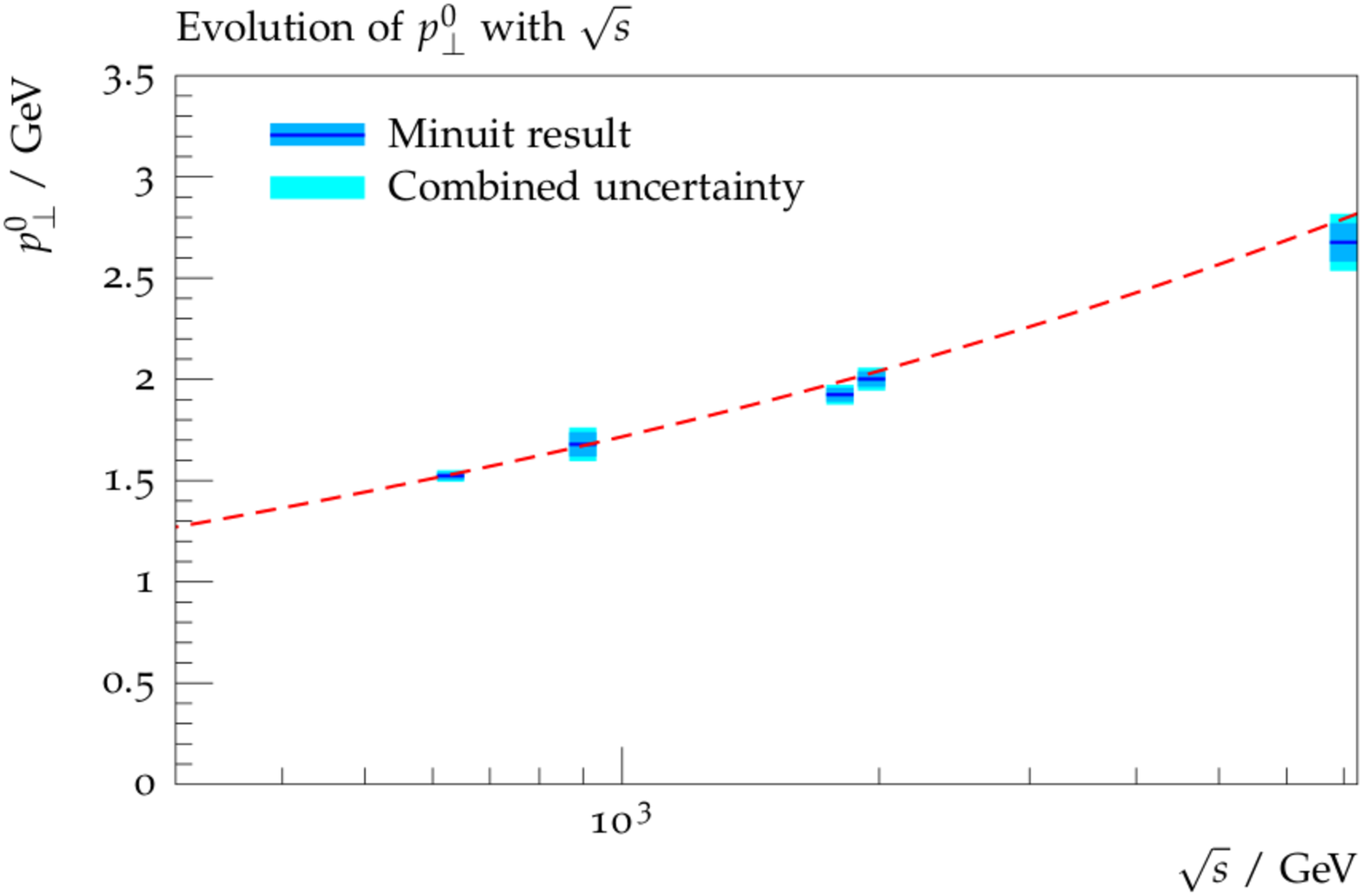}
  \caption{The evolution of tuned values of \ptmin in the \pythiasix MPI model
    as a function of collider centre-of-mass energy $\sqrt{s}$, for minimum bias
    data with a fiducial charged multiplicity cut of $N_\text{ch} \ge 1$. A
    power law form of evolution, as used by the pomeron-inspired ansatz in the
    Pythia family of event generators, gives a remarkable description of
    empirical evolution of the MPI cutoff.}
  \label{fig:schulzskands}
\end{figure}

We should also mention the successfulness of the DL Pomeron-inspired energy
evolution ansatz for minimum bias data. This has been studied by Schulz and
Skands~\cite{Schulz:2011qy}, by using the same tuning machinery as used by ATLAS
to fit values of \ptmin for \pythiasix against all available minimum bias data
at a range of energies. There is no assumption of an energy evolution form in
this study, yet the usual ansatz does give a strikingly high quality fit to this
evolution form, as shown in Figure~\ref{fig:schulzskands}. However, it is not
clear whether underlying event observables are also compatible with this form,
as they certainly do not appear to be compatible with minimum bias description
at the \emph{same} energy with current models.

Finally, we note that the ATLAS measurements of inelastic and diffractive
cross-sections, discussed in Section~\ref{sec:xsec}, have also had an impact on
MPI modelling, although not directly in ATLAS studies. The ``4C'' author tune of
\pythiaeight specifically adds a damping of diffractive cross-section evolution
into its improved inclusive diffraction model, to better describe this and other
data, and this tune is in use in ATLAS production for the simulation of pile-up
minimum bias events.

\section{Conclusions}

We have described the modelling of multi-parton interactions in Monte Carlo
generators, and the measurements and generator tuning activities in the ATLAS
experiment which have been used to increase our understanding and modelling
ability for multiple scattering effects at the LHC. Tunes have been constructed
using a wide range of PDFs for three event generators, \pythiasix, \pythiaeight,
and \herwigjimmy, in addition to ATLAS' use of author-supplied tunes of the
\herwigpp and Sherpa generators. Interesting dependences on PDF details have
been observed, in particular several strong and anomalous effects of MC-adapted
(mLO) PDFs on MPI observables. No model currently describes MPI-dominated data
well below a particle \pt cut of 500~MeV.

The increased constraint on the energy evolution of inclusive MPI will also
assist the LHC program when the centre-of-mass energy is increased to 8, 9,
and/or 14~TeV in the coming years. This improved description of \atlas data has
not been entirely without cost: particularly where ATLAS underlying event data
has been used, there are signs of tension with CDF data and between MB and UE
observables, forcing a split into MB and UE tune families. This matter, and that
of improved hadronisation and jet structure description, will be taken up in the
next set of ATLAS tunes -- including data with greater hard-scattering scales in
UE and including flavour constraints via identified particle data -- and model
developments.

\section*{Acknowledgements}

AB wishes to thank the organisers of the Krakow Theoretical Physics Summer
School for the invitation and hospitality in Zakopane, other members of the
ATLAS soft QCD and MC tuning subgroups for their many contributions to the work
described here. Many thanks also to the Edinburgh University PPE group for
additional support, and the Durham University IPPP and Scottish Universities
Physics Alliance for an Associateship grant and Advanced Fellowship
respectively.

%%%%%%%%%%%%%%%%%%%%%%%%%%%%%%%%%%%%%%%%%%%%%%%%%%%%%%%%%%%%%%%%%%%%%%%%%%%%%%%
% Bibliography
%%%%%%%%%%%%%%%%%%%%%%%%%%%%%%%%%%%%%%%%%%%%%%%%%%%%%%%%%%%%%%%%%%%%%%%%%%%%%%
%
\bibliographystyle{atlasnote}
\bibliography{refs}

\end{document}